\documentclass[fleqn,10pt]{wlscirep}
\usepackage[utf8]{inputenc}
\usepackage[T1]{fontenc}

\usepackage{tabu}

\newcommand {\zz}  { {\bf z} }
\newcommand {\bgg}  { {\bf g} }

\newcommand {\xx}  { {\bf x} }
\renewcommand {\aa}  { {\bf a} }

\newcommand {\yy}  { {\bf y} }

\newcommand {\mA}  { {\bf A} }

\renewcommand{\Pr}{\hbox{\bf{Pr}}}

\newcounter{comment}

\newcommand{\quotes}[1]{``#1''}
\usepackage{physics}
\usepackage{subcaption}
\usepackage[short]{optidef}
\usepackage{svg}
\usepackage{siunitx}
\usepackage{rotating}

\usepackage{setspace}

\usepackage{color, colortbl}
	
\definecolor{LightCyan}{rgb}{0.88,1,1}
\usepackage[linesnumbered,ruled,vlined,onelanguage]{algorithm2e}

\newcommand{\beginsupplement}{%
        \setcounter{table}{0}
        \renewcommand{\thetable}{S\arabic{table}}%
        \setcounter{figure}{0}
        \renewcommand{\thefigure}{S\arabic{figure}}%
     }

\title{Identifying and Analyzing Sepsis States: A Retrospective Study on Patients with Sepsis in ICUs.}

\author[1*]{Chih-Hao Fang}
\author[2]{Vikram Ravindra}
\author[3]{Salma Akhter}
\author[4]{Mohammad Adibuzzaman}
\author[5]{Paul Griffin}
\author[6]{Shankar Subramaniam}
\author[1]{Ananth Grama}
\affil[1*]{Department of Computer Science, Purdue University, West Lafayette, IN, USA}

\affil[2]{Department of Computer Science, University of Cincinnati, Cincinnati, OH, USA}
\affil[3]{Regenstrief Center for Healthcare Engineering, Purdue University, West Lafayette, IN, USA}

\affil[4]{Department of Medical Informatics and Clinical Epidemiology, Oregon Health \& Science University, Portland, Oregon, USA}

\affil[5]{Department of Industrial Engineering, Penn State University, University Park, PA, USA}

\affil[6]{Department of Bioengineering, University of California, San Diego, La Jolla, CA, USA}

\affil[*]{fang150@purdue.edu (CHF); ayg@cs.purdue.edu (AG)}

\begin{abstract}
Sepsis accounts for more than 50\% of hospital deaths, and the associated cost ranks the highest among hospital admissions in the US. Improved understanding of disease states,  progression, severity, and clinical markers has the potential to significantly improve patient outcomes and reduce cost. We develop a computational framework that identifies disease states in sepsis and models disease progression using clinical variables and samples in the MIMIC-III database. We identify six distinct patient states in sepsis, each associated with different manifestations of organ dysfunction. We find that patients in different sepsis states are statistically significantly composed of distinct populations with disparate demographic and comorbidity profiles. Our progression model accurately characterizes the severity level of each pathological trajectory and identifies significant changes in clinical variables \textcolor{black}{and treatment actions} during sepsis state transitions. Collectively, our framework provides a holistic view of sepsis, and our findings provide the basis for future development of clinical trials, prevention, and therapeutic strategies for sepsis.
\end{abstract}
\begin{document}

\flushbottom
\maketitle
\thispagestyle{empty}

\clearpage

\section*{Author Summary}
\textcolor{black}{
Sepsis is a potentially life-threatening condition that occurs when the body's response to an infection damages its own tissues. Sepsis may be misdiagnosed because the patient is not thoroughly assessed or the symptoms are misinterpreted, which can lead to serious health complications or even death. Improved understanding of disease states, progression, severity, and clinical markers has the potential to significantly improve patient outcomes and reduce cost. In this work, we identified distinct states in sepsis in terms of their extremal clinical manifestations, also known as archetypes. We identified six states, each characterized by a unique set of pathological responses that can be mapped back to organ function(s), along with an association between patient attributes and sepsis states. We also find that these states manifest distinct comorbidity profiles before infection. Modeling sepsis progression as a Markov chain, we provide estimates of treatment actions (average amount of fluids, dosage of vasopressors, usage of mechanical ventilators) and the expected state transitions. Overall, by analyzing the relationship between pre-existing comorbidities and sepsis states, changes in clinical measurements, and treatment actions during disease progression, one can prognosticate individuals' outcomes and devise better prevention and therapeutic strategies.}

\begin{figure*}[ht]
   \centering
   \hspace{-2.5cm}
   
    \begin{subfigure}[h]{1.0\textwidth}
        \centering
        \includegraphics[width=18cm]{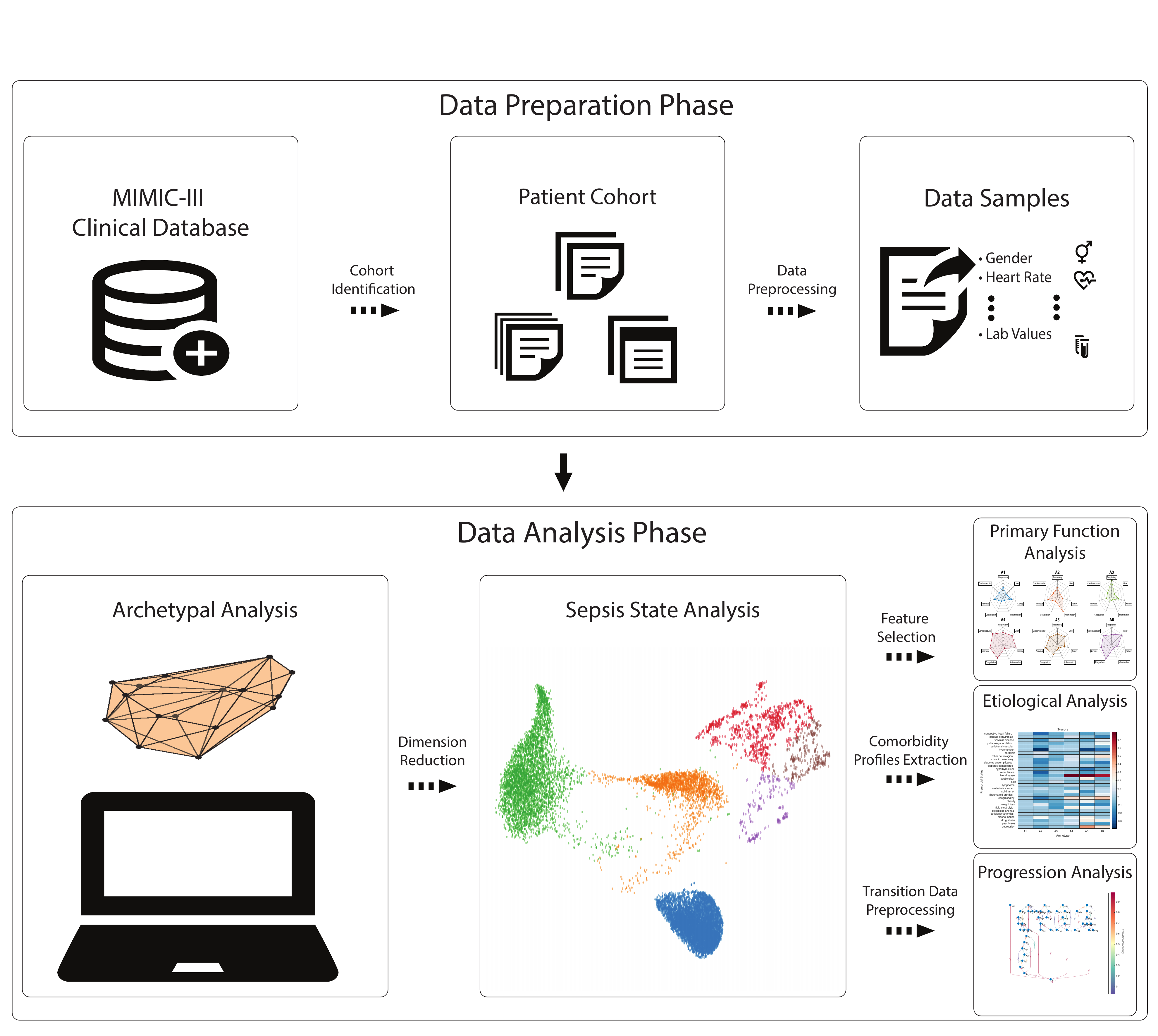}
    \end{subfigure}
    
    \caption{ Illustration of proposed framework: The data preparation phase extracts 42 variables (demographic profiles, vital signs, laboratory tests, mechanical ventilation status of the patients, and the comorbidity profiles) from 16,546 distinct sepsis patients admitted to Beth Israel Deaconess Medical Center from the MIMIC-III database. In the data analysis phase, we use archetypal analysis to find distinct states in sepsis. 
    %We then use a dimension reduction method (UMAP) to visualize the identified sepsis states. 
    We validate that each state corresponds to patient clusters that are statistically distinct from the distribution of the cohort as a whole, and the SOFA score, SIRS score, and mortality rate are calculated to characterize each sepsis state. In primary function analysis, selected features from archetypes are used to identify the \textit{primary functions} (namely, \textit{nervous system, inflammation and infection, liver function, kidney function, coagulation, respiratory function, and, cardiovascular function}) of each sepsis state. In etiological analysis, we find correlation between pre-existing comorbidity profiles (30 types) and sepsis states. Finally, in progression analysis, we use higher-order Markov chains to model the dynamics of pathological processes of sepsis. We then use archetypal analysis to identify distinct types of sepsis state transitions and use z-score analysis to find representative clinical markers of each state transition.} \label{fig:overview}
\end{figure*}
\clearpage
\section{Introduction}

Sepsis accounts for more than 50\% of hospital deaths~\cite{liu2014hospital}, and the cost of sepsis management ranks the highest among hospital admissions for all illnesses in the United States \cite{paoli2018epidemiology}. Key factors in improving patients' outcomes are the early diagnosis of sepsis, and subsequent timely and appropriate treatment actions. While significant progress has been made towards the former\cite{dellinger2013surviving,rhodes2017surviving}, 
with recent development of the Sequential Organ Failure Assessment (SOFA or Quick-SOFA) measure outside the Intensive Care Unit (ICU)\cite{seymour2016assessment,williams2017sirs}, the latter continues to be a significant challenge~\cite{iskander2013sepsis,gotts2016sepsis,marshall2014have}. There have been a number of efforts aimed at classifying  disorders that broadly comprise sepsis, which have resulted in categories such as Systemic Inflammatory Response Syndrome (SIRS), severe sepsis, and septic shock. These, in turn, have resulted in treatment strategies with limited success. More recently, these categorizations have been abandoned, in favor of a more broadly accepted definition of sepsis as a `life-threatening organ dysfunction caused by a dysregulated host response to infection' \cite{singer2016third}. \iffalse This dysfunction is characterized by a SOFA score of two points or more. \fi
Although SOFA score is a more comprehensive measure of the severity of health status of patients with sepsis (Table \ref{tab:SOFA}), and a good predictor of mortality\cite{jentzer2018predictive}, the diverse mechanisms underlying sepsis and how they map to SOFA scores are still not fully understood. This is in spite of significant efforts aimed at developing and deploying new and improved treatments. As a result, current approaches to sepsis treatment are primarily guideline-based, as opposed to relying on clinicians' decision-making capability, when presented with a patient's unique set of clinical variables\cite{dugar2020sepsis}.

A personalized decision process for sepsis must be capable of differentiating heterogeneous response from diverse groups of patients, and understanding the etiology of disease to minimize errors and maximize treatment efficacy. With the goal of motivating research on such personalized decision processes, the Medical Information Mart for Intensive Care version III (MIMIC-III)\cite{johnson2016mimic} database released de-identified clinical data from approximately 46,000 patients admitted to Beth Israel Deaconess Medical Center in Boston, Massachusetts between 2001 and 2012. We use clinical variables in the MIMIC-III database, along with a range of novel algorithmic and statistical constructs for our retrospective study of sepsis states and response (\textcolor{black}{Fig} \ref{fig:overview} ).

\textcolor{black}{We first identify a patient cohort 
using AI Clinician\cite{komorowski2018artificial}
%that satisfies the sepsis-3 criteria\cite{singer2016third} 
from five tertiary ICUs in Boston.} 
%This results in a sample of 16,546 distinct patients with 20,944 ICU admissions. 
%We then extract clinical variables for the cohort, including demographic data, vital signs, lab results, and other information such as the use of a ventilator, and various comorbidities before sepsis infection, to characterize the health status of the patients with sepsis during the ICU stay.} 
\textcolor{black}{AI Clinician \cite{komorowski2018artificial} defines sepsis as the presence of suspicion of infection in conjunction with evidence of organ dysfunction (SOFA score $\geq$ 2) 48 hours before and up to 24 hours after onset of infection. We then extract clinical variables, including vital signs, lab results, and the use of a ventilator, from 24 hours before and up to 48 hours after the onset of infection, as well as demographic data and various comorbidities before sepsis infection, to characterize the health status of the patients with sepsis during the ICU stay. This results in a sample of 16,546 distinct patients from five tertiary ICUs in Boston  with 20,944 ICU stays.} We summarize this data in Table \ref{tab:variables}.

\begin{table}[]
\centering
\vspace{-1cm}
\large
\caption{Description of the cohort.} \label{tab:variables}

\begin{tabular}{llll}

\hline
\textbf{Demographic} & \textbf{Type or Unit} & \textbf{Normal Range}                              & \textbf{ \textcolor{black}{Mean (std) Value} } \\ \hline
Age                  & years                 & N/A                                                & 64.57 (16.67)                     \\ \hline
Gender               & binary                & 1 = Female, 0 = Male                               & 0.44 (0.50)                      \\ \hline
\textbf{Vitals}      & \textbf{}             & \textbf{}                                          & \textbf{}                            \\ \hline
HR                   & bpm                   & 60 - 100                                           & 87.21 (16.84)                     \\ \hline
SysBP                & mmHg                  & $\leq$ 120                            & 119.92 (20.35)                   \\ \hline
MeanBP               & mmHg                  & 70 - 100                                           & 78.21 (13.48)                    \\ \hline
DiaBP                & mmHg                  &  80                             & 57.12 (13.32)                    \\ \hline
Temp                 & Celsius               & 36.5 - 37.5                                        & 36.91 (2.01)                     \\ \hline
RR                   & bpm                   & 12 - 20                                            & 20.21 (5.19)                     \\ \hline
\textbf{Lab Values}  & \textbf{}             & \textbf{}                                          & \textbf{}                            \\ \hline
GCS                  & N/A                   & 15                                                  & 12.57 (3.49)                     \\ \hline
SpO2                 & percent               & 95 - 100                                           & 96.91 (2.65)                     \\ \hline
FiO2                 & fraction              & 21\% inhaled from natural air                     & 0.46 (0.18)                      \\ \hline
Potassium            & mEq/L                 & 3.5 - 5.0                                          & 4.08 (0.56)                      \\ \hline
Sodium               & mEq/L                 & 135 - 145                                          & 138.69 (4.89)                    \\ \hline
Chloride             & mEq/L                 & 96 - 106                                           & 104.72 (6.24)                    \\ \hline
Glucose              & mg/dL                  & 80 - 130                                           & 138.96 (51.18)                   \\ \hline
BUN                  & mg/dL                  & 7 - 20                                             & 29.25 (22.56)                    \\ \hline
Creatinine           & mg/dL                  & 0.6 (0.5) -1.2 (1.1) \textcolor{black}{M (F)}      & 1.49 (2.16)                      \\ \hline
Magnesium            & mg/dL                  & 1.5 - 2.5                                          & 2.06 (0.35)                       \\ \hline
Calcium              & mg/dL                  & 8.8 - 10.7                                         & 8.31 (0.80)                      \\ \hline
Ionised Ca           & mmol/L                  & 1.16 - 1.32                                        & 1.13 (0.12)                      \\ \hline
CO2                  & mEq/L                 & 23 - 29                                            & 25.82 (5.66)                      \\ \hline
SGOT                 & u/L                   & 5 - 40                                             & 155.69 (583.57)                  \\ \hline
SGPT                 & u/L                   & 7 - 56                                             & 583.57 (466.35)                  \\ \hline
Total Bilirubin      & mg/dL                 & 0.1 - 1.2                                          & 2.41 (5.15)                      \\ \hline
Albumin              & g/dL                  & 3.4 - 5.4                                          & 3.00 (0.68)                      \\ \hline
Hb                   & g/dL                  & 13.5 (12.0) - 17.5 (15.5) \textcolor{black} {M (F)} & 10.30 (1.74)                     \\ \hline
WBC                  & $\times$ $10^9$ /L                  & 4.5 - 11.0                                         & 12.27 (8.29)                     \\ \hline
Platelets            & $\times$ $10^9$ /L                  & 150 - 450                                          & 228.70 (139.19)                  \\ \hline
aPTT                  & s                     & 30 - 40                                            & 37.81 (19.34)                    \\ \hline
PT                   & s                     & 11 - 13.5                                          & 16.19 (6.75)                      \\ \hline
INR                  & N/A                   & $\leq$ 1.1                            & 1.51 (0.84)                       \\ \hline
Arterial PH          & N/A                   & 7.35 - 7.45                                        & 7.39 (0.07)                      \\ \hline
PaO2                 & mmHg                  & 80-100                                             & 125.15 (72.45)                   \\ \hline
PaCO2                & mmHg                  & 35-45                                              & 41.95 (10.81)                    \\ \hline
Arterial BE          & mEq/L                 & -2 - +2                                            & 0.35 (5.02)                      \\ \hline
Arterial lactate     & mmol/L                & 0.5-1                                              & 2.05 (1.66)                      \\ \hline
HCO3                 & mEq/L                 & 22 - 28                                            & 24.67 (5.09)                     \\ \hline
Shock Index          & bpm/mmHg              & 0.5 - 0.7                                          & 0.75 (0.20)                      \\ \hline
PaO2/FiO2            & mmHg                  & $>$ 500 at sea level                      & 311.00 (223.71)                  \\ \hline
\textbf{Others}      & \textbf{}             & \textbf{}                                          & \textbf{}                            \\ \hline
Weight               & kg                    & N/A                                                & 83.23 (24.65)                    \\ \hline
Mechvent             & binary                & 0 = False, 1 = True                                & 0.37 (0.48)                      \\ \hline
Comorbidity Count             & Integer                & 0 - 30                               & 4.01 (2.17)                      \\ \hline
\end{tabular}\\
\footnotesize{ All of the normal ranges presented apply to adults. HR, Heart Rate; SysBP, Systolic Blood Pressure; MeanBP, Mean Blood Pressure; DiaBP, Diastolic Blood Pressure; Temp, Temperature; GCS, Glasgow Coma Scale; RR, Respiratory Rate; BUN, Blood Urea Nitrogen; SGOT, Serum Glutamic-Oxaloacetic Transaminase; SGPT, Serum Glutamic Pyruvic Transaminase; Hb, Hemoglobin; WBC, White Blood Cells; PTT, Partial Thromboplastin Time; PT, Prothrombin Time; INR, International Normalized Ratio, Arterial BE, Arterial Base Excess. }

\end{table}

\clearpage

We develop a novel mathematical framework that: (i) identifies distinct sepsis states using archetypal analysis; (ii) extracts representative sets of features from clinical variables to differentiate sepsis states and identifies associated biomarkers that can be mapped back to organ function(s); (iii) analyzes relationships between sepsis states, demographic variables, and comorbidities pre- and post-infection; and (iv) models sepsis progression using a higher-order Markov chain and identifies significant changes in clinical variables \textcolor{black}{and treatment actions} during sepsis state transitions. \textcolor{black}{We demonstrate that our framework identifies distinct sepsis states -- each state characterized by a unique set of pathological responses that can be mapped back to organ function(s) and an association between patient attributes and sepsis states. We also find that these states manifest distinct comorbidity profiles before infection. Moreover, our computational framework also provides insights into understanding the pathological processes in sepsis. Our state transition graphs provide an estimate of treatment actions (average amount of fluids, the dosage of vasopressors, the usage of mechanical ventilators), and the expectation of state transitions. Finally, our computational framework  identified distinct types of sepsis state transitions, each characterized by a different set of clinical transition biomarkers. By analyzing the relationship between pre-existing comorbidities and sepsis states, changes in clinical measurements \textcolor{black}{ and treatment actions} during disease progression, one can prognosticate individuals' outcomes and devise prevention and therapeutic strategies.
}

\section*{Results}
\subsection*{Identifying distinct sepsis states from the cohort.}

\subsubsection*{Archetypal analysis of sepsis cohort.} \label{sec:AA_cohort}

We pose the following important question: \textit{do there exist distinct states of sepsis with different clinical manifestations, recovery rates, demographic and pathological characteristics, and is it possible to identify these states from patient clinical measurements?} We formulate this problem as one of finding archetypes (representatives of states) of sepsis, and design powerful mathematical models and methods for solving this problem. A geometric interpretation of our approach is to view each \textcolor{black}{patient as a point} \textcolor{black}{characterized by clinical manifestations} in a high dimensional space of attributes, and archetypes as corners of a convex hull in this high dimensional space. Within this representation, each data point can be approximated as a linear combination of the archetypes. Since archetypes form a convex polytope, the coefficients in the linear combinations sum to one (convex combinations). 

This formulation has several advantages over traditional clustering techniques (e.g., $k$-means). Archetypes represent extremal or pure states -- to this end, they have clear clinical interpretations. Second, each convex combination has a well-characterized interpretation as a mixture of pure states. Finally, descriptors of archetypes may themselves be processed to identify clinical markers of pure states. \textcolor{black}{The elbow method was used to determine the number of archetypes in the dataset. Specifically, we measured how well the archetypes and the coefficient matrix approximate the original data with respect to the Frobenius norm and chose the elbow point as the optimal number of archetypes. Using this procedure, shown in Fig \ref{fig:elbow} and Fig \ref{fig:archetypes} in Appendix, we discover six distinct states in sepsis among our cohort. Since archetypes represent extreme sepsis states, in the rest of this discussion, we use the terms ``sepsis state" and ``archetype" interchangeably.}

\begin{figure}[h]
 %\vspace{3.5cm}
   \centering
   
    \begin{subfigure}[h]{1.0\textwidth}
    \hspace{0.5cm}
    \includegraphics[width=18cm]{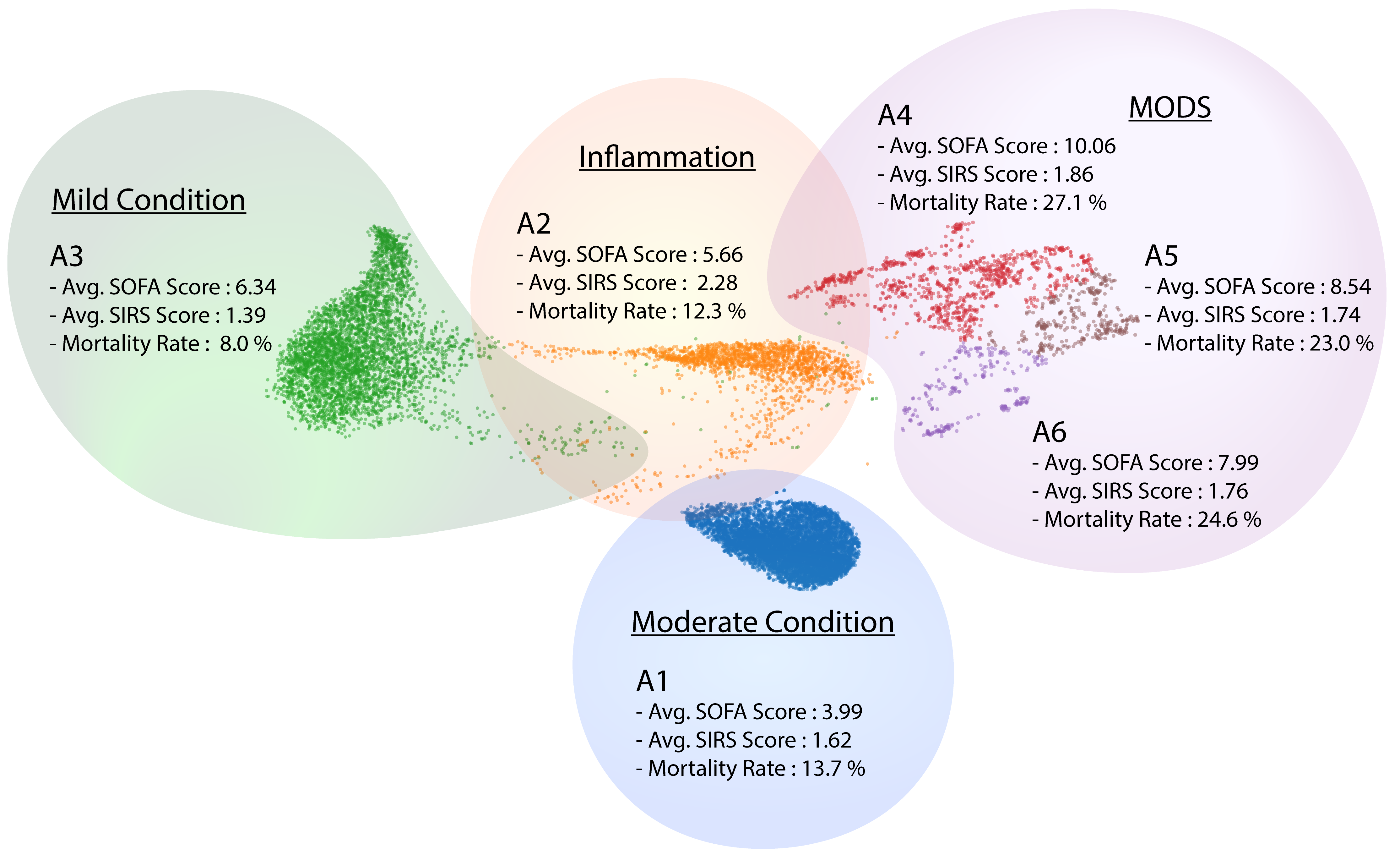}
    \end{subfigure}
    
    \caption{ Visualization (using low-dimensional UMAP embedding) of the six derived sepsis states. Colors represent different sepsis states. \textcolor{black}{The average SOFA score, average SIRS score, and mortality rate are used to characterize sepsis states.} Based on these scores, we characterize states A1 (blue), A2 (orange), and A3 (green) as `moderate condition', `inflammation', and `mild condition', respectively, and we characterize states A4 (red), A5 (brown), and A6 (black) as `Multiple Organ Dysfunction Syndrome (MODS)'. }
    \label{fig:embed}
\end{figure}

\textcolor{black}{Fig} \ref{fig:embed} shows a uniform manifold approximation and projection (UMAP) embedding of the data points, along with the archetypes (represented by colors). Note that archetypes (A1 through A6) do not appear as corners of the convex polytope since this is a two-dimensional embedding of a higher dimensional attribute space.

\textcolor{black}{To ascertain that the identified clusters are distinct, we apply a statistical test to validate that the probability distributions corresponding to these groups are significantly different from the distribution of the cohort as a whole. We also test to ensure that the probability distribution of each group is significantly different from others, as characterized by a multivariate analysis of variance (MANOVA) procedure (please see Methods Section \ref{sec:significant-testing} for more details ). Statistics of pairwise Hotelling t-square test across sepsis states and two-sample t-test for each variable for each sepsis state compared to overall populations are shown in \textcolor{black}{Table} \ref{fig:t2-test} and \textcolor{black}{Table} \ref{fig:t-test} in Appendix, respectively. The pairwise Hotelling t-square test across sepsis states demonstrates that sepsis states are significantly different from each other, and the two-sample t-test demonstrates that most of the clinical variables are significantly different from the overall populations. Since there is no apparent separation between the MODS group in the embedding space, shown in  \textcolor{black}{Fig} ~\ref{fig:embed}, we further applied two-sample t-tests for each variable between the MODS group, shown in Table \ref{fig:t-test2} in Appendix. In general, SGOPT and SGPT are significantly different, which are the dominant signals in the MODS group. A distinct subset of clinical variables is significantly different between the MODS group. Specifically, besides SGOT and SGPT, age, meanBP, potassium, calcium, total bilirubin, Hb, white blood cell count, INR, arterial lactate are significantly different (p-values $<$ 0.001) between A4 and A5; SGOT, potassium, glucose, and white blood cell count are significantly different (p-values $<$ 0.001)  between A5 and A6; SGOT, SGPT, gender, GCS, FiO2, glucose, BUN, Magnesium, Calcium, PT, and INR are significantly different (p-values $<$ 0.001)  between A4 and A6.}

\begin{table*}[]
\caption{Statistics of the clinical  variables for each sepsis state. \textcolor{black}{Clinical variables that are selected from our feature selections methods are highlighted with light cyan.} \textcolor{black}{Note that the average age, gender, and comorbidity count for each state are calculated as the average age, gender, and comorbidity count among the patients who have passed through the sepsis state of interest.} }\label{tab:archetype-stats}
%\centering
\scriptsize

\hspace{-0.4cm}
\begin{tabular}{llllllll}
\hline
\textbf{Demographic} & \textbf{Type or Unit} & \textbf{A1 Mean (std)} & \textbf{A2 Mean (std)} & \textbf{A3 Mean (std)} & \textbf{A4 Mean (std)} & \textbf{A5 Mean (std)} & \textbf{A6 Mean (std)} \\ \hline
  \rowcolor{LightCyan} Age                  & years                 & \textcolor{black}{64.79 (16.65)}        & \textcolor{black}{59.55 (17.85)}      & \textcolor{black}{65.53 (16.67)}      & \textcolor{black}{64.09 (16.82)}      & \textcolor{black}{58.66
  (19.82)}       & \textcolor{black}{60.16 (20.12)}        \\ \hline
Gender               & binary                & \textcolor{black}{0.44 (0.50)}        & \textcolor{black}{0.38 (0.49)}        & \textcolor{black}{0.47 (0.50)}        & \textcolor{black}{0.46 (0.50)}        & \textcolor{black}{0.47 (0.50)}        & \textcolor{black}{0.55 (0.50)}         \\ \hline
\textbf{Vitals}      & \textbf{}             & \textbf{}              & \textbf{}              &                        &                        &                        &                        \\ \hline
\rowcolor{LightCyan} HR                   & bpm                   & 87.17 (16.80)      & 95.57 (18.36)       & 84.94 (16.36)      & 88.73 (19.21)      & 90.94 (19.25)      & 89.28(19.37)       \\ \hline
SysBP                & mmHg                  & 119.90 (20.35)     & 121.42 (19.55)     & 120.77 (19.93)     & 119.37 (22.40)       & 119.83 (21.39)     & 119.58 (20.94)      \\ \hline
MeanBP               & mmHg                  & 78.19 (13.47)      & 80.05 (14.24)      & 78.87 (13.91)      & 77.22 (15.29)      & 81.16 (15.25)      & 78.13(13.36)       \\ \hline
\rowcolor{LightCyan} DiaBP                & mmHg                  & 57.07 (13.29)       & 59.55 (13.86)       & 58.31 (13.83)       & 58.07 (15.11)      & 60.46 (15.31)      & 56.99 (13.89)       \\ \hline
Temp                 & Celsius               & 36.9064 (1.78)       & 37.37 (11.04)       & 36.75 (1.42)       & 36.62 (3.08)       & 36.99 (0.98)       & 36.86 (0.94)         \\ \hline
RR                   & bpm                   & 20.2032 (5.18)       & 21.89 (6.29)        & 19.50 (4.74)       & 20.67 (6.00)       & 21.00 (5.60)        & 20.90 (5.19)        \\ \hline
\textbf{Lab Values}  & \textbf{}             & \textbf{}              & \textbf{}              &                        &                        &                        &                        \\ \hline
\rowcolor{LightCyan} GCS                  & N/A                   & 12.56 (3.45)        & 12.24 (3.53)       & 13.96 (2.39)       & 10.67 (4.76)       & 11.20 (4.44)       & 11.77 (4.03)       \\ \hline
SpO2                 & percent               & 96.91 (2.64)        & 97.18 (2.48)       & 97.08 (2.17)        & 95.98 (4.80)       & 96.23 (4.31)        & 96.00 (5.94)       \\ \hline
\rowcolor{LightCyan} FiO2                 & Fraction              & 0.46 (0.18)          & 0.47 (0.19)        & 0.28 (0.07)        & 0.52 (0.21)        & 0.48 (0.22)         & 0.47 (0.22)        \\ \hline
Sodium               & mEq/L                 & 138.70 (4.89)      & 138.40 (5.06)      & 138.15 (4.40)      & 138.94 (5.48)      & 138.55 (4.53)      & 138.56 (5.50)      \\ \hline
Chloride             & mEq/L                 & 104.75 (6.25)      & 103.46 (5.86)       & 104.46 (5.75)      & 102.49 (7.53)      & 103.45 (7.39)      & 103.79 (7.07)      \\ \hline
Potassium            & mEq/L                 & 4.08 (0.57)        & 4.22 (0.60)        & 4.14 (0.63)        & 4.31 (0.78)        & 4.06 (0.67)        & 4.34 (0.70)        \\ \hline
\rowcolor{LightCyan} Glucose              & mg/dl                  & 139.0001 (51.01)     & 133.47 (45.00)     & 137.65 (55.98)     & 150.89 (77.09)      & 138.89 (63.89)       & 120.54 (38.26)     \\ \hline
\rowcolor{LightCyan} BUN                  & mg/dl                  & 29.31 (22.62)      & 24.96 (19.83)       & 26.68 (20.66)       & 33.63 (20.58)      & 26.26 (17.48)      & 29.35 (18.57)      \\ \hline
\rowcolor{LightCyan} Creatinine           & mg/dl                  & 1.49 (2.18)        & 1.02 (0.87)        & 1.50 (1.62)         & 2.05 (1.59)        & 2.00 (1.85)         & 2.01 (2.00)        \\ \hline
Magnesium            & mg/dl                  & 2.06 (0.35)        & 2.07 (0.31)        & 2.02 (0.33)        & 2.10 (0.39)        & 2.04 (0.42)        & 2.29 (1.09)        \\ \hline
Calcium              & mg/dl                  & 8.31 (0.80)        & 8.39 (0.76)        & 8.38 (0.77)        & 8.50 (1.18)        & 8.13 (0.87)        & 8.20 (0.86)         \\ \hline
Ionised Ca           & mmol/L                  & 1.1317 (0.1223)        & 1.1315 (0.09)        & 1.14 (0.10)        & 1.08 (0.13)        & 1.07 (0.11)          & 1.09 (0.13)        \\ \hline
CO2                  & mEq/L                 & 25.85 (5.67)       & 26.41 (5.46)       & 24.81 (4.64)       & 24.17 (6.83)       & 24.70 (6.49)       & 23.24 (6.10)       \\ \hline
\rowcolor{LightCyan} SGOT                 & u/L                   & 121.1987 (319.22)    & 115.25 (328.74)    & 97.13 (281.49)      & 6.56 $\times$ $10^3$ (1.49 $\times$ $10^3$)    & 2.01 $\times$ $10^3$ (1.76 $\times$ $10^3$)    & 7.66 $\times$ $10^3$ (1.43 $\times$ $10^3$)    \\ \hline
\rowcolor{LightCyan} SGPT                 & u/L                   & 104.71 (293.00)      & 103.37 (292.21)    & 81.68 (254.66)     & 3.02 $\times$ $10^3$ (1.20 $\times$ $10^3$)    & 6.39 $\times$ $10^3$ (1.52 $\times$ $10^3$)    & 6.69 $\times$ $10^3$ (1.15 $\times$ $10^3$)    \\ \hline
Total Bilirubin      & mg/dL                 & 2.41 (5.16)        & 1.93 (4.48)        & 1.95 (4.22)        & 5.37 (5.77)        & 3.37 (2.20)        & 4.56 (6.35)        \\ \hline
Albumin              & g/dL                  & 3.00 (0.68)                 & 2.87 (0.67)        & 3.20 (0.70)         & 2.97 (0.62)         & 3.02 (0.69)         & 2.95 (0.57)        \\ \hline
Hb                   & g/dL                  & 10.30 (1.73)        & 9.49 (1.51)         & 10.49 (1.87)       & 10.42 (1.75)         & 10.90 (1.96)       & 10.75 (1.93)       \\ \hline
\rowcolor{LightCyan} WBC                  & $\times$ $10^9$ /L                  & 12.23 (8.22)        & 20.70 (15.47)       & 10.95 (6.33)       & 13.12 (7.94)       & 11.24 (4.94)       & 13.01 (6.89)       \\ \hline
\rowcolor{LightCyan} Platelets            & $\times$ $10^9$ /L                  & 223.79 (126.04)    & 905.58 (158.96)    & 229.67 (130.25)    & 185.09 (121.61)    & 189.93 (104.53)     & 194.62 (127.32)    \\ \hline
\rowcolor{LightCyan} aPTT                  & s                     & 37.78 (19.30)      & 37.52 (18.74)      & 37.84 (20.27)       & 44.14 (23.64)      & 41.61 (23.91)      & 43.14 (21.21)      \\ \hline
\rowcolor{LightCyan} PT                   & s                     & 16.1654 (6.70)       & 15.91 (5.10)       & 15.88 (6.33)       & 20.68 (10.91)      & 21.93 (11.96)      & 25.94 (19.01)      \\ \hline
\rowcolor{LightCyan} INR                  & N/A                   & 1.51 (0.83)        & 1.46 (0.61)        & 1.48 (0.79)        & 2.08 (1.32)         & 2.50 (2.01)        & 2.96 (2.86)        \\ \hline
Arterial PH          & N/A                   & 7.39 (0.07)        & 7.40 (0.08)         & 7.39 (0.07)        & 7.35 (0.10)        & 7.37 (0.10)         & 7.36 (0.10)        \\ \hline
PaCO2                & mmHg                  & 41.97 (10.83)       & 41.78 (10.70)      & 40.87 (8.69)        & 41.27 (11.78)      & 39.21 (10.65)      & 39.86 (11.64)       \\ \hline
\rowcolor{LightCyan} PaO2                 & mmHg                  & 120.83 (64.14)     & 121.77 (60.94)     & 381.02 (79.38)     & 126.95 (72.86)      & 120.98 (70.09)      & 117.92 (50.43)     \\ \hline
Arterial BE          & mEq/L                 & 0.36 (5.024)        & 0.80 (5.21)        & 0.37 (3.95)         & -1.92 (6.64)        & -1.94 (5.88)       & -2.46 (6.65)       \\ \hline
\rowcolor{LightCyan} Arterial lactate     & mmol/L                & 2.04 (1.60)        & 1.88 (1.46)         & 2.10 (1.57)        & 5.50 (5.36)         & 3.95 (3.97)        & 4.58 (4.59)         \\ \hline
HCO3                 & mEq/L                 & 24.68 (5.08)        & 25.06 (5.00)       & 24.47 (4.74)       & 22.79 (6.13)       & 23.08 (5.66)       & 21.93 (5.77)       \\ \hline
Shock Index          & bpm/mmHg              & 0.75 (0.20)        & 0.81 (0.20)        & 0.72 (0.19)         & 0.77 (0.24)        & 0.79 (0.24)        & 0.77 (0.23)        \\ \hline
\rowcolor{LightCyan} PaO2/FiO2            & mmHg                  & 292.9807 (172.17)     & 294.59 (179.67)    & 1.38 $\times$ $10^3$ (300.05)    & 285.03 (205.49)     & 318.13 (272.09)    & 301.85 (164.02)    \\ \hline
\textbf{Others}      & \textbf{}             & \textbf{}              & \textbf{}              &                        &                        &                        &                        \\ \hline
\rowcolor{LightCyan} Weight               & kg                    & 83.29 (24.65)      & 81.95 (24.44)       & 79.28 (22.80)       & 85.18 (28.70)       & 89.30 (35.93)       & 82.31 (24.22)      \\ \hline
\rowcolor{LightCyan} Mechvent             & binary                & 0.37 (0.48)         & 0.33 (0.47)        & 0.09 (0.29)         & 0.56 (0.50)         & 0.50 (0.50)        & 0.41 (0.49)        \\ \hline
\rowcolor{LightCyan} Comorbidity Count           & Integer                   & \textcolor{black}{3.93 (2.16)}        & \textcolor{black}{3.36 (2.00)}        & \textcolor{black}{3.75 (2.07)}         & \textcolor{black}{4.32 (2.18)}         & \textcolor{black}{4.04 (2.10)}         & \textcolor{black}{4.08 (1.99)}        \\ \hline
\end{tabular}
\end{table*}
%\end{sidewaystable}

\subsubsection*{Characterizing sepsis states : SOFA score, SIRS score, and mortality rate. \label{sec:standard-metric}}

\textcolor{black}{
We use the SOFA score, SIRS score, and mortality rate as measures to characterize sepsis states (please see Table \ref{tab:archetype-stats} in the Appendix for detailed statistics for each clinical variable in each sepsis state.). Since SOFA (and SIRS) scores vary over time, we use the average SOFA (and SIRS) scores for each state. 
%as the ratio of the summation of SOFA (and SIRS) scores in each state to the total samples in each state to characterize the sepsis state. 
We define mortality rate for each state as the rate of mortality among the patients who have passed through the sepsis state of interest. Stated otherwise, if a patient transitions from state A1 to A2, back to A1, and then dies, we only attribute this once to states A1 and A2, even though the patient visited state A1 twice.}

Based on the SIRS criteria, a patient with SIRS score higher than two is diagnosed with sepsis infection (please see full discussion in Appendix section \ref{sec:sofa_n_sirs}.). This definition of sepsis is mainly focused on signs of inflammation exhibited by patients. We find that, among the sepsis states, only state A2 satisfied the SIRS criteria. Consequently, we identify state A2 as primarily representing inflammatory response. \textcolor{black}{According to the sepsis-3 criteria \cite{singer2016third}, sepsis is defined as having a change of SOFA score higher than two in the presence of clinical suspicion of infection (as indicated by the ordering of IV antibiotics and blood cultures), and the higher the score, the more severe the condition.} 
%\textcolor{black}{SOFA score measures the degree of organ dysfunction in the body, and the higher the score is, the more severe the condition is.} 
It is reported that patients who developed Multiple Organ Dysfunction Syndrome (MODS) display significantly higher mortality rate \cite{sterling2017impact,gotts2016sepsis}.We observe that the mortality rate, as well as the SOFA scores of state A4, A5, and A6 are significantly higher than the other types. Thus, we hypothesize that the A4, A5, and A6 represent MODS with heterogeneous organ dysfunction. Compared to the MODS states, states A1 and A3 display lower mortality rates and SOFA scores, with A3 having the lowest mortality rate and SOFA score. Therefore, we characterize states A1 and A3 as `moderate condition' and `mild condition', respectively.  

%{\bf ayg: We call A1 mild and A3 better. What does "better" mean? Is A3 better than A1? If so, can we call A3 mild and A1 moderate, as Adib suggests?}
%\textcolor{black}{CHF: done.}

%\subsubsection{Importance of characterizing sepsis and its states.}

A restrictive definition of sepsis has significant adverse implications for diagnosis and treatment. 
%Sepsis was first defined as systemic inflammatory response syndrome (SIRS) \cite{bone1992definitions}. However, the 
The SIRS metric has been criticized for its inability to identify all possible host responses for sepsis since the SIRS criteria focuses solely on inflammatory excess; hence it is an inaccurate predictor for mortality. This diagnostic metric for sepsis was replaced by sepsis-2, and eventually by sepsis-3. Sepsis-3 uses the SOFA score to characterize the health status of patients with sepsis. It has been shown to be a more accurate predictor of mortality \cite{ferreira2001serial}, compared to SIRS and sepsis-2. Our results support the arguments against the SIRS metric, and reinforce the use of SOFA scores for severity of sepsis infections. As mentioned earlier, only state A2 in our cohort qualified as a sepsis infection based on SIRS scores -- leading to potentially inadequate care for other sepsis states. Our analysis demonstrates that SOFA scores correlate well with our identified states, and that the severity and mortality rate for identified states correlates well with their SOFA sores. However, as we show in the rest of this study, SOFA score alone does not capture the diversity of sepsis states -- motivating our multidimensional approach based on archetypal analysis.

\subsubsection*{On the generalizability of archetypes.}

\textcolor{black}{We demonstrate the generalizability of archetypal analysis. We first generate ``ground truth'' sepsis states by computing corresponding archetypes using the entire set of samples. We then characterize the state of each sample by assigning it to the closest archetype (or the highest coefficient in the convex combination). Next, we divide the entire set of samples into training sets (90 \%) and test sets (10 \%). We show that: i) the computed archetypes using training sets are very close to the ``ground truth'' archetypes; and ii) the cluster assignment for samples in the test sets using the archetypes from training sets are very similar to the cluster labels using ``ground truth'' archetypes. We run archetypal analysis with 1000 iterations and 20 repetitions for both cases, where the entire samples and training sets are used for statistically stable results. We report: i) relative errors of sepsis states, as measured by $\sum^{6}_{i=1} \frac{\norm{\mA_i -\mA^{\textit{train}}_{i}}_{2}}{\norm{\mA_i}}$, where $\mA$ and $\mA^{\textit{train}}$ denote the ``ground truth'' archetypes and the archetypes computed from training set; ii) cluster assignment accuracy on the test set; and iii) errors in SOFA scores, SIRS scores, and mortality rates on the test set. We find that the computed archetypes using training sets are very close to the ``ground truth'' archetypes (averaged relative error over 20 repetitions is 0.0180), that the cluster assignment using archetypes from the training set is consistent with the ground truth assignment ( averaged cluster assignment accuracy over 20 repetitions on tests sets is 99.88 \% ), and that the computed SOFA scores, SIRS scores and mortality rates  using archetypes from the training set are consistent with those using ground truth archetypes (averaged errors on the SOFA scores, SIRS scores and mortality rates on the test sets are 0.1887, 0.0256, and 0.0104, respectively.)}

%Thus, if the SOFA score had not replaced the SIRS measure, patients belonged to the types other than A2, which comprised the types displaying significantly higher mortality rates and would have missed the treatments for sepsis, leading to a worse end. Conversely, the SOFA metric covers a broader range of the conditions of sepsis than the SIRS. Besides, the order of the SIRS score, which can be interpreted as the severity level of sepsis infections, and the mortality rate of these sepsis types are isomorphic.

\subsection*{Selecting distinguishing features of sepsis states.}

\begin{figure}[h]
   \centering
   
    \begin{subfigure}[h]{0.5\textwidth}
        \centering
        \includegraphics[width=10cm]{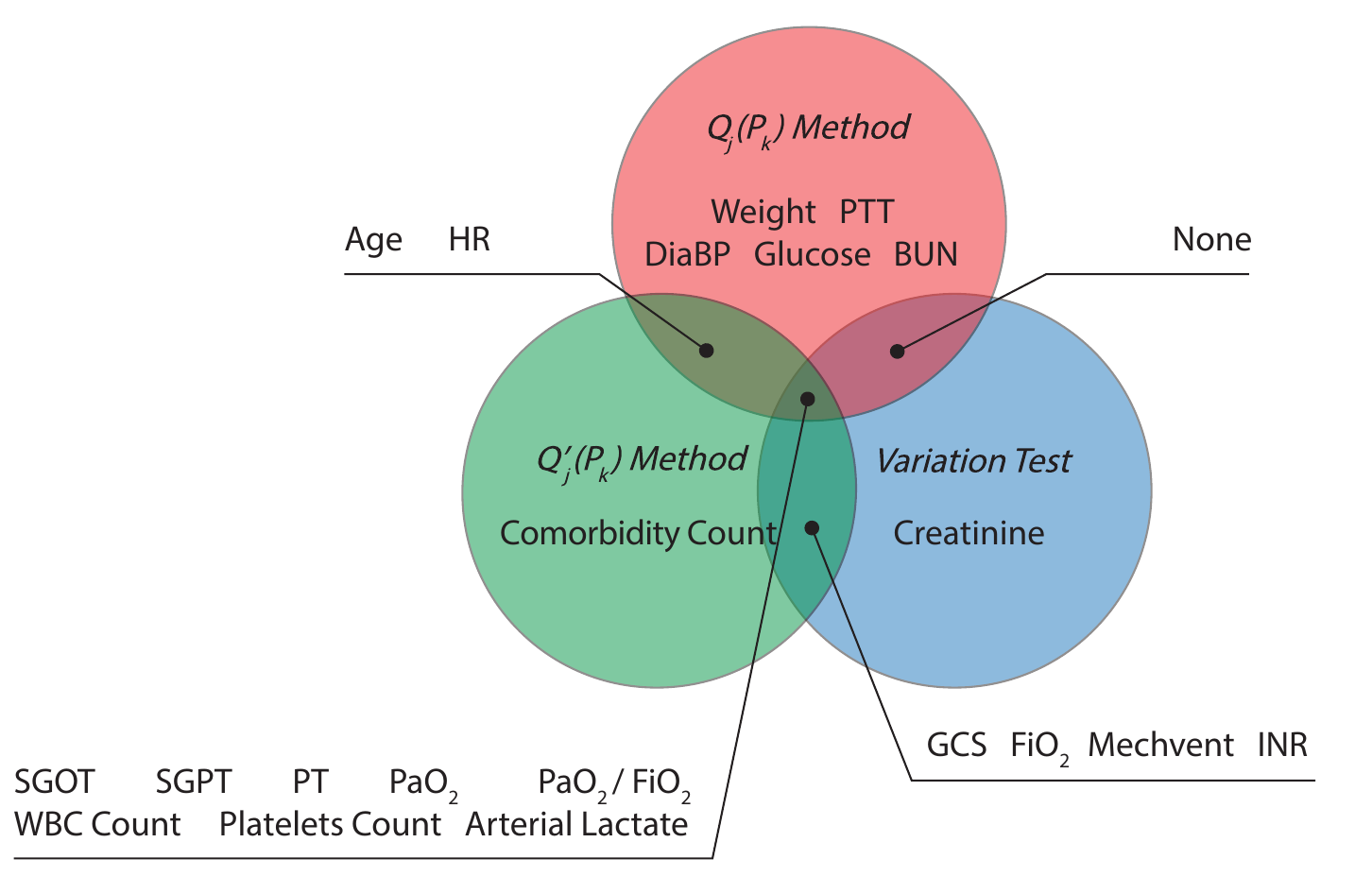}
    \end{subfigure}
    
    \caption{Visualization of the selected features (21 features in total) by $Q_j(P_K)$ , $Q'_j(P_K)$ , and $Variation$ $test$.  $Q_j(P_K)$ calculates the discriminative power of feature $i$ for a given clustering as the ratio of inter-cluster inertia to the total inertia computed using feature $i$. $Q'_j(P_K)$ calculates the discriminative power of feature $i$ as the ratio of inter-cluster inertia computed using feature $i$ to total inter-cluster inertia computed using all features. Variation test selects features that have the lowest probability of overlap across clusters (please see Methods section \ref{sec:feature-selection} for more details). Note that there is a significant overlap between features chosen by these selection criteria. However, each criterion yields a distinct set of features significantly associated with different sepsis states.}\label{fig:ven_diag}
\end{figure}

\textcolor{black}{ To identify discriminative attributes for each state, we use three criteria. The first two criteria are based on the Huygens-Steiner theorem to measure the inertia (\textit{i.e.}, the tendency of a physical object to remain still or continue in motion) of the points in Euclidean space. The third criterion finds the most distinct features across populations and prunes them for each state (please see method section \ref{sec:feature-selection}  for more details.) The top 15 ranked features for each feature selection method were selected, shown in \textcolor{black}{Fig \ref{fig:ven_diag}  (See Table \ref{tab:features} for the ranking.)}. As shown in \textcolor{black}{Fig}  \ref{fig:ven_diag} , 8 features, \textit{i.e.}, SGOT, SGPT, PT, PaO2, PaO2/FiO2, WBC Count, Platelets Count, and Arterial lactate, were selected by all three criteria; 6 features, \textit{i.e.}, Age, HR, GCS, FiO2, Mechvent, and INR , were selected by two; and 7 features, \textit{i.e.}, Weight, PTT, DiaBP, Glucose, BUN, Creatinine and Comorbidity  Count were selected by one. We use these features to analyze the primary profiles for each sepsis state in the next section.}

\subsection{Analyzing the primary attributes for sepsis states.}
\label{sec:expression}

\iffalse
\begin{table}[h]
\caption{Summary of the biomarkers for each primary function.}\label{tab:biomarker-tab}
\centering
\begin{tabular}{lllll}
\hline
\textbf{Nervous System}             & GCS        &                 &                  &   \\      
\hline
\textbf{Inflammation/ Infection}    & WBC Count  & Platelets Count & HR               &                \\ \hline
\textbf{Liver function tests}       & SGPT       & SGOT            & Arterial Lactate &                 \\ \hline
\textbf{Kidney function tests}      & Creatinine & BUN             &  &          \\ \hline
\textbf{Coagulation tests}          & PT         & aPTT            & INR              & Platelets Count \\ \hline
\textbf{Respiratory function tests} & PaO2       & FiO2            & PaO2/FiO2        & Mechvent        \\ \hline
\textbf{Cardiovascular tests}       & DiaBP      & Vassopressor    &                 &                 \\     
\hline
\end{tabular}

\end{table}
\fi

\begin{figure}[h]
%\vspace{-1.5cm}

 \begin{subfigure}[h]{0.33\textwidth}
        \centering
        \includegraphics[width=5cm]{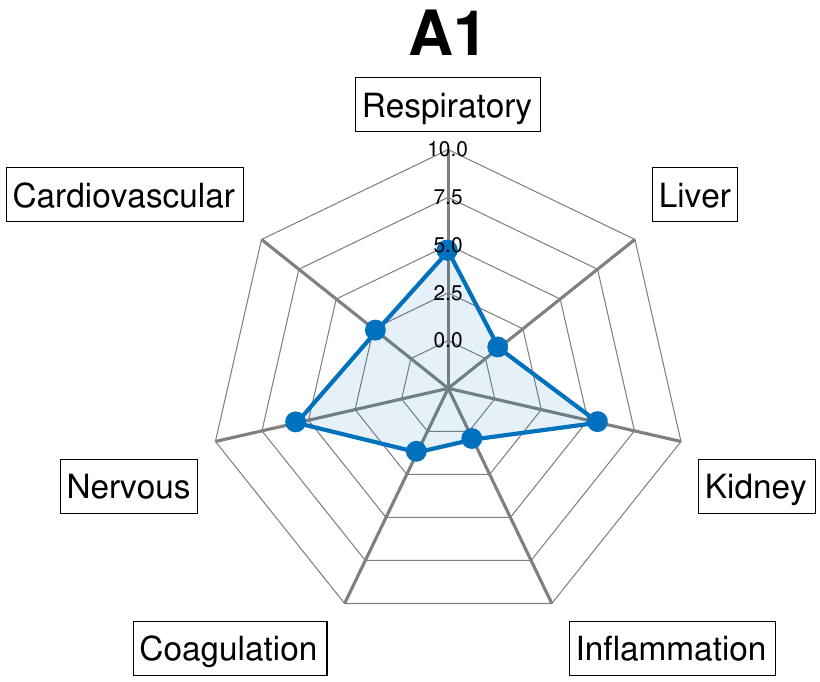}
        \end{subfigure}~
        \begin{subfigure}[h]{0.33\textwidth}
        \centering
        \includegraphics[width=5cm]{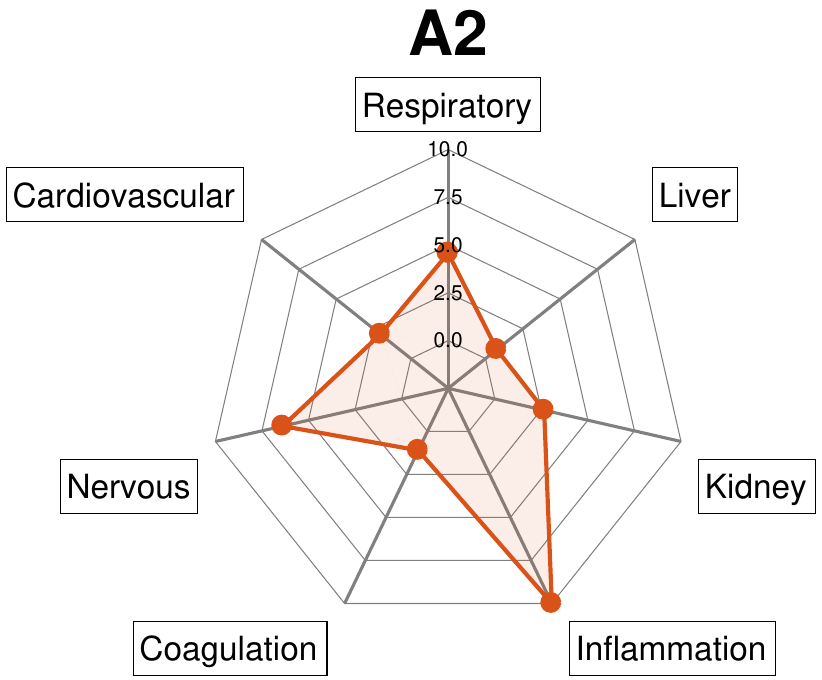}
        \end{subfigure}~
        \begin{subfigure}[h]{0.33\textwidth}
        \centering
        \includegraphics[width=5cm]{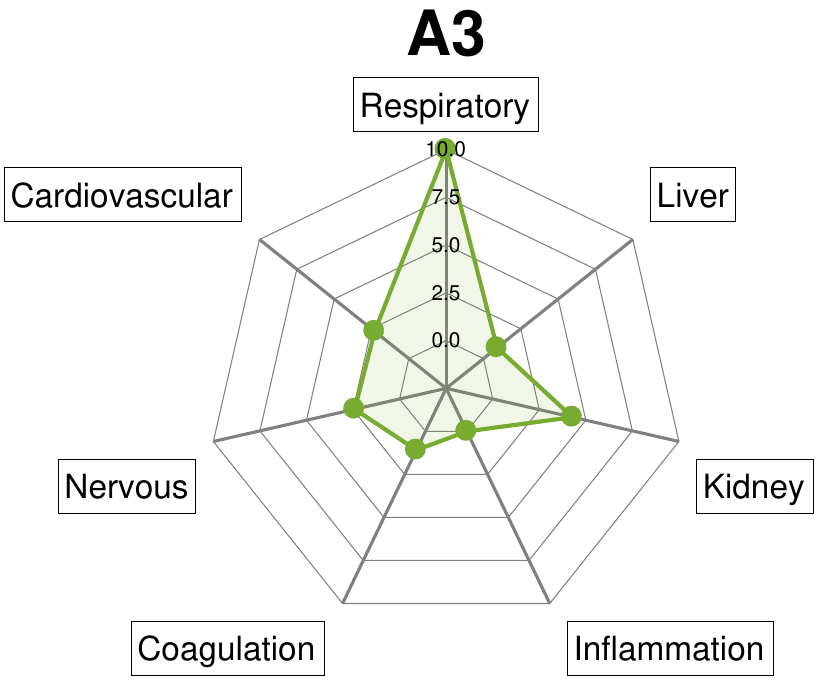}
        \end{subfigure}\\
        \begin{subfigure}[h]{0.33\textwidth}
        \centering
        \includegraphics[width=5cm]{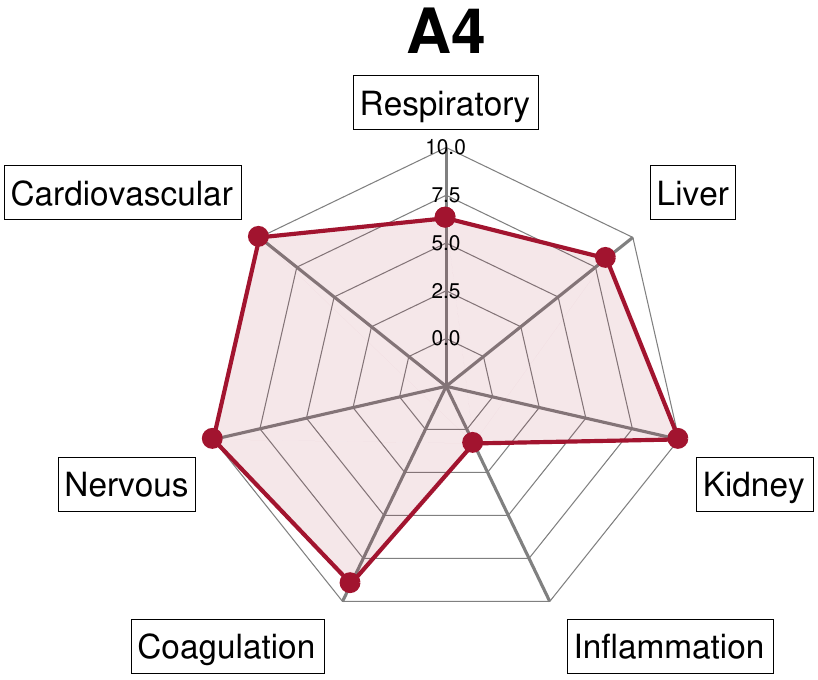}
        \end{subfigure}~
        \begin{subfigure}[h]{0.33\textwidth}
        \centering
        \includegraphics[width=5cm]{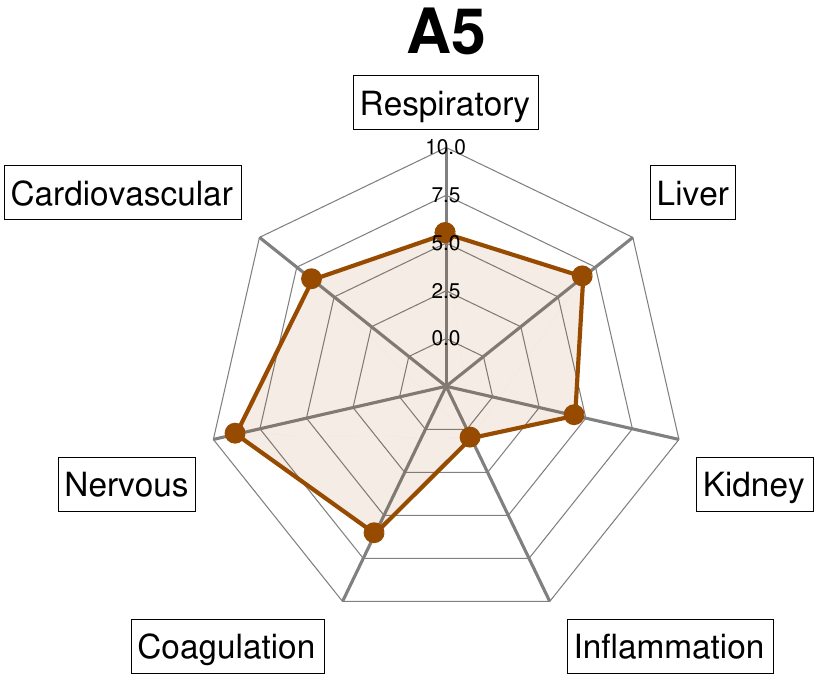}
        \end{subfigure}~
        \begin{subfigure}[h]{0.33\textwidth}
        \centering
        \includegraphics[width=5cm]{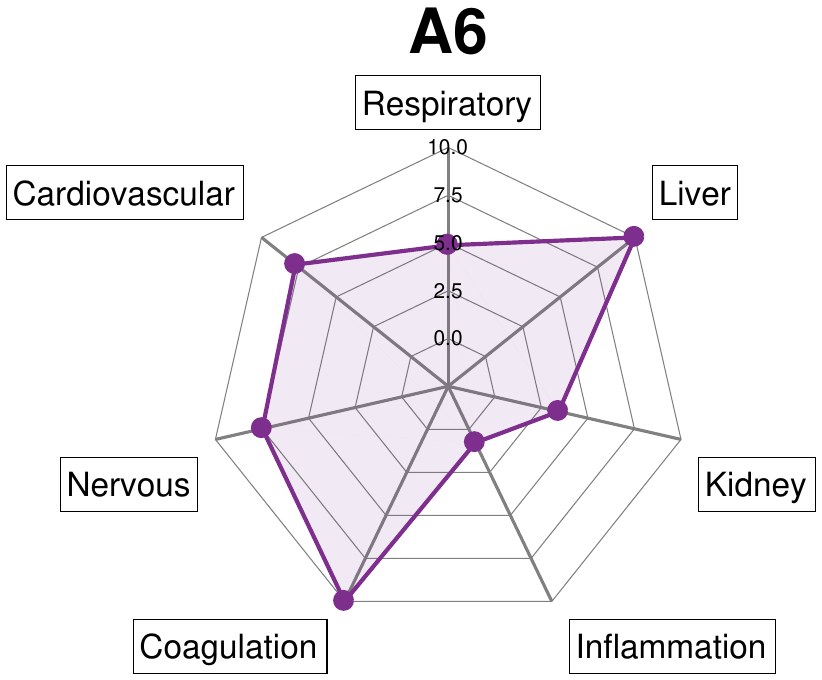}
        \end{subfigure}\\
    
    \caption{Spider-plot of primary functions affected in each sepsis state (represented by corresponding colors). There are seven different dimensions of primary function for each sepsis sate. The measured dimensions are \textit{nervous system, inflammation and infection, liver function, kidney function, coagulation, respiratory function, and cardiovascular function}, respectively. The scale of each dimension ranges from 0 to 10, with higher values indicating higher affect on the primary function.}
    \label{fig:primary_function}
\end{figure}

\textcolor{black}{
Among the 21 features selected by our methods, 18 are vitals and lab results that are known biomarkers of organ functions or other aspects of overall health. From this set of features, we identify associated primary health indicators corresponding to the \textit{nervous system, inflammation and infection, liver function, kidney function, coagulation, respiratory function, and cardiovascular function}. We refer to these seven as {\em primary functions}. We used these selected features to calculate the expression of these primary functions (please see method Section \ref{sec:primary_function_expression} for more a detailed calculation.). The spider-plot of primary functions affected in each sepsis state is shown in \textcolor{black}{Fig } \ref{fig:primary_function}.}

We find that each sepsis state manifests distinct expressions of organ dysfunction. We provide a detailed examination of the expression level of the biomarkers for each primary function next. We provide: (i) the biomarkers that are used in assessment of organ function; (ii) the severity level of biomarkers to assess organ function, and (iii) the expressions of these biomarkers in each sepsis state. \textcolor{black}{A summary of the statistics for these biomarkers highlighted in color light cyan can be found in Table \ref{tab:archetype-stats}.}

\subsubsection*{Nervous System. }

\iffalse 
GCS is commonly used for bedside assessment of brain injury, and for assessing the consciousness level of critically ill patients, including those with sepsis.
\fi

We use the Glasgow Coma Scale (GCS) to evaluate the state of the nervous system. GCS estimates coma severity based on eye, verbal, and motor criteria, and classifies the patient into mild (score = 13 -- 15), moderate (score = 9 -- 12), severe (score = 3 -- 8), and vegetative state (score  less than 3)~\cite{teasdale1974assessment}. 
We find that the average GCS scores for states A1 through A6 are 12.56, 12.24, 13.96, 10.67, 11.20, and 11.77, respectively. This indicates that the non-MODS states -- A1, A2, and A3, are mild GCS states, while the MODS states -- A4, A5, and A6 states, are within the range of moderate GCS state. Among all of the sepsis states, state A3 displays the highest level of consciousness, while the A4 corresponds to the lowest level of consciousness. 

\textcolor{black}{We find that the GCS scores correlate well with  SOFA scores and mortality rate of sepsis states, indicating that the GCS score is a good predictor of severity level or mortality for patients with sepsis.}

\subsubsection*{Inflammation and Infection.}

White Blood Cell (WBC) count, Heart Rate (HR), and platelet count are used to evaluate inflammation and infection response in the body. Among these, WBC count and HR are also used in the SIRS criteria to characterize systemic inflammation~\cite{bone1992definitions}. In acute inflammatory conditions, an increase in HR is often observed.  HR in sepsis increased when patients suffer from hypovolemia and hypoperfusion. The WBC count increases from a normal value of 4.5 to 11.0 $\times$ $10^9$ /L to 15.0 to 20.0 $\times$ $10^9$ /L, with WBC levels higher than 11.0 defined as leukocytosis. 
%{\bf ayg: What are these increases in the previous sentence due to?} \textcolor{black}{chf: Dr. Salma, I would appreciate hearing your thoughts on this.}
While not considered in the SIRS criteria, the elevation of platelet count is an important indicator for inflammation and infection\cite{rose2012etiology}. Inflammatory conditions such as bacterial infection, sepsis, malignancy, and tissue damage, motivate a reactive response that elevates platelet count, namely secondary thrombocytosis (platelet count higher than 500 $\times$ $10^9$ /L)\cite{vora1993secondary}. 

We find that leukocytosis and secondary thrombocytosis are observed in state A2. This state displays the highest average WBC count, platelet count, and HR with an average of 20.70 $\times$ $10^9$ /L, 905.58 $\times$ $10^9$ /L, and 95.57, respectively. The average WBC count in state A3 is within, but close to the maximum of the normal range (10.95 $\times$ $10^9$ /L). Slightly elevated WBC count is observed in states A1, A4, A5, and A6, with an average of 12.22 $\times$ $10^9$ /L, 13.12 $\times$ $10^9$ /L, 11.24 $\times$ $10^9$ /L, and 13.01 $\times$ $10^9$ /L, respectively. The average platelet counts in states  A1, A3, A4, A5, and A6 are within the normal range, with averages of 223.79 $\times$ $10^9$ /L, 229.67 $\times$ $10^9$ /L, 185.09 $\times$ $10^9$ /L, 189.93 $\times$ $10^9$ /L, and 194.64 $\times$ $10^9$ /L, respectively. 

In summary, states A1, A3, A4, A5, and A6 show few signs of inflammation. In contrast, state A2 reveals high inflammatory response, as all the inflammatory biomarkers -- WBC count, platelet count, and HR, are significantly elevated.

\subsubsection*{Liver function.}

SGOT, SGPT, and arterial lactate are used to characterize liver function. An increase in SGOT and SGPT levels indicates damage to the liver. In general, the severity of liver dysfunction can be classified as mild, moderate, or severe if elevation of SGOT and SGPT levels is less than 5 times, 5-10 times, and 10-50 times the upper reference limit. In addition to SGOT and SGPT, arterial lactate is a biomarker for liver dysfunction. Arterial lactate is primarily cleared by the liver, with a small amount of additional clearance by the kidneys. Thus, arterial lactate is elevated when liver function is compromised. In a healthy body, the lactate level is usually less than two mmol/L. Patients with hyperlactatemia usually have lactate levels higher than two mmol/L. Lactate levels higher than four mmol/L are considered to be in a severe state of hyperlactatemia.

We find that non-MODS states --  A1, A2, and A3, reveal mild liver damage, with only a mild increase in SGOT and SGPT levels (less than 5 times the upper reference limit), as well as a mild increase in arterial lactate. The average SGOT levels in states A1, A2, and A3 are 121.20 u/L, 115.25 u/L, and 97.13 u/L,  respectively; the average levels of SGPT in states A1, A2, and A3 are 104.71 u/L, 103.37 u/L, and 81.68 u/L, respectively; and the average arterial lactate levels in states A1, A2, and A3 are 2.04 mmol/L, 1.88 mmol/L, and 2.10 mmol/L, respectively. In contrast to non-MODS states, SGOT, SGPT, and arterial lactate are all in severe levels in MODS states -- A4, A5, and  A6. The average SGOT levels in states A4, A5, and A6 are 6.56 $\times$ $10^3$ u/L, 2.01 $\times$ $10^3$ u/L, and 7.66 $\times$ $10^3$ u/L, respectively; the average SGPT levels in states A4, A5, and  A6 are 3.02 $\times$ $10^3$ u/L, 6.39 $\times$ $10^3$ u/L, and 6.69 $\times$ $10^3$ u/L, respectively; and the average arterial lactate levels in states A4, A5, and  A6 are 5.50 mmol/L, 3.95 mmol/L, and 4.59 mmol/L, respectively. Not identified by our feature selection methods, but also a representative biomarker, high levels of bilirubin are often associated with liver damage \cite{cui2018elevated}. Patients with sepsis having (i) bilirubin $\geq$ 4.0 mg/dL, or (ii) SGPT levels of twice the upper limit of normal for age are considered to have a sepsis-associated liver injury (SALI)~\cite{goldstein2005international}. A high level of bilirubin ( $\geq$ 2.5 to 3 mg/dL) can cause jaundice. In our study, the average bilirubin levels in states A4, A5, and A6  are 5.37 mg/dL, 3.37 mg/dL, and 4.56 mg/dL, respectively, indicating common occurrence of jaundice in MODS states.

 In summary, non-MODS states reveal mild liver damage, reflected in a mild increase in SGOT, SGPT, and arterial lactate levels. In contrast, MODS states display severe liver dysfunction, with SGOT, SGPT, and arterial lactate all at severe levels. This is generally accompanied with jaundice. Finally, we note that state A6 potentially develops ischemic injury, since we observe that: (i) both SGOT and SGPT are more than 50 times higher than the upper reference limit; and (ii) SGOT is greater than SGPT \cite{giannini2005liver}. 
 %\textcolor{black}{Remove: We show in the next section that the development of ischemic injury is related to a set of comorbidities associated with state A6 before sepsis infection. 
%}

\subsubsection*{Kidney function.}

 Blood Urea Nitrogen (BUN) test and serum creatinine, identified by our feature selection methods, are common biomarkers of Acute Kidney Injury (AKI). AKI, defined as a sudden episode of kidney failure or kidney damage that happens within a few hours or a few days, is a common complication in sepsis patients. It is associated with high morbidity and mortality\cite{peerapornratana2019acute}. BUN measures the amount of urea nitrogen in the blood. Urea nitrogen is removed from the blood by the kidneys; consequently,
 high BUN levels indicate potential kidney damage. 
 A serum creatinine test provides an estimate of filtration efficiency of kidneys (glomerular filtration rate). An increased level of creatinine in blood is indicative of potential impaired kidney function. \textcolor{black}{Our feature selection methods also identify glucose. Higher glucose levels are often observed in patients with compromised kidney function, such as Chronic Kidney Disease (CKD).}
 
We find that state A3 exhibits relatively better kidney function when compared to the other sepsis states since both serum creatinine (1.02 mg/\textcolor{black}{d}L) and BUN (24.96 mg/\textcolor{black}{d}L), though slightly elevated, are the lowest. In the rest of the states, damage to kidneys is observed, with state A4 being the worst \textcolor{black}{(highest average glucose level (150.89 mg/dL))}. The average levels of serum creatinine in states A1, A2, A4, A5, and A6 are 1.489 mg/\textcolor{black}{d}L, 1.450 mg/\textcolor{black}{d}L, 2.05 mg/\textcolor{black}{d}L, 2.0 mg/L, and 2.0 mg/\textcolor{black}{d}L, respectively; the BUN levels in states A1, A2, A4, A5, and A6 are 29.31 mg/\textcolor{black}{d}L, 26.68 mg/\textcolor{black}{d}L, 33.63 mg/\textcolor{black}{d}L, 26.26 mg/\textcolor{black}{d}L, and 29.35 mg/\textcolor{black}{d}L, respectively.

\subsubsection*{Coagulation function.}

Activated Partial Thromboplastin Time (aPTT), Prothrombin Time (PT), and International Normalized Ratio (INR) are identified by our feature selection methods. These are measures of coagulation function. Sepsis-associated coagulopathy (SAC) is typically diagnosed by PT prolongation or elevated INR, in conjunction with reduced platelet count\cite{walborn2018international}. aPTT is also used as a test for coagulation in patients with sepsis. Increased aPTT and PT above normal values, and decreased platelet count below normal value indicate long clotting time (DIC) and bleeding in sepsis patients~\cite{simmons2015coagulopathy}. 

We find that in non-MODS states, aPTT is within the normal range, and that INR and PT are slightly elevated. The average aPTT values in states A1, A2, and A3 are 37.78 s, 37.52 s, and 37.84 s, respectively; the average INR values in states A1, A2, and A3 are 1.51, 1.46, and 1.48, respectively; and the average PT values in states A1, A2, and A3 are 16.16 s, 15.91 s, and 15.88s, respectively. In contrast to non-MODS states, an increase in values of INR, PT, and aPTT is observed in MODS states. The average aPTT values in states A4, A5, and A6 are 44.14 s, 41.61 s, and 43.14 s, respectively; the average INR values in states A4, A5, and A6  are 2.08, 2.49, and 2.96, respectively; and the PT values in states A4, A5, and A6 are 20.68 s, 21.92 s, and 25.93 s, respectively.  We also examine the platelet count in these states. Although the average platelet count in all sepsis states is within the normal range, a higher percentage of the cases with a platelet count below the normal range (150 $\times$ $10^9$ /L) are observed in MODS states. The percentage of cases with platelet count lower than normal from states A1 to A6 are 30\%, 0\%, 27.7\%, 44.9\%, 40.3\%, and 44.9\%, respectively. 

In summary, nearly one-third of the cases in states A1 and A3 develop a mild condition of SAC or DIC, while more than 40\% of cases in the MODS group have worse SAC or DIC.

\subsubsection*{Respiratory function.} 

PaO2, FiO2,  PaO2/FiO2, and the use of a mechanical ventilator are identified by our feature selection methods. These are commonly used to measure respiratory function. PaO2 measures the pressure of oxygen dissolved in blood, and how well oxygen can move from the airspace of the lungs into the blood.  FiO2 is defined as the concentration of oxygen that a person inhales. Patients experiencing difficulty in breathing are provided with oxygen-enriched air. Therefore, higher FiO2 is observed if the respiratory function is compromised. PaO2/FiO2 is a known measure for the assessment of respiratory dysfunction, such as Acute Respiratory Distress Syndrome (ARDS). Under the Berlin ARDS definition, patients with PaO2/FiO2 levels in the range of 200--300, 100--200, and less than 100 are classified as mild, moderate, and severe ARDS. The SOFA metric also incorporates PaO2/FiO2 as a  parameter in assessing respiratory function. According to the SOFA score, a normal person has a PaO2/FiO2 ratio of approximately 500 and a patient with PaO2/FiO2 ratio between 300 -- 400, 200 -- 300, 100 -- 200, and less than 100 would have SOFA scores 1, 2, 3, and 4, respectively. Thus, a lower PaO2/FiO2 ratio indicates worse respiratory condition. Conversely, high PaO2/FiO2 ratio (PaO2 $>$ 300 mmHg) indicates that the lung is exposed to hyperoxia. Mechanical ventilators are often used in ICUs to assist or replace spontaneous breathing, indicating compromised respiratory function. 

We find that patients in state A3 display excessive amounts of PaO2 and a slight increase in FiO2 (0.07 higher than the normal value of 0.21, on average). This indicates that patients in state A3 are less prone to lung dysfunction, but that state A3 manifests hyperoxia. The lower fraction of patients on ventilators in state A3 also indicates better lung function, compared to other states. The fraction of patients on a ventilator in state A3 is the lowest, at 0.09.  The \textcolor{black}{PaO2/FiO2} parameter also indicates that state A3 does not develop ARDS. Distinct from state A3, respiratory functions are compromised to varying extents in other states. We find that both PaO2 and FiO2 in states A1, A2, A4, A5, and A6  are slightly elevated. The average values of PaO2 in states A1, A2, A4, A5, and A6 are 20.83 mmHg, 21.77 mmHg, 26.95 mmHg, 20.98 mmHg, and 17.92 mmHg, respectively, and the average values of FiO2 in states A1, A2, A4, A5, and A6 are 0.46, 0.47, 0.52, 0.48, and 0.47, respectively. A higher rate of patients on ventilators is observed in states A1, A2, A4, A5, and A6, with the mean value of 0.37, 0.33 0.56, 0.50, and 0.41, respectively. We observe that states A1, A2, A4, A5, and A6 correspond to mild ARDS. The average values of \textcolor{black}{PaO2/FiO2} in states A1, A2, A4, A5, and A6  are, respectively, 292.98 mmHg, 294.59 mmHg, 285.03 mmHg, 318.13 mmHg, and 301.8 5 mmHg, which is close to the boundary of normal value of 300mmHg in the Berlin ARDS definition and close to SOFA score  of 2. We highlight that among these states, state A4, one of the MODS states that displays the highest SOFA score and mortality rate, shows the highest FiO2, the lowest PaO2/FiO2, and the highest rate of ventilator use. 

In summary, states A1, A2, A4, A5, and A6 manifest mild respiratory dysfunction with state A4 being the worst. State A3 shows better respiratory function, as the average value of FiO2, and rate of ventilator use is the lowest. According to both the Berlin ARDS definition and SOFA score, state A3 type does not manifest ARDS. However, state A3 manifests hyperoxia, since the average value of PaO2 in state A3 is higher than 300 mmHg.

\subsubsection*{Cardiovascular function.}

\iffalse
DiaBP, identified by our feature selection method, is a measure of potential hypotension (systolic blood pressure $\leq$ 90 mmHg or diastolic $\leq$ 60 mm Hg) and vasopressor use, and can be used to identify potential septic shock~\cite{rhodes2017surviving}.

Severe hypotension can deprive the brain and other vital organs of oxygen and nutrients, leading to a life-threatening condition called shock. Patients with septic shock can be identified with a clinical construct of sepsis with persisting hypotension requiring vasopressors to maintain mean arterial pressure (MAP) of 65 mmHg, and having a serum lactate level higher than two mmol/L despite adequate volume resuscitation~\cite{rhodes2017surviving}. 
\fi

\textcolor{black}{DiaBP and serum lactate are identified by our feature selection method.
DiaBP is indicative of potential hypotension (systolic blood pressure $\leq$ 90 mmHg or diastolic $\leq$ 60 mm Hg), and serum lactate is an important biomarker of septic shock ~\cite{rhodes2017surviving}.} Patients with septic shock can be identified with a clinical construct of sepsis with persisting hypotension requiring vasopressors to maintain mean arterial pressure (MAP) of 65 mmHg, and having a serum lactate level higher than two mmol/L despite adequate volume resuscitation~\cite{rhodes2017surviving}.

We observe that all sepsis states except A5 show lower blood pressure. The average DiaBP of state A5 is 60.46, and the average values of DiaBP in states A1, A2, A3, A4, and A6 are 57.07, 59.55, 58.31, 58.07, and 57.00, respectively. \textcolor{black}{MODS states show significantly higher serum lactate levels. The average serum lactate values in states A1, A2, and A3 are 2.04 mmol/L, 1.88 mmol/L, and 2.10 mmol/L, respectively. In contrast, the average serum lactate values in states A4, A5, and A6 are 5.50 mmol/L, 3.95 mmol/L, and 4.58 mmol/L, respectively.}

We also find that the dosage of vasopressin in MODS states is significantly higher than non-MODS states. The average dosage of vasopressin in states A1, A2, and A3 are 0.06 mcg/kg/min, 0.08 mcg/kg/min, and 0.01 mcg/kg/min, respectively. In contrast, the average dosage of vasopressin in states  A4, A5, and A6 are 0.26 mcg/kg/min, 0.13 mcg/kg/min, and 0.13 mcg/kg/min, respectively.

In summary, non-MODS states show mild hypotension, while MODS states are potentially in septic shock, with state A4 being the worst. It has been shown that the development of septic shock is an accurate predictor of mortality~\cite{vincent2019frequency,mohamed2017predictors}. 
\textcolor{black}{Our results are consistent with these studies since states A4, A5, and A6 consist of patients with higher SOFA scores and correlate with higher mortality rates.}

\subsection*{Correlation of demographic variables and comorbidities with sepsis states.}

Variations in patients' demographics, such as gender, age, and medical comorbidities, present additional considerations for classifying sepsis states~\cite{iskander2013sepsis}. 
%We analyze the relative contributions of the development of sepsis types by these factors. Previously, not only the vital signs and lab values were selected by the feature selection methods, but these 
Our feature selection methods identify age, weight, and comorbidities, indicating that these attributes are strongly correlated with sepsis states. 

%development of these sepsis types is highly dependent on these factors. Thus, in what follows, we examine the effect of age, weight, and the previously developed comorbidities before sepsis infection on the development of sepsis types separately.

\begin{figure*}[h]
   \centering
   
        \begin{subfigure}[h]{0.5\textwidth}
        \caption{ }\label{fig:age-pop}
        \includegraphics[width=8.0cm]{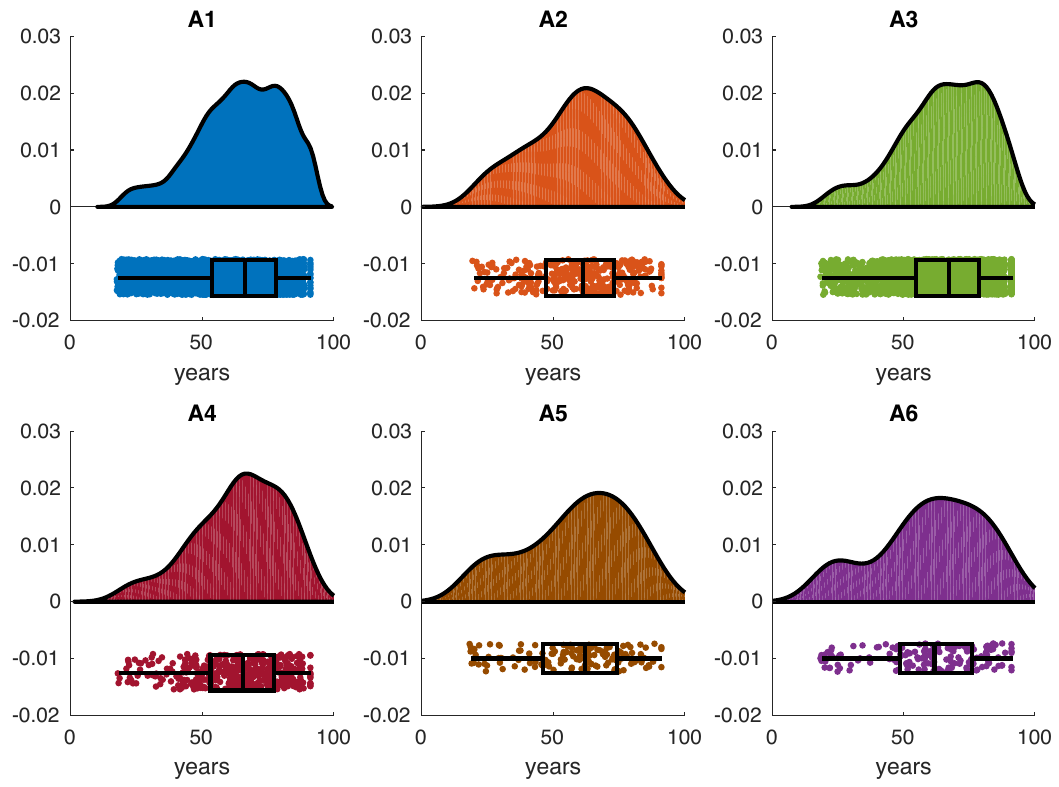}
        \end{subfigure}~
        \begin{subfigure}[h]{0.5\textwidth}
        \caption{}\label{fig:weight-pop}
        \includegraphics[width=8.0cm]{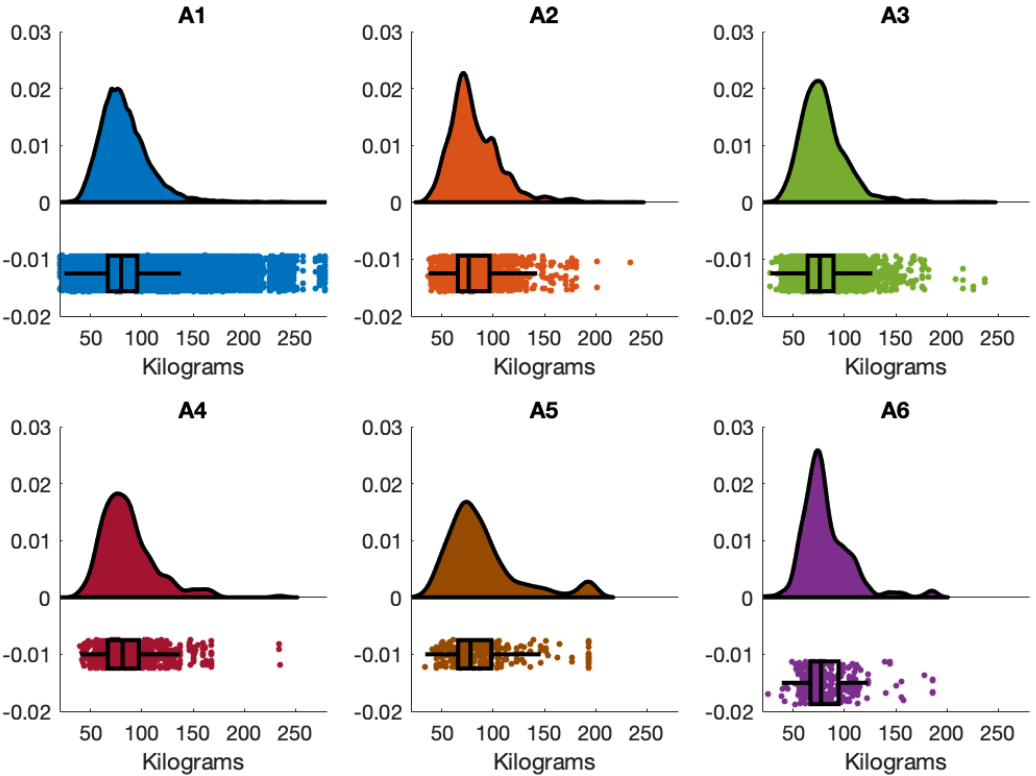}
        \end{subfigure}\\
        \begin{subfigure}[h]{0.5\textwidth}
        \centering
        \caption{}\label{fig:Comorbidity -pop}
        \includegraphics[width=8.0cm]{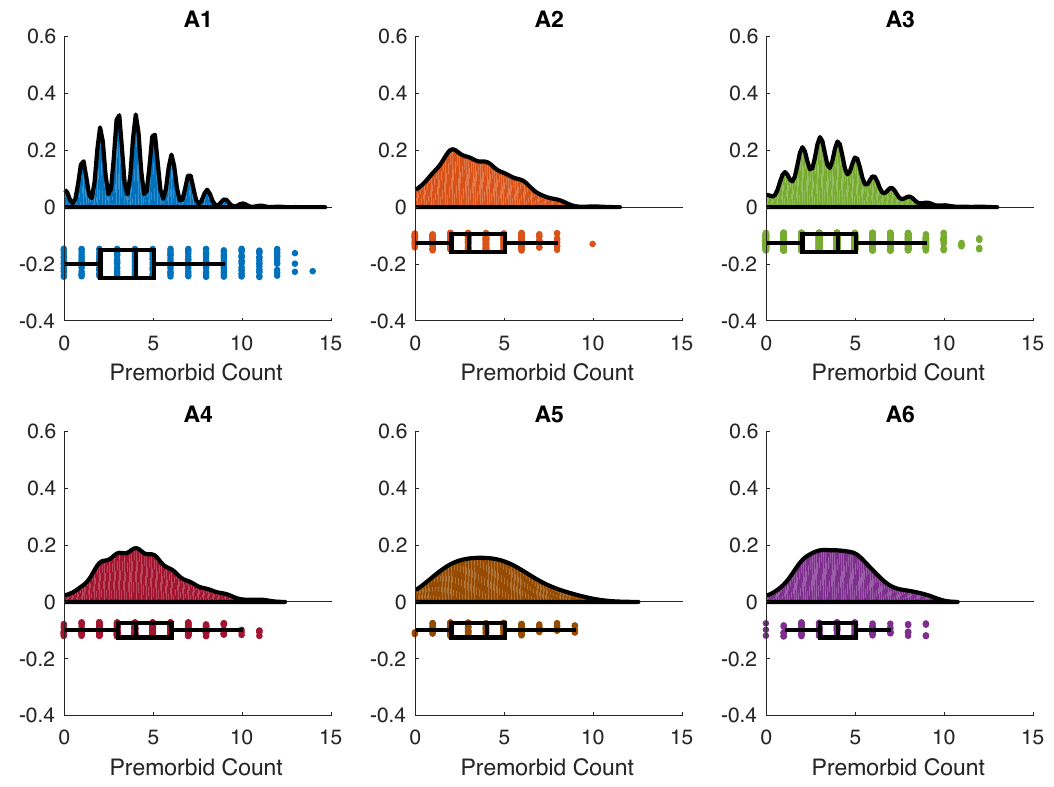}
        \end{subfigure}~
        \begin{subfigure}[h]{0.5\textwidth}
        \centering
        \caption{}\label{fig:z-score}
        \includegraphics[width=8cm]{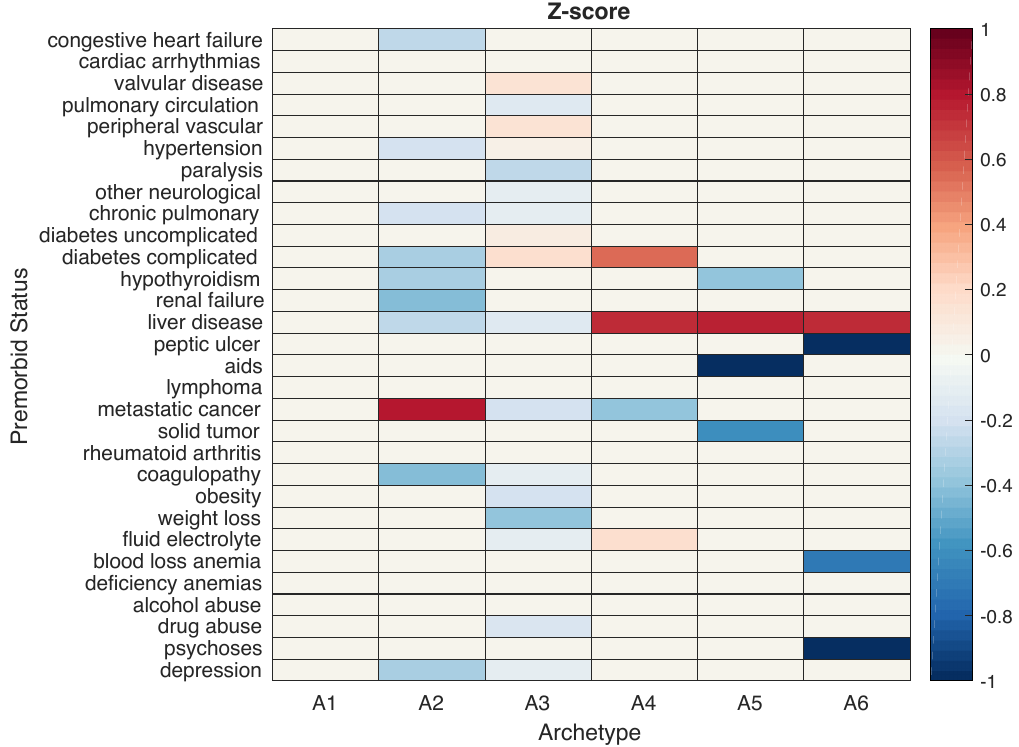}%
    \end{subfigure}
    
    \caption{ \textbf{ (\textcolor{black}{A})} The population distribution for each sepsis type stratified by age. \textbf{(\textcolor{black}{B})} The population distribution for each sepsis type stratified by weight. \textbf{(\textcolor{black}{C})} The population distribution for each sepsis type stratified by the number of comorbidities before infection. \textbf{(\textcolor{black}{D})} Z-score analysis of the comorbidity profiles (row) of each sepsis type (column). Entries approaching red in intensity indicate that the comorbidity profiles are expressed in the corresponding sepsis states, and entries closer to blue indicate that the comorbidity profiles are suppressed in corresponding sepsis states. \textcolor{black}{Entries with p-value $> 0.05$. are set to 0.} }
    
\end{figure*}

\vspace{-0.2cm}
\subsubsection*{Correlation of demographic variables with sepsis states.} 

The distributions of sepsis states in terms of patient age are shown in \textcolor{black}{Fig}~\ref{fig:age-pop} . We observe that while the average ages of patient in states A1, A3, and A4 are close to the average age of the entire cohort, the average age patients in states A2, A5, and A6 are significantly lower than the average age of the entire cohort -- \textcolor{black}{the average age of the entire cohort is 64.57 years, and the average ages of states A1 to A6 are 64.79 years, 59.55 years, 65.53 years, 64.09 years, 58.66 years, and 60.16 years}, respectively. Several studies have been shown that advanced age has been associated with worse outcomes\cite{rowe2017sepsis,nasa2012severe}. We also find that worse outcomes are observed in older people for severe sepsis types. Specifically, in MODS states, we observe that state A4, shown to be associated with the highest mortality, also has the highest average age. On the other hand, we observe that the sepsis state that demonstrates notable expression of inflammatory response, \textit{i.e.,} A2, is associated with lower average age.

 The distributions of sepsis states in terms of patient weight are shown in \textcolor{black}{Fig} ~\ref{fig:weight-pop}. We observe that the average weight of the entire cohort and the average weight of all sepsis states except A5 are within 4 percent. The average weight of patients in state A5 is 7 percent higher than the average weight of the entire cohort. The average weight of the entire cohort is 83.27 kilograms and the average weight of patients in states A1 to A6 is 83.29 kilograms, 84.95 kilograms, 79.28 kilograms, 85.18 kilograms, 89.30 kilograms, and 82.31 kilograms, respectively.

 \subsubsection*{Comorbidity profiles and their association with sepsis states.}
 
 %In Section \ref{sec:expression}, we observed that sepsis states manifest heterogeneous expression of organ functions. 
 %Not only the demographic variables could affect the health outcomes, but it is possibly due to the unique comorbidity profiles the patients had before sepsis infection that leads to entirely different pathways of the sepsis.\cite{valderas2009defining} Thus, we 
 
 We investigate the association of different comorbidity profiles with sepsis states. First, we construct distributions of sepsis states by the total number of pre-existing comorbidities,  shown in \textcolor{black}{Fig} \ref{fig:Comorbidity -pop}. We observe that, compared to non-MODS states, the MODS group has a higher number of comorbidities -- \textcolor{black}{the average comorbidity count for states A1 to A6 is 3.93, 3.36, 3.75, 4.32, 4.04, and 4.08, respectively.} The higher the number of comorbidities, the worse the outcomes of sepsis. %I included the number of comorbidities as a measure since there are many papers are interested in studying the association between the number of comorbidities and the survival rate. (such as the paper "A high burden of comorbid conditions leads to decreased survival in breast cancer". The main results from this paper is " The average number of comorbidities was 2.2 with hypertension and obesity being the most common. Significant differences were found in the number of comorbidities between African Americans (2.61) and Caucasians (1.78) (P<0.005). )") 
%We note that states in the MODS group have higher number of comorbidities and correspondingly worse survival rate. 
%These comorbidities are different so we should treat each independently. Thus I also use z-score to analize them in the next paragraph.
 
 Our results show clear association between comorbidities and patient outcomes.
 %Thus far, we have confirmed that the more comorbidities the patient had developed before infection, the worse outcomes of the sepsis infection. However, the question of \textit{which type of comorbidity profiles the patients had that leads to the development of particular sepsis type} remained unanswered. Thus, we analyze the relationship between sepsis types and various types of comorbidity profiles. 
 We next analyze the relationship between specific comorbidity profiles and their association (positive or negative) with sepsis states. We use the z-score to measure the distance between the observed condition (comorbidity) and its average over the entire cohort. If the z-score of a condition is positive in a state, we note that the condition is expressed in the state; conversely, if the z-score is negative, the condition is suppressed in the state. To ensure that a diverse range of conditions is covered, a comprehensive set of comorbidity measures, 30 types in all, are included in the z-score analysis (please see \textcolor{black}{S6 Table} \iffalse \ref{tab:Comorbidity} \fi in the appendix for more detail for each type).

 \textcolor{black}{ Once the z-scores are computed, we apply the two-sample t-test to ensure that the computed values are statistically significantly different. The statistically significant z-scores (p-value $\leq$ 0.05) are shown in \textcolor{black}{Fig}~\ref{fig:z-score}. (\textcolor{black}{Fig} \ref{fig:etiological} in Appendix shows the original z-scores and the corresponding p-values for the pairwise two-sample t-test for the comorbidity profiles of each sepsis type.)}
 
 \textcolor{black}{As shown in \textcolor{black}{Fig}~\ref{fig:z-score}, we find that none of the conditions are significantly differentially expressed from the overall cohort in state A1. This is explained by the fact that state A1 represents a mild sepsis state. We find that the z-score of metastatic cancer (0.81) is significantly expressed in the inflammation state (A2 state). A slightly increased differential expression of valvular disease, peripheral vascular, hypertension, and diabetes (uncomplicated and complicated) is observed in state A3, with values of 0.14 and 0.13, 0.05, 0.08, and 0.18, respectively. We observe a strong association between MODS states and liver disease. Therefore, we observe that z-scores of liver disease in MODS states are statistically higher than average. The z-scores of liver disease in states A4, A5, and A6 are 0.73, 0.78, and 0.74, respectively. Individuals with poor kidney health manifest fluid and electrolyte imbalances. We observe the z-score of fluid and electrolyte imbalances in state A4 is statistically higher than average, with a value of 0.17. Lastly, we find a strong association between complicated diabetes and state A4, with a z-score value of 0.53.}

\clearpage
   \begin{figure}
   \centering
    \begin{subfigure}[h]{1.0\textwidth}
        \centering
        \caption{ }\label{fig:3-order-Markov}
        \includegraphics[width=12.0cm]{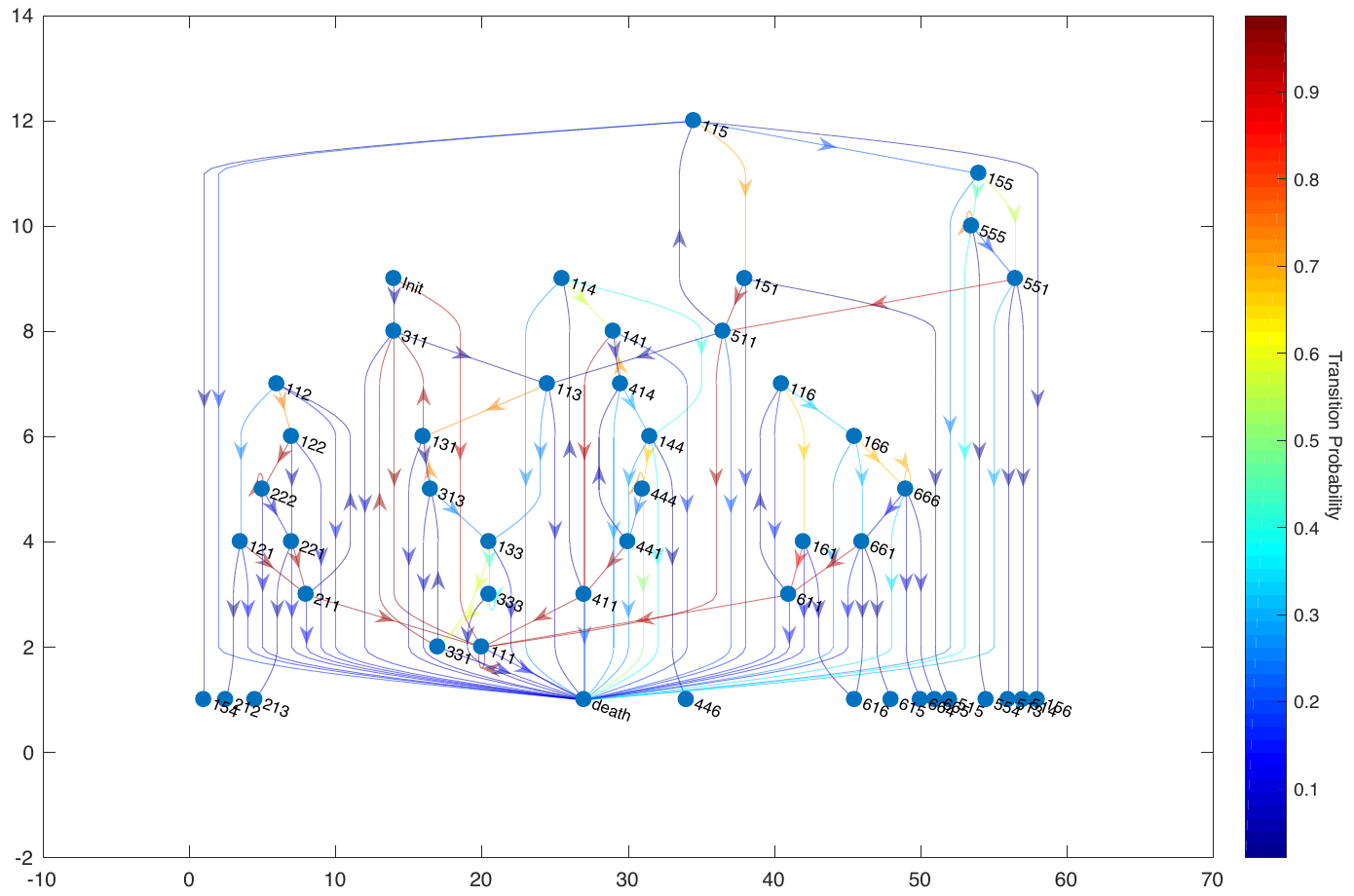}
        \end{subfigure}\\
        \begin{subfigure}[h]{1.0\textwidth}
        \centering
        \caption{}\label{fig:first-order-treatment}
        \includegraphics[width=12.0cm]{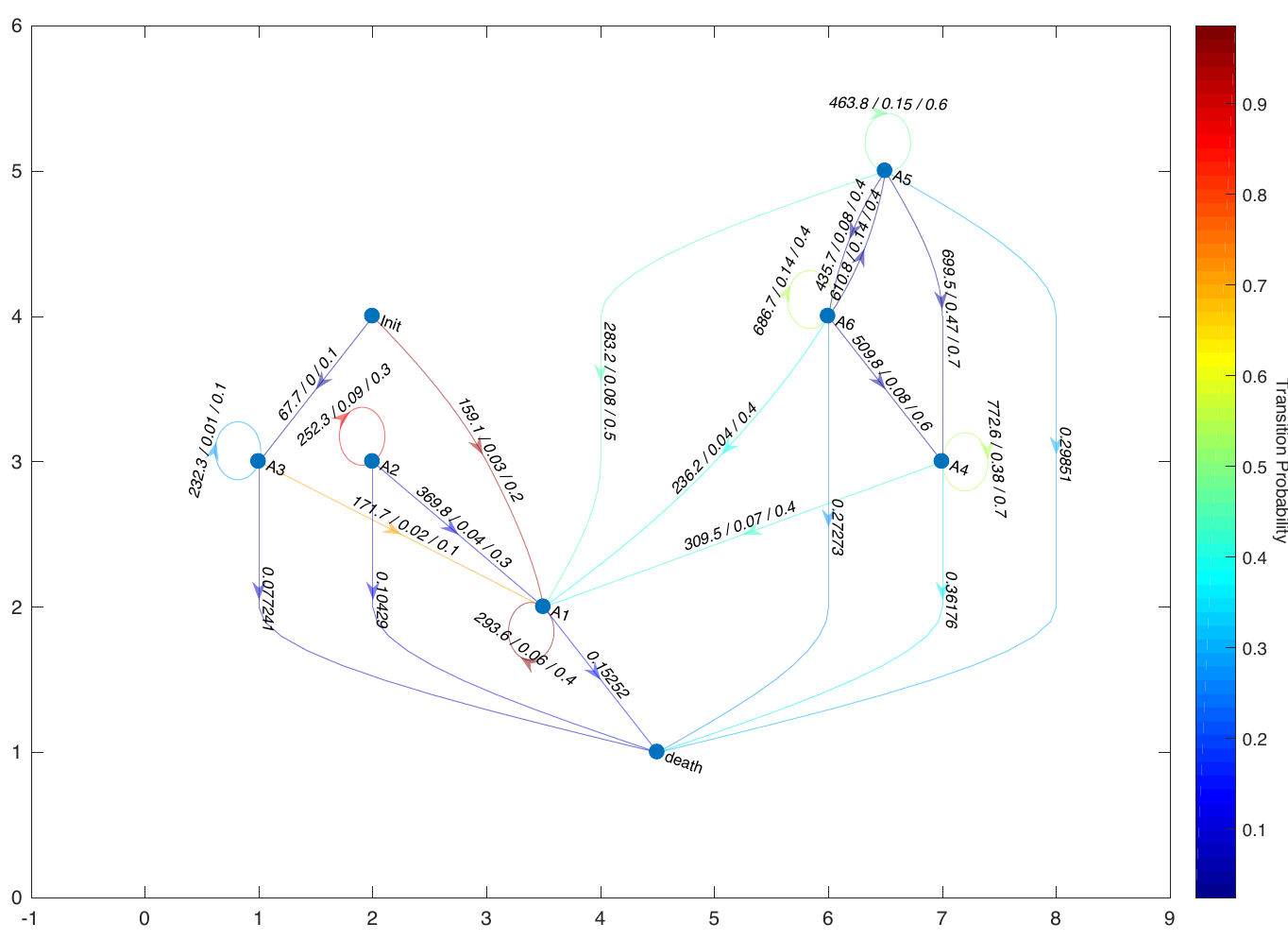}
        \end{subfigure}
    \caption{{\textbf{ (\textcolor{black}{A})} 
    \textcolor{black}{
    Third-order transition graph: Edges approaching red in color indicate higher transition probabilities, and edges approaching black indicate lower transition probabilities.}
    \textbf{(\textcolor{black}{B})} \textcolor{black}{First-order transition graph. Averaged fluids, maximum dosage of vasopressor, the probability of using mechanical ventilator, and the mortality rate are shown on the transitions.}   } }
\end{figure}

 \clearpage
\subsection*{Analyzing pathological processes of sepsis.}
 
 Understanding the progression of pathological process of sepsis is essential for designing disease management protocols. First, we model sepsis patients' trajectories over identified states and analyze the association between pathological trajectories and mortality rates. Specifically, we model transition across sepsis states using higher-order Markov chains and compute the mortality rate of each node, which corresponds to a particular pathological trajectory in a Markov chain. 
 \textcolor{black}{Here, a third-order Markov chain, represented as directed de Bruijn graph, in conjunction with the association between third-order transitions and mortality rate is shown in \textcolor{black}{Fig}} ~\ref{fig:3-order-Markov} (See \textcolor{black}{Fig} \ref{fig:higher-order-markov} in Appendix for the second-order counterparts.)
 
 In \textcolor{black}{Fig} \ref{fig:3-order-Markov}, we observe that nearly all sepsis patients begin in state $A1$ and that the conditional probability of remaining in the same state given they were in $A1$ state in the previous three time points is approaching one \textcolor{black}{(98.74\%)} -- the state \quotes{111} of the third-order Markov chain is nearly an absorbing state. Also, we observe that if patients were in the states other than state $A1$ and enter state $A1$, they remain in state $A1$. This indicates that most of the sepsis patients remain in moderate condition. Lastly, we find that there are very few transitions across MODS groups. This suggests that patients in different MODS groups are composed of distinct sub-populations.
 
 Combined with knowledge of the severity level of sepsis states (Section \ref{sec:standard-metric}, \textcolor{black}{Fig}~\ref{fig:embed}), we can characterize the mortality trend for each pathological trajectory. (Note that we had shown in Section \ref{sec:standard-metric} that MODS states (state A4, A5, and A6) consist of patients with higher mortality rates than non-MODS states (state A1, A2, and A3).) As shown in \textcolor{black}{Fig} ~\ref{fig:3-order-Markov}, we observe a higher mortality rate in trajectories that traversed MODS states. Also, we observe that the longer a patient stays in MODS states, the higher the mortality rate. Finally, we find that mortality rate increases if there is a transition from a non-MODS state to a MODS state. Conversely, if there is a transition from a MODS state to a non-MODS state, the mortality rate decreases.

\textcolor{black}{
Transition dynamics across states are functions of treatments, patient characteristics, and sepsis states. The identification of sepsis states guides physicians to monitor a set of variables from patient covariates to assess the status of sepsis patients or the severity level of organ dysfunctions. Based on this information, a set of treatment actions are performed to manage sepsis. To find treatment actions during transitions, we identified the amounts of fluids infused, the dosage of vasopressor, and mechanical ventilators as treatment actions. We kept track of the averaged fluids, maximum dosage of vasopressor, and the probability of using a mechanical ventilator between transitions. The First-order transition graph and treatment actions are shown in \textcolor{black}{Fig} \ref{fig:first-order-treatment}. We find that a different set of treatment actions was imposed for each state with a general trend of amount of fluids, vasopressors, and mechanical ventilators on the MODS group. In addition, amount of fluids, vasopressors, and the usage of mechanical ventilators are reduced during the transitions from MODS states to non-MODs states (A1). }

Next, we identify distinctive clinical transition makers of sepsis progression. We construct a state transition dataset as a collection of gradients of clinical measurements associated with transitions from one sepsis state to another, to quantify dominant gradient trends using archetypal analysis. Formally, given a transition dataset $\mathcal{G}=\{g_1,\dots,g_{m'}\}$, we find a set of archetypes of gradients so that each gradient is a convex combination of archetypes and each archetype is a convex combination of the gradients. Following the procedure described in Section \ref{sec:AA_cohort}, we identified six archetypes of gradients, labeled G1 to G6.

\begin{figure}[ht!]
  \vspace*{-1cm}
   \centering
   
        \begin{subfigure}[h]{0.5\textwidth}
        \centering
        \caption{}\label{fig:gradient_umap}
        \includegraphics[width=8.0cm]{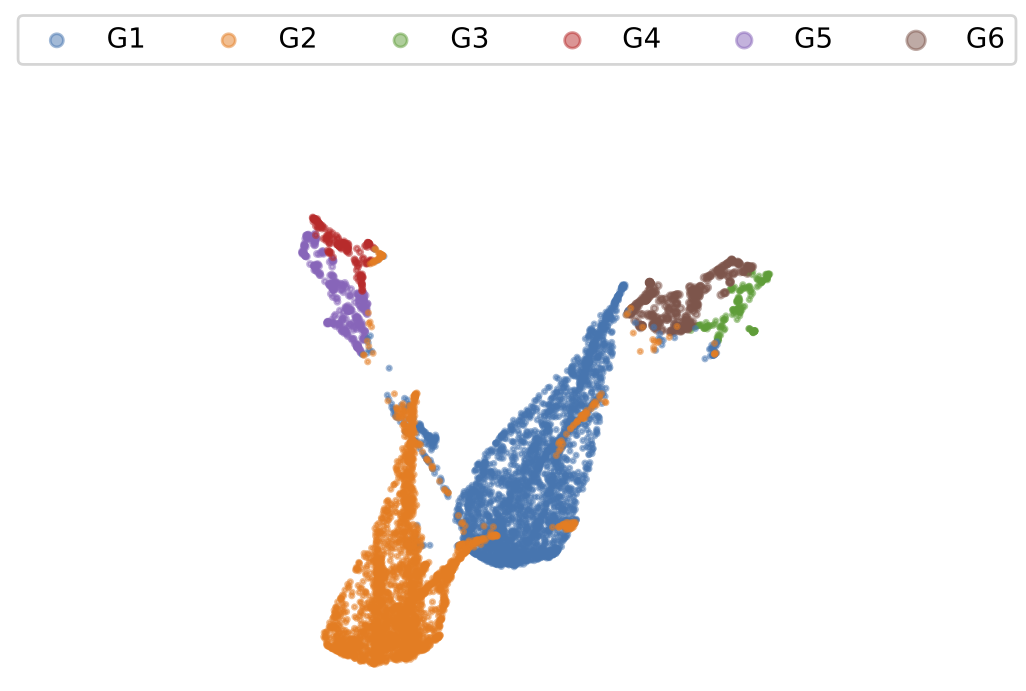}
        \end{subfigure}~
        \begin{subfigure}[h]{0.5\textwidth}
        \centering
        \caption{}\label{fig:gradient_z_score}
        \includegraphics[width=8cm]{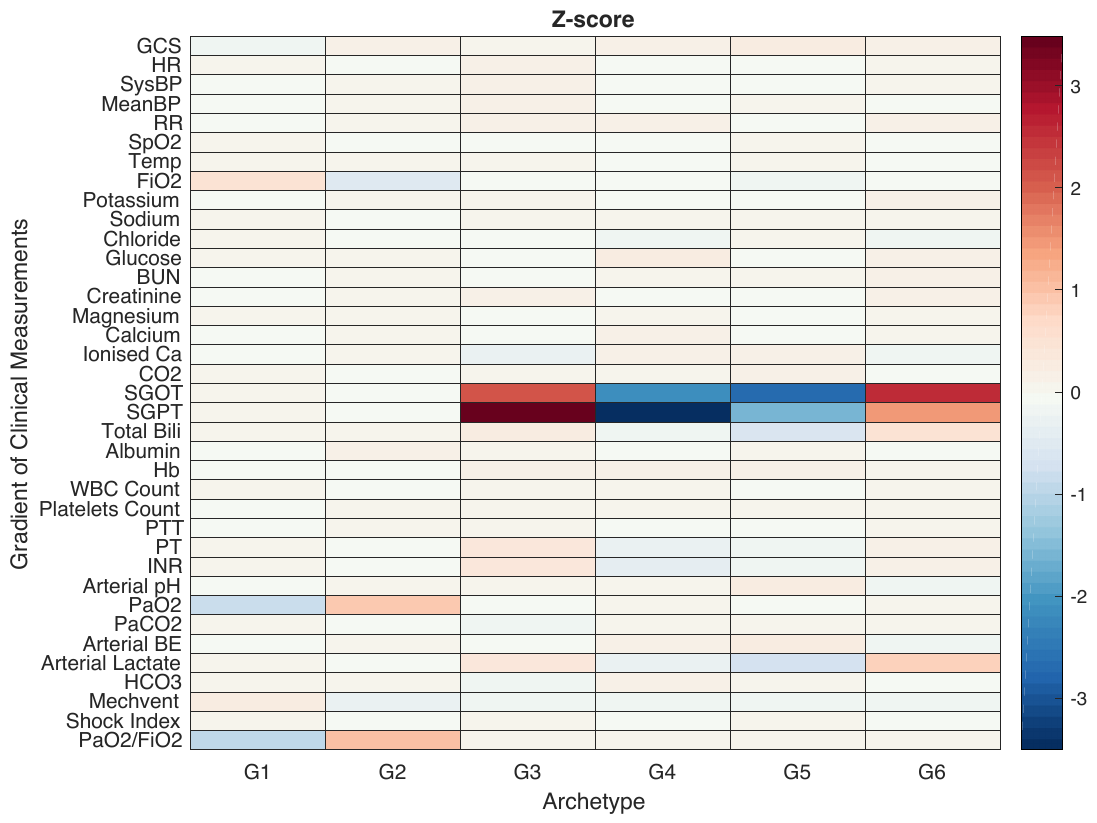}%
    \end{subfigure}\\
    \begin{subfigure}[h]{1.0\textwidth}
        \centering
        \caption{}\label{fig:gradient_histo}
        \includegraphics[width=14cm]{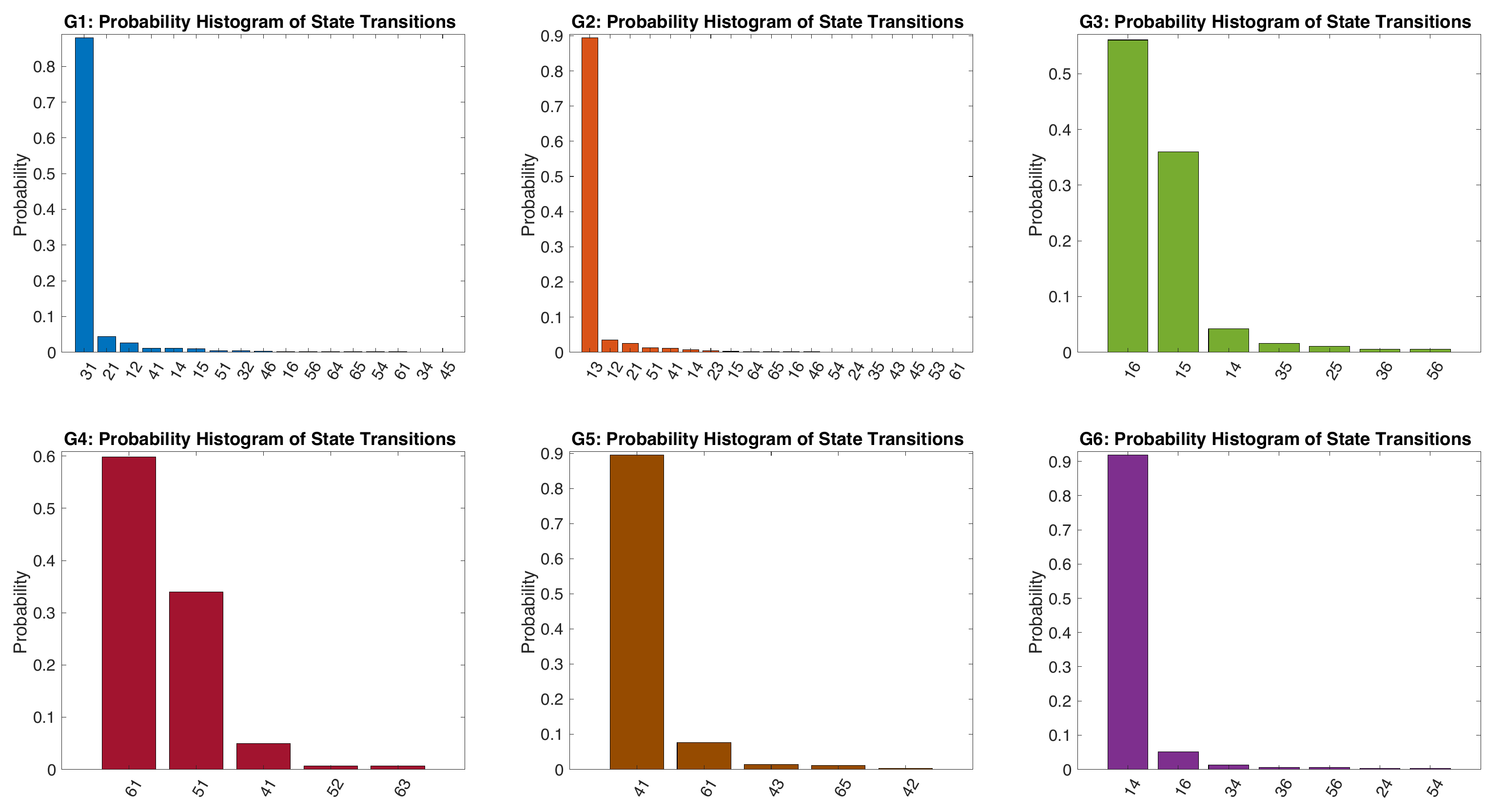}%
    \end{subfigure}\\
    
    \caption{{\textbf{(\textcolor{black}{A})} Visualization (using low-dimensional UMAP embedding) of the six derived gradient groups. \textbf{(\textcolor{black}{B})} Z-score analysis of the gradients of clinical measurements (row) of each gradient group (column). Entries approaching red indicate that the gradient of the clinical readings are expressed in the corresponding gradient groups, and entries closer to blue indicate that the gradient of the clinical readings are suppressed in corresponding gradient groups. \textbf{(\textcolor{black}{C})} Probability histogram of state transitions of each gradient group. } }
\end{figure}

\textcolor{black}{Fig}~\ref{fig:gradient_umap} shows a UMAP embedding of the gradients, along with the archetypes of gradients (represented by colors).

To identify discriminating components across gradient groups, we compute the z-score of each component in each gradient group and identify components that diverse significantly. \textcolor{black}{Fig}~\ref{fig:gradient_z_score} shows the z-score analysis of the gradients of clinical measurements of each gradient group. (Fig \ref{fig:pval-gradient} in Appendix shows  the corresponding p-values for the pairwise two-sample t-test for the gradients of clinical measurements of each gradient group.) We observe decreases in PaO2 and PaO2/FiO2 and an increase in FiO2 in the G1 group, indicating a decrease in respiratory function expression. We observe increases in PaO2 and PaO2/FiO2 and decrease of FiO2 in the G2 group, indicating an increase in respiratory function expression. \textcolor{black}{Finally, biomarkers for liver function decrease in the G4 and G5 groups (SGOPT and SGPT in the G4 group, and SGOT, SGPT, bilirubin, and arterial lactate in the G5 group), indicating milder damage to liver function.}

Finally, we examine if the identified gradient groups consist of disparate components of state transitions. \textcolor{black}{Fig} ~\ref{fig:gradient_histo} shows the probability histogram of state transitions of each gradient group. We observe strong connections between the change of biomarkers and state transitions; \textcolor{black}{G1 group consists of a large portion of transitions from state A3 to state A1; G2 group consists of a large portion of transitions from state A1 to A3; G3 and G6 groups consist of a large portion of transitions from non-MODS states to MODS states; G4 and G5 groups consist of a large portion of transitions from MODS states to non-MODS states.}

\section*{Discussion}
We present a computational framework to identify disease states and model pathological processes of sepsis from 16,546 distinct patients collected from the MIMIC-III database \cite{johnson2016mimic}. We identified six sepsis states based on the measurement of 42 variables (demographic profiles, vital signs, laboratory tests, mechanical ventilation status of the patients, and information on pre-existing clinical conditions) from these sepsis patients. \textcolor{black}{Among these states, State 1 manifests a moderate condition of sepsis, State 2 primarily represents inflammation and infection with evident signs of inflammatory response, State 3 corresponds to the highest survival rate, but is typically associated with hyperoxia. The last three states show signs and symptoms of Multiple Organ Dysfunction Syndrome (MODS) with diverse manifestations of organ failures. }

Our framework identified the most discriminating attributes for each sepsis state and showed that each state manifests a unique set of pathological responses, which correspond to different extents of organ dysfunction. These observations have two significant implications: (i) in contrast to the SOFA metric, our method identifies a larger number of attributes to provide a comprehensive view of sepsis symptoms, allowing for a more detailed diagnostic criterion; and (ii) it is possible to focus on a smaller set of attributes to differentiate sepsis symptoms, potentially reducing the associated diagnostic time and associated cost. Our identification of three distinct MODS states associated with a higher mortality rate, provides insight into advanced management of sepsis in ICU environments.

We also analyzed the association of different demographics and comorbidity profiles with identified sepsis states. Our results revealed that these sepsis states are composed of distinct populations with different demographics and comorbidity profiles, some of which have been supported in prior results. We find that a higher percentage of patients in MODS states had developed liver disease before the onset of sepsis, validating that patients with liver disease are more prone to developing severe sepsis \cite{gustot2009severe}. \textcolor{black}{Studies have shown inflammatory response during the tumor progression \cite{liu2015inflammation,zhao2021inflammation}. We find that patients with metastatic cancer are over-represented in state A2. The effect of diabetes on the outcome of sepsis remains controversial \cite{costantini2021type, schuetz2011diabetes, chang2012diabetic, esper2009effect}. We find that patients with uncomplicated diabetes are over-represented in state A3 and that patients with complicated diabetes are over-represented in state A4 (highest mortality rate). Fluid and electrolyte imbalances are commonly observed in critically ill patients with compromised kidney function\cite{lee2010fluid, jung2016electrolyte}. We find that patients with fluid and electrolyte imbalances are over-represented in state A4. Our etiological analysis, presented in  \textcolor{black} {Fig} \ref{fig:etiological} in Appendix, also shows other associations between comorbidity profiles and sepsis states. However, these associations might not be statistically significant. Further virtual clinical trials \cite{hernan2010causal} are needed to validate these associations. }

Although we used the most distinguishing attributes to analyze organ function, other attributes, including ABG, electrolytes, albumin, shock index, and hemoglobin, are also crucial in analyzing various aspects of the health status of sepsis patients. We also investigated these variables and found that these variables provide significant insight into patients' health status in different sepsis states; many of which are consistent with current literature. However, there are a few cases that are at odds with the current literature. We highlight a few cases here, and refer readers to the detailed description in the \textcolor{black}{Appendix} section: Analyses of other variables \ref{sec:other_var}. Metabolic acidosis often occurs in sepsis patients with organ failure, and metabolic alkalosis has been noticed in sepsis patients\cite{kreu2017alkalosis}. We also find that a higher percentage of metabolic acidosis occurs in MODS types. However, we observe fewer cases of metabolic alkalosis in our cohort. We also find that Arterial BE is a significant predictor of metabolic acidosis for sepsis patients. Hypocalcemia, measured by the concentration of ionized calcium or serum calcium in the body, might be observed in critically ill patients, especially in those with sepsis\cite{muller2000disordered,burchard1992hypocalcemia}, and is reported to be associated with increased severity of illness and increased mortality\cite{muller2000disordered}. We find that low ionized calcium concentrations coincide with worse outcomes, while low serum calcium concentrations do not.

By comparing the SIRS scores, SOFA scores, and mortality rates across the states, we confirm that the SIRS metric only identifies a subset of sepsis states \cite{rhodes2017surviving}. Furthermore, our results show that the SOFA metric covers a broader spectrum of disease states and is a more accurate predictor of mortality for sepsis patients\cite{ferreira2001serial}.

Finally, our framework provides insight into the complex states of sepsis and the pathological processes underlying state transitions. By analyzing the relationship between pre-existing comorbidities and sepsis states, changes in clinical measurements \textcolor{black}{treatment actions} during disease progression, and severity of each pathological trajectory, one can prognosticate individuals' outcomes and devise prevention and therapeutic strategies. \textcolor{black}{As a narrative example, for patients with suspected sepsis, we examine 21 features selected from our feature selection methods that differentiate sepsis states. Once the sepsis state is identified, finer management of sepsis disease can be applied to each sub-group based on the associated clinical variables and comorbidities that are uniquely expressed in these states. Using the transition graph, we predict patients' transition path and corresponding treatment actions. For example, in the first-order transition graph, if the patient were in state A4, the patient would be less likely to transition to state A5 or A6 and stay in state A4 in the next 4 hours. In addition, the average amounts of fluids, the dosage of vasopressors, and the usage of mechanical ventilators are presented as a reference point to guide treatment decisions.}

\textcolor{black}{In our patient cohort (AI clinician cohort), patients are defined as sepsis when the SOFA score is higher than two within the time window of 48 hours before and up to 24 hours after the onset of infection. However, Sepsis-3 requires a {\em change} in the SOFA score of two or more, consequent to infection. The baseline SOFA score can be assumed to be zero only when the presence of preexisting organ dysfunction is unknown. Therefore, although the patients in our cohort have a suspicion of infection, a subset of patients in our cohort have SOFA scores higher than two but do not manifest change in SOFA score of two or more over the baseline. Here, we analyze the number of non-sepsis patients included in our cohort if it was constructed under strict Sepsis-3 criteria. We define the change in SOFA score consequent to infection as the maximum SOFA score after onset (within the time window of up to 24 hours) minus the baseline SOFA score (within the time window of 48 hours before the onset) and identify the patient as sepsis if the difference is two or more. Baseline SOFA is measured as the minimum SOFA score among the available time points prior to the onset. If no available SOFA scores are present in the 48 hour time window, the baseline SOFA score is assumed to be zero. Our AI Clinician cohort includes 21,329 ICU stays. Of these 21,329 stays, 20,944 are included as our final cohort for the subsequent analyses since we focus on identifying sepsis states within the time window of 24 hours before and up to 48 hours after the onset of infection. The cohort constructed under the Sepsis-3 criteria includes 16,018 ICU stays, and all 16,018 of these ICU stays are also included in our AI Clinician cohort. Although there are 4,926 stays in our AI Clinician cohort that were excluded from the Sepsis-3 criteria since the change in SOFA score for these stays is smaller than 2, only 1,278 stays can be classified as non-Sepsis-3, as only these stays have available SOFA scores up to 48 hours before the onset to define baseline SOFA scores, as well as the available SOFA scores within 24 hours after the onset of sepsis to estimate the change in SOFA score. The other 3,648 stays are unknown with respect to their inclusion under Sepsis-3.}

\textcolor{black}{The current work characterizes correlation between the treatment actions and the state transitions. These correlations can be used to generate hypothesis for subsequent trials to establish causal relationships.
Virtual clinical trials \cite{hernan2010causal} or counterfactual queries \cite{schulam2017reliable,sundin2018using} are needed to estimate causal effects of treatment actions on the state transitions or clinical outcomes. Ongoing research focuses on designing virtual clinical trials and construction of causal models to further analyze sepsis states based on the pathophysiology and on learning personalized intervention strategies.
}

\iffalse
Our computational framework also provides insight into understanding the pathological processes in sepsis. A third-order Markov model reveals that most patients with sepsis begin in State 1 (moderate condition) and remain in State 1. We also find that patients who transition to State 1 from other states are likely to remain in State 1. Importantly, this analysis shows that identified Sepsis states are not developmental (i.e., one transitions from one state to the other as the disease progresses), rather that the states correspond to different `types' or manifestations. Specifically, there are only low-probability paths that transition across MODS type, suggesting that patients in different MODS type are composed of distinct groups of sub-populations. Our transition model successfully reflects the trend of the severity level of each pathological trajectory. Our analysis also reveals that the longer a patient remains in MODS states, the higher the mortality rate. Furthermore, if there is a transition from a non-MODS type to a MODS type, the mortality rate increases; conversely, if there is a transition from a MODS type to a non-MODS type, the mortality rate decreases.

Finally, our computational framework identifies distinct types of sepsis state transitions, each characterized by a distinct set of clinical transition biomarkers. 

\fi

\section*{Methods}

\subsection*{Patient cohort.}

\textcolor{black}{We use the sepsis cohort defined by AI Clinician \cite{komorowski2018artificial}. Patient samples were collected from five distinct ICUs in Boston, Massachusetts, stored in the Medical Information Mart for Intensive Care version III (MIMIC-III) database. Patients that are diagnosed with sepsis when they had both suspicion of infection, defined as the presence of the prescription of antibiotics and sampling of bodily fluids for microbiological culture (as also used in prior work of Nemati et al. \cite{nemati2018interpretable}, Johnson et al. \cite{johnson2018comparative}, and recommended by the Sepsis-3 criteria \cite{singer2016third}), and the evidence of organ dysfunction, defined as the total SOFA score higher than 2 (baseline SOFA scores are assumed to be zero.\cite{seymour2016assessment, raith2017prognostic}), within the time window of 48 hours before and up to 24 hours after onset of infection. As recommended by the Sepsis-3 criteria \cite{singer2016third}, the time of the onset is defined as the earliest event of the following two conditions: i) if the antibiotic was given first, the microbiological sample must have been collected within 24 hours; ii) if the microbiological sampling occurred first, the antibiotic must have been administered within 72 hours. Finally, sepsis patients whose age was less than 18 years old at the time of ICU admission, whose mortality was not documented, and withdrawal of treatment (vasopressors) were excluded from our cohort.}

\subsection*{Data preprocessing.}
Data were included from up to 24 hours prior to the estimated onset of sepsis and 48 hours after the onset to capture the characteristics of the early stages of sepsis. \textcolor{black}{The resulting cohort includes 16,546 distinct patients with 20,944 stays in ICUs.} Patients’ data are modeled as discrete multivariate time series with 4-hour time steps. Clinical variables associated with the patients are demographics, vital signs, lab values, severity measures such as SIRS and SOFA scores, and other information relating to use of ventilator and the number of comorbidities before sepsis infection. We also track whether the patient survived for 48 hours and extract the history of 30 types of comorbidities \cite{elixhauser1998comorbidity} before infection, as supplementary variables. Outliers were removed when the clinical variables were out of normal range (e.g., weight larger than 300 kg and blood pressure below 0). If there are multiple measurements within the 4 hour time period, either the max values are taken (e.g., vasopressor), are averaged (e.g., heart rate), or are accumulated (e.g., urine output and fluids). To address the issue of missing values,  we use time-limited sample-and-hold method \cite{hug2009detecting} for each variable with a different maximum hold time to identify missing values. Linear interpolation and k-nearest-neighbor methods were then used for the imputations \cite{tutz2015improved}.

\subsection*{Archetypal analysis.\cite{cutler1994archetypal}}
\label{sec:arch_analysis_formulation}

Archetypal analysis views each point in a dataset as a mixture (convex combination) of ``pure types'', or ``archetypes''. The convexity constraint here implies that in contrast to traditional clustering techniques that aim to identify ``typical'' representatives, archetypal analysis aims to identify ``extremal'' points in the dataset. The archetypes  are themselves mixtures (convex combinations) of the points in the data set. Archetypes can be learned by minimizing the squared error in representing each point as a mixture of archetypes. Specifically, let $\xx_{1},\dots,\xx_{n}$ be the data points in $\mathcal{R}^m$. The problem is to find a set of archetypes $\{\zz_{1},\dots,\zz_{K}\}$ so that each archetype $\zz_k$ is a convex combination of the data points, \textit{i.e.,} $\sum^n_{j=1} \beta_{kj}\xx_j$, with the constraints of: (i)  $\beta_{kj} \geq 0 \quad \forall j $; and (ii)      $\sum^n_{j=1}\beta_{kj}=1$ (convexity constraint), and that each data point $\xx_i$ can be best approximated by a convex combination of the archetypes, \textit{i.e.}, $\sum^K_{k=1}\alpha_{ik}\zz_k$, with the constraints: (i) $\alpha_{ik} \geq 0 \quad \forall i$; and (ii) $\sum^p_{k=1}\alpha_{ik}=1$. 

We can then define the following optimization problem: 
\begin{align}
    \min_{\{\alpha_{ik},\beta_{kj}\}}\sum^n_{i=1} \norm{\xx_i - \sum^p_{k=1}\alpha_{ik}\sum^n_{j=1}\beta_{kj}\xx_j }^2 \label{eq:AA_obj},
\end{align}
and the archetype problem is to find $\alpha$'s and $\beta$'s to minimize the objective \ref{eq:AA_obj} subject to the aforementioned constraints. This problem can be solved using general-purpose constrained nonlinear least squares methods, the alternating minimizing algorithm, or the projected gradient procedure. The learned archetypes (for $K$ $>$ 1) form a convex hull of the data set such that all of the points can be well-represented as a convex mixture of the archetypes. In our study, we first treat patient measurements as points in a high-dimensional space, and find archetypes for our cohort. These archetypes represent extreme states in sepsis, and each patient can be expressed as a convex combination of these states. We also use archetypal analysis to identify the archetypes of the transitions of clinical measurements. Please see Section \ref{sec:transition} for a detailed discussion.

\subsection*{Testing statistical significance of states.}\label{sec:significant-testing}

We characterize the statistical significance of each sepsis state based on the data points (patient records) mapped to the corresponding archetype from the cohort. Each point is a mixture of sepsis states, and we can assign the point to the closest sepsis state (the corresponding archetype). 
A statistical interpretation of this formulation views data points as mixtures of samples from the six distinct multivariate distributions. To ascertain that these distributions are indeed distinct, we first validate that the probability distributions corresponding to these groups are significantly different from the distribution of the cohort as a whole. We also test to ensure that the probability distribution of each group is significantly different from others, as characterized by a multivariate analysis of variance (MANOVA) procedure. Two-sample testing is sensitive to the homogeneity of covariance matrices from the compared populations. We use the Box test to compare variation in multivariate samples. We then use a Hotelling T-Squared testing variant to compare the mean vectors from two populations. We note that covariance matrices and mean vectors from the compared pairs are significantly different, with a 95 \% confidence interval. \textcolor{black}{Statistics of pairwise Hotelling t-square test across sepsis states and two-sample t-test for each variable for each sepsis state are shown in \textcolor{black}{Table} \ref{fig:t2-test} and \textcolor{black}{Table} \ref{fig:t-test}, respectively.} 

\subsubsection*{Comparing mean vectors from two populations.}

We use Two-sample Hotelling's T$^2$ tests to characterize significant differences between the mean vectors of two multivariate distributions (in reality, datasets drawn from these distributions). Two-sample Hotelling's T$^2$ tests are sensitive to violations of the assumption of equal variances and covariances. Different approximation of the sample variance is needed when the covariance matrices of the two populations are significantly different. We use Box’s M test for significant differences between covariance matrices.

\paragraph{Testing homogeneity of covariance matrices : Box’s M Test.}

Consider a sample set $\{\xx_{11},\dots,\xx_{1n_1}\}$ in $\mathcal{R}^m$ sampled from population $\Theta_1$ and a sample set $\{\xx_{21},\dots,\xx_{2n_2}\}$ in $\mathcal{R}^m$ sampled from population $\Theta_2$. Assuming that the sample sizes $n_1$ and $n_2$ are sufficiently large, Box’s M Test tests the null hypothesis that the population covariance matrices are equal, \textit{i.e.}, $H_0 : \Sigma_1=\Sigma_2$. Let $S_1$ and $S_2$ be the sample covariance matrices from the populations $\Theta_1$ and $\Theta_2$, where each $S_j$ is based on $n_j$ independent observations, we define the pool variance $S_{pooled}$ as follows:
\begin{align*}
    &S_{pooled} = \frac{1}{n_1+n_2-2} (n_1-1)S_1 +(n_2 -1)S_2,\\
\end{align*}
and the value of $M$ is given by:
\begin{align*}
    &M = (n_1+n_2 -2)\ln|S_{pooled}| - \left( (n_1 -1)\ln|S_1| +(n_2-1)\ln|S_2| \right).
\end{align*}
Then, $M(1-c)$  has an approximate $\chi^2_{df}$-distribution, where: 
\begin{align*}
    & c= \frac{2m^2+3m-1}{6(m+1)(n_1+n_2-1)}\left( 
    \frac{1}{n_1-1}+\frac{1}{n_2-1}-\frac{2}{n_1+n_2-2}\right),\\
    & df= \frac{m(m+1)(n_1+n_2-1)}{2}.
\end{align*}
The null hypothesis $H_0$ is rejected when $M(1 - c ) > \chi^2_{df}(\alpha)$ (or p-value < $\alpha$) .

\paragraph{Testing homogeneity of mean vectors : Hotelling’s T-Squared test.}

We test the equality of vector means from populations $\Theta_1$ and $\Theta_2$. The null hypothesis is that the population means are equal, \textit{i.e.,} $H_0 : \boldsymbol{\mu_1}=\boldsymbol{\mu_2}$. If Box’s M test indicates that the two covariance matrices are not significantly different, we can assume $\Sigma_1=\Sigma_2$ and:
\begin{align*}
    &T^2 = (\bar{\xx}_1-\bar{\xx}_2)^\intercal \left( (\frac{1}{n_1+n_2}) S_{pooled} \right)^{-1}(\bar{\xx}_1-\bar{\xx}_2).\\
\end{align*} 
If Box’s M test concludes that $\Sigma_1 \neq \Sigma_2$, 
\begin{align*}
    &T^2 = (\bar{\xx}_1-\bar{\xx}_2)^\intercal \left( \frac{1}{n_1}S_1 + \frac{1}{n_2}S_2\right)^{-1}(\bar{\xx}_1-\bar{\xx}_2)
\end{align*}
In either case, $T^2$ approximates chi-square distribution with $m$ degrees of freedom, \textit{i.e.,} $\chi^2_{m}$. The null hypothesis is rejected when $T^2> \chi^2_{m}(\alpha)$ (or p-value < $\alpha$) .

\subsection*{Low-dimensional embeddings of dataset.}

We use Uniform Manifold Approximation and Projection (UMAP)\cite{mcinnes2018umap} to compute a mapping from a dataset $X=\{\xx_1,\dots,\xx_n\}$ in $\mathcal{R}^m$ to its corresponding lower-dimensional representation $Y=\{\yy_1,\dots,\yy_n\}$ in $\mathcal{R}^d$ that preserves as much of the local and the global structure from the original space. UMAP assumes that the dataset $X$ is uniformly drawn from a Riemannian manifold $M$. With this assumption, the goal is to reconstruct $M$ and to find a mapping from $M$ into $\mathcal{R}^d$. To do so, UMAP first approximates the manifold and finds a  fuzzy simplicial set that captures all topological properties of the manifold $M$. Similarly, given a current lower-dimensional representation in $Y'$ of the data $X$ in $\mathcal{R}^m$, it can also construct a fuzzy simplicial set from $Y'$. Having the two fuzzy simplicial sets, one constructed from $X$ and the other constructed from $Y'$, UMAP then measures how good $Y'$ is as a representation of $X$ using cross-entropy $C$ of two fuzzy sets:
\begin{align*}
    C( (A,\mu),(A,\upsilon)) = \sum_{a\in A} \left(\mu(a)\log(\frac{\mu(a)}{\upsilon(a)}) + (1-\mu(a))\log(\frac{1-\mu(a)}{1-\upsilon(a)}) \right)
\end{align*}

The above objective function can be minimized using first-order optimization methods or second-order methods\cite{kylasa2020parallel,fang2018newton}.
%Once the objective is minimized, the corresponding solution $Y'$ is close to $Y$.

\subsection*{Feature selection methods.\cite{dudek2007cluster}}\label{sec:feature-selection}

\textcolor{black}{
Three criteria are developed to identify discriminative attributes for each state. The first method, which we refer to as the $Q_j(P_K)$, calculates the discriminative power of feature $i$ for a given clustering as the ratio of inter-cluster inertia to the total inertia computed using feature $i$. Intuitively, this method quantifies the heterogeneity of feature $i$ across clusters. The second method, which we refer to as $Q'_j(P_K)$, calculates the discriminative power of feature $i$ as the ratio of inter-cluster inertia computed using attribute $i$ to total inter-cluster inertia computed using all attributes. Intuitively, this method computes the relative heterogeneity of feature $i$ with respect to all other features. The third method uses a variation test that selects features with the lowest probability of overlap across clusters.}
\subsubsection*{Quality-index based approach.}

%{\bf ayg: can you clean the following paragraph? Are $x$'s the weights or $w$s the weights? If $x$s are the weights (as it would appear from the sentence following "Define ....", they what are $w$s?}
%\textcolor{black}{chf: just editted directly.}

Given a set $S = \{\xx_{1},\dots, \xx_{N}\}$ of $N$ points in $\mathcal{R}^m$ that is partitioned as $P_K= \{C_1,\dots,C_K \}$, where for each cluster pair $C_i, C_j$ $, 1\leq i,j,\leq K$ and $i\neq j$, $C_i\bigcap C_j = \Phi$, define $\bgg_k$ to be the mean values of the instances in cluster $C_k$, $\bgg$ be the average values over all the instances in $S$, the total inertia $T = \sum^N_{i=1} d^2(\xx_i,\bgg)
$ measures the dispersion of the points in the set $S$, where $d^2(\xx_i,\xx_{i'}) = \sum^m_{j=1} (x_{ij}-x_{i'j})^2 $ is the squared Euclidean distance. According to Huygens-Steiner Theorem, the total inertia $T$ can be decomposed into inter-cluster inertia $B$ and within-cluster inertia $W$:
\begin{align*}
    T &= B+W \\
       &= \sum^K_{k=1} d^2(\bgg_k,\bgg) + \sum^K_{k=1} I(C_k)\\
       &= \sum^K_{k=1} d^2(\bgg_k,\bgg) + \sum^K_{k=1}\sum_{i\in C_k} d^2(\xx_i,\bgg_k)
\end{align*}
Here, inter-cluster inertia $B$ measures the separation between the clusters, $I(C_k)$ is the inertia for cluster $C_k$, and the within-cluster inertia $W$ is the summation of the inertia of the clusters that measures the heterogeneity within the clusters.

\iffalse
\paragraph*{Quality Index.}
The quality index the ratio of the homogeneity value of each cluster and the corresponding homogeneity value associated with the partition $P_0 = S$, \textit{i.e.,} the entire dataset. This can be interpreted as the gap between the null hypothesis, \textit{i.e.,} partition into one cluster, and the given partition in $k$ clusters.

\paragraph{Partition Quality.} Given a set of points $S$ and a partition $P_k$, the null hypothesis is the total inertia $T$. Since the total inertia is a constant, the smaller the within-cluster inertia $W$, the larger the inter-cluster inertia $B$. The optimal strategy of assigning clusters is to minimize the inertia for each cluster. This leads to the following ratio as a measure of partition quality: 

\begin{align*}
    Q(P_k)=\frac{B}{T} = 1- \frac{W}{T}= 1 - \frac{I(W)}{I(T)}.
\end{align*}

\fi

\paragraph{Variable quality.} The \textit{quality index} is given by the ratio of the homogeneity value of each cluster and the corresponding homogeneity value associated with the partition $P_0 = S$. This can be interpreted as the gap between the null hypothesis, \textit{i.e.,} partition into one cluster, and partition into $k$ clusters. We can use the quality index at each variable (in our case, patient feature) $j$ to find the importance of the features.  Given a set of points $S$ partitioned by $P_K$, the null hypothesis is that the total inertia of the system is $T_j$. Since the partition $P_K$ is given, the total inertia of the system is fixed at $T_j$ and the optimal strategy of assigning clusters is to select minimal within-cluster inertia, $W_j$. This leads to the following ratio as a measure of variable quality: 
%
\iffalse
We can also define a quality measure in the variable level to compute the contribution of each variable  into the resulting clusters. Similar to the partition quality $Q(P_k)$, the variable quality can be defined as: 
\fi
%
\begin{align*}
    &Q_j(P_k)=\frac{B_j}{T_j} = 1- \frac{W_j}{T_j},\\
    &W_j= \sum_{i\in C_k} (x_{ij}-g_{kj})^2,\\
    &T_j= \sum^N_{i=1} (x_{ij}-g_j)^2.
\end{align*}
Alternatively, since the partition $P_K$ is given, we can ignore the variability of within-cluster inertia in the variable quality measure. That is, we only consider how each variable contributes to the total  inter-cluster inertia:
\begin{align*}
    &Q'_j(P_k)=\frac{B_j}{\sum^p_{i=1}B_i}.
\end{align*}

In our study, we use both  $Q_j(P_K)$ and $Q'_j(P_K)$ variable quality measures for feature selection. We also include a third feature selection criteria based on a variation test.

\subsubsection*{Variation test.}

Given a partition $P_K=\{C_1,\dots,C_K\}$ of a set of points $S$, we can regard the points in cluster $C_i$ as being sampled from a distinct multivariate probability distribution $\Theta_i$. To find the $j$-th feature that can distinguish the clusters, there exist at least two pairs of marginal distributions such that the difference of the mean value at the $j$-th feature sampled from the compared marginal distributions is larger than some threshold $\theta$ with probability $1-\delta$, where $\delta$ is sufficiently small.  We set $\theta=\sigma_{ij}$ when comparing clusters $C_i$ and $C_l$ ($l \neq i$) and collect the union of the variables from each case as the selected features. Although this feature selection method treats each dimension independently, we find that the selected features are similar to those using quality-index based approaches. See \textcolor{black}{S1 Text} for a detailed comparison.

\subsection*{Expression of primary functions.} \label{sec:primary_function_expression}

\textcolor{black}{
 We measure the overall expression of each of these primary functions for each sepsis state, shown as a  spider-plot of primary functions affected in each sepsis state in \textcolor{black} {Fig} \ref{fig:primary_function}, using the level of corresponding biomarkers, weighted by the confidence level (as measured by the number of overlaps between feature selection methods). For sepsis state $i$, we calculate the distance $d_i$ between readings from the biomarkers to the boundary of the normal range for each primary function. We then normalize each $d_i$ by dividing  by the maximum of the distance across sepsis states, \textit{i.e.}, $d_i/\max\{d_1,\dots,d_K\}$. Since each primary function can be tested by more than one biomarker, each of which has a different confidence level, we define the final expression of the primary function as a summation of normalized distances. The final value is linearly normalized into the range from 0 to 10, with higher values indicating higher expression of primary function. }

\textcolor{black}{\subsection*{Z-score analysis for comorbidity profiles.} We assess whether the comorbidity profile of interest is uniquely expressed or suppressed in a certain sepsis state. We compute the corresponding Z-score:}
\begin{align*}
    \textcolor{black}{z^i_{j} = \frac{w^i_{j}-\mu_j}{\sigma_j},}
\end{align*} 
\textcolor{black}{where $\mu_j$ is the rate of presence of comorbidity $j$ among all the patients in the cohort, $\sigma_j$ is its corresponding standard deviation, and $w^i_{j}$ is the reference point, calculated as the rate of the presence of comorbidity $i$ among patients who have passed through sepsis state $j$. Here, $z^i_{j}$ measures how far the reference point $w^i_{j}$ is from the population mean for the sepsis state $i$. If $z^i_{j}$ is positive, $w^i_{j}$ is expressed in sepsis state $i$, and vice versa.}

\subsection*{Analysis of Sepsis progression.}

\paragraph{Higher-order Markov chains.}
\label{sec:markov_chain}

Higher-order Markov chains are used to model transitions across sepsis states \cite{raftery1985model}. Formally, given a dataset (observational traces) $\mathcal{D}$ from m patients.
\begin{align*}
    \mathcal{D}&\triangleq \Big\{ \{\textbf{h}_i =  \{(t_{ij},\xx_{ij},a_{ij})\}_{j=1}^{n_i}  \Big\}_{i=1}^{m} \\ 
&= \left\{\begin{array}{lr}
        \textbf{h}_1=\{(t_{1j},\xx_{1j},a_{1j})\}_{j=1}^{n_1}\\
         \quad\quad\quad\quad\quad \vdots\\
        \textbf{h}_m=\{(t_{mj},\xx_{mj},a_{mj})\}_{j=1}^{n_m}
        \end{array}\right\} 
\end{align*} 
where $\textbf{h}_i$ denotes patient $i$'s \textit{trace} with $n_i$ time ordered points, and each point is defined as $(t,\xx,a)$, meaning that at time $t$, clinical measurements $\xx$, and sepsis state $a$ are observed. We model the transition of disease through these states using a higher-order Markov chain. The transition probability of $l$-th order Markov model can be written as:
\begin{align*}
  \Pr[\mathcal{S}_t=i_0|\mathcal{S}_{t-1}=i_1,\dots,\mathcal{S}_{t-1}=i_l] ,
\end{align*} where $l \geq 1$ denotes the order of a Markov chain and $i_j$ denotes the realization of sepsis state at time point $t-j$. The higher-order Markov model can be represented as a directed valued De Bruijn graph, denoted as $\mathcal{B}(K,l)$, where the set of vertices is given by:
\begin{align*}
  V &= \mathcal{A}^K \\ &=\{(a_1,\dots,a_1,a_1),(a_1,\dots,a_1,a_2),\dots,(a_1,\dots,a_1,a_K),(a_1,\dots,a_2,a_1),\dots,(a_K,\dots,a_K,a_K)\},
\end{align*}
and the set of edges is given by:
\begin{align*}
  E=\{((\hat{a}_1,\hat{a}_2,\dots,\hat{a}_l),(\hat{a}_2,\dots,\hat{a}_l,a_j)): a_i \in \mathcal{A}, 1\leq i \leq l , 1\leq j\leq K\},
\end{align*} where each edge takes the value of its corresponding transition probability.  In our study, we analyze sepsis transitions up to third order Markov models.

\paragraph{Identification of transition markers.}
\label{sec:transition}

Archetypal analysis and z-score analysis are used to identify transition markers across sepsis states. Formally, let transition dataset $\mathcal{G}=\{g_1,\dots,g_{m'}\}$ represent gradients of clinical measurements on transition from one sepsis state to another, collected from dataset $\mathcal{D}$ of $m$ patients. We first find distinct groupings of gradients using archetypal analysis. That is, given a transition dataset $\mathcal{G}=\{g_1,\dots,g_{m'}\}$, we aim to find a set of archetypes of gradients so that each gradient is a convex combination of archetypes and each archetype is a convex combination of the gradients. Once archetypes are identified, we define gradient states by mapping the gradient points to the corresponding archetype. We then compute z-scores to find the transition markers by finding the subset of features (components) of each cluster that are over-represented or suppressed. The z-score is computed as follows:
\begin{align*}
    z^{i}_{j} = \frac{\mu^{i}_{j}-\bar{\mu_j}}{\bar{\sigma_j}}.
\end{align*} 
Here, $\bar{\mu_j}$ is the mean value of $j$-th feature over the entire population, $\bar{\sigma_j}$ is its corresponding standard deviation, and $\mu^{i}_{j}$ is the mean value of $j$-th feature for gradient group $i$. The z-score $z^{i}_{j}$ measures how far the reference point $\mu^{i}_{j}$ is from the population mean for the gradient group $i$. If $z^{i}_{j}$ is positive, $\mu^{i}_{j}$ is expressed in gradient group $i$, and vice versa.

\section*{Acknowledgments}
The authors thank Dr. Poching DeLaurentis and Dr. Ping Huang for initial discussions.

%Acknowledgements should be brief, and should not include thanks to anonymous referees and editors, or %effusive comments. Grant or contribution numbers may be acknowledged.

\section*{Author contributions statement}

Conceptualization, C.H.F., V.R.; Formal Analysis: C.H.F., V.R. S.A., A.G.; Investigation, C.H.F., V.R. S.A., M.A., A.G.; Data Curation, C.H.F.; Writing – Original Draft, C.H.F., A.G.; Writing – Review \& Editing, C.H.F., V.R., S.A., M.A., P.G., S.S., and A.G., Visualization, C.H.F.; All authors have read and approve the manuscript.

\section*{Financial Disclosure Statement}

\textcolor{black}{This research is supported by grants from the US National Science Foundation (awards number 1908691, AG, 2019263, AG, and 0939370, AG).  C-HF and AG were funded by the grants from the US National Science Foundation. The funders had no role in study design, data collection and analysis, decision to publish, or preparation of the manuscript.}

\section*{Data Availability}
Researchers can follow the instructions from the website at \url{https://mimic.mit.edu/docs/gettingstarted/} to access the MIMIC-III database. Source code for data extraction from the MIMIC-III database, along with implementations of other methods, can be downloaded at \url{https://github.com/fang150/Sepsis\_Plos}.

%Must include all authors, identified by initials, for example:
%A.A. conceived the experiment(s),  A.A. and B.A. conducted the experiment(s), C.A. and D.A. analysed the %results.  All authors reviewed the manuscript. 

%\section*{Additional information}

%To include, in this order: \textbf{Accession codes} (where applicable); \textbf{Competing interests} (mandatory statement). 

%The corresponding author is responsible for submitting a \href{http://www.nature.com/srep/policies/index.html#competing}{competing interests statement} on behalf of all authors of the paper. This statement must be included in the submitted article file.

\bibliography{sepsis}

\newpage

\beginsupplement

\clearpage

\section*{Supplementary Material}
\subsection*{Analyses of other variables.}\label{sec:other_var}
\paragraph{Arterial blood gas (ABG).}
ABG is a blood test that assesses the gas exchange and acid-base balance of the body. It is an essential marker for critical patients admitted to ICUs. Arterial PH, PaO2, PaCO2, HCO3, and Arterial base excess are the main components of ABG. The presence of acidosis (arterial PH less than 7.35) or alkalosis (arterial PH higher than 7.45) is assessed by measuring arterial PH in the blood. Combined with arterial pH and PaCO2, one can measure the existence of respiratory acidosis or respiratory alkalosis in the body. Respiratory acidosis occurs when PaCO2 is higher than 45 mmHg, with an arterial PH less than 7.35. Respiratory alkalosis occurs when PaCO2 is less than 35 mmHg, with an arterial PH higher than 7.45. Combined with arterial PH and HCO3, one can measure the existence of metabolic acidosis or metabolic alkalosis. Metabolic acidosis occurs when HCO3 is less than 22 mEq/L, with an arterial PH less than 7.35. Metabolic alkalosis occurs when HCO3 is higher than 28 mmHg, with an arterial PH higher than 7.45.  Among four types of acid-base disorders (respiratory acidosis, respiratory alkalosis, metabolic acidosis, and metabolic alkalosis ), metabolic acidosis is common in sepsis patients with organ failure \cite{ kellum2004metabolic}. Metabolic alkalosis has been noted in sepsis patients\cite{kreu2017alkalosis}.
In our cohort, we find that the average values of arterial PH and HCO3 in MODS states are lower than non-MODS states, indicating that a higher portion of patients with metabolic acidosis is observed in MODS states. The average values of arterial PH for states A1 through A6 are 7.39, 7.40, 7.39, 7.35, 7.37, and 7.36, respectively. The average values of HCO3 for states A1 through A6 are 24.68, 25.06, 24.47, 22.79, 23.08, and 21.93, respectively. The percentage of \textcolor{black}{cases} with metabolic acidosis in states A1 through A6 are 8.7\%, 6.6\%, 6.7\%, 22.8\%, 16.4\%, and 20.5\%, respectively. On the other hand, we observe fewer cases of metabolic alkalosis in our cohort. The percentage of \textcolor{black}{cases} with metabolic alkalosis in states A1 through A6 are 4.5\%, 6.2\%, 3.9\%, 4.7\%, 2.6\%, and 1.1\%, respectively.

\paragraph{Arterial base excess (Arterial BE).}
Arterial BE reflects the metabolic component of the acid-base balance. Arterial BE measures the amount of H+ required to return the blood PH to normal when PaCO2 is within the normal range, with the normal range from -2 to +2.  The base excess increases in metabolic alkalosis and decreases (or becomes negative) in metabolic acidosis. Metabolic acidosis is common in sepsis patients with organ failure \cite{ kellum2004metabolic}, and metabolic alkalosis can also occur in sepsis patients \cite{kreu2017alkalosis}. In our study, we find that the average value of arterial BE in MODS states is negative, indicating an inclination towards metabolic acidosis in MODS states. The average values of arterial BE for states A1 through A6 are 0.35, 0.80, 0.37, -1.9, -1.94, and -2.5, respectively, and the percentage of \textcolor{black}{cases} with metabolic acidosis in states A1 through A6 are 8.7\%, 6.6\%, 6.7\%, 22.8\%, 16.4\%, and 20.5\%, respectively. Therefore, arterial BE is an important predictor of metabolic acidosis for sepsis patients.

\paragraph{Albumin.} 
 Albumin is one of the essential proteins, with a normal range of 3.4 to 5.4 g/dl. Albumin is responsible for plasma colloid osmotic pressure, acting as a major binding protein for endogenous and exogenous compounds (drugs), with antioxidant and anti-inflammatory properties, and operates as a buffer to balance acid-base status of the body. A lower albumin level (Hypoalbuminemia) is often observed when: (i) patients have nutritional deficiencies \cite{keller2019nutritional}; (ii) patients develop chronic liver disease, advanced hepatic cirrhosis, or end-stage renal disease \cite{gatta2012hypoalbuminemia}; or (iii) an inflammation is present \cite{don2004poor}. Albumin, in addition to crystalloids, is often used for initial resuscitation and subsequent intravascular volume replacement in patients with sepsis and septic shock \cite{rhodes2017surviving}. Although albumin administration is widely used in the management of sepsis, the benefit of the use of albumin for resuscitation in this population remains controversial -- while several meta-analyses have shown that the administration of albumin in ICU patients has beneficial effects on health outcomes\cite{vincent2016fluid,xu2014comparison}, other studies have shown contradictory results\cite{patel2014randomised}. However, the results are less conclusive when the included studies have differing experimental design and comparison groups. As a result, albumin administration is suggested with weak confidence \cite{rhodes2017surviving}. In our study, we find hypoalbuminemia in all sepsis states, and the albumin level in various sepsis states is not significantly different. Future development of clinical trials may focus on comparing the effects of albumin administration on the health outcomes for different sepsis states.

\paragraph{Hemoglobin (Hb).} The normal range of hemoglobin for males and females is 13.5 to 17.5 grams per deciliter and 12.0 to 15.5 grams per deciliter, respectively. A lower hemoglobin level in the body indicates a low red blood cell (RBC) count (Anemia). Anemia is common in sepsis due to inflammation, liver and renal impairment, and cancer. RBC transfusion is strongly recommended for patients with sepsis when hemoglobin concentration falls to less than 7.0 g/dL in adults in the absence of extenuating circumstances, such as myocardial ischemia, severe hypoxemia, or acute hemorrhage\cite{rhodes2017surviving}. We find that the average Hb values in all sepsis states are lower than the normal range, with A2 (Inflammation state) having the lowest Hb values. The average Hb values for states A1 through A6 are 10.30, 9.49, 10.49, 10.42, 10.90, and 10.75, respectively.

\paragraph{Shock index (SI).}
Shock index (SI) is a bedside assessment defined as heart rate (HR) divided by systolic blood pressure (SysBP), with a normal range of 0.5 to 0.7 in healthy adults.  SI is suggested as a measure in the triage and management of critically ill patients. Its use is also suggested as a predictor of clinical outcomes, such as the serum lactate level in the body, the risk of mortality, and other markers of morbidity\cite{tseng2015utility}. However, a retrospective database review shows that SI does not correlate with the mortality rate in emergency room patients\cite{liu2012modified}. Our results support this claim. We find that SI is not highly associated with worse outcomes. While the MODS states display higher SI than A1 and A3 states, A2 (Inflammation state) displays the highest SI, which is the state associated with lower SOFA score and mortality rate. The average SI values for states A1 through A6 are 0.75, 0.81, 0.72, 0.77, 0.79, and 0.77, respectively. Furthermore, SI is not an independent predictor of hyperlactatemia (serum lactate $\geq$ 4.0 mmol/L): while A2 state manifests the highest average SI, it displays the lowest arterial lactate. The average arterial lactate levels for states A1 through A6 are 2.04, 1.88, 2.10, 5.50, 3.95, and 4.58.

\paragraph{Ionized calcium.}
In health, serum ionized calcium concentration is maintained between approximately 1.16 and 1.32 mmol/L. Ionized hypocalcemia (ionized calcium levels $<$ 1.16 mmol/L) are common in critically ill patients with sepsis, cardiac failure, pulmonary failure renal failure, post-surgery or burns \cite{muller2000disordered}. Recent studies show that low ionized calcium concentrations coincide with increased severity of illness and increased mortality\cite{muller2000disordered}. We find that ionized hypocalcemia occurs in all sepsis states, and the average ionized calcium concentration in MODS states is lower than non-MODS states. The average ionized calcium concentration for states A1 through A6 are 1.13, 1.13, 1.14, 1.08, 1.09, and 1.09, respectively.

\paragraph{Calcium.}
The normal range of total serum calcium concentration is 8.8 mg/dL to 10.7 mg/dL. Hypocalcemia, defined as serum calcium concentration less than 8.8 mg/dL or serum ionized calcium concentration less than 4.7 mg/dL, is common in critically ill patients, especially in those with sepsis\cite{muller2000disordered}. We find that hypocalcemia occurs in all sepsis states. However, low serum calcium concentrations do not coincide with the increased severity of illness and increased mortality. The serum calcium concentrations for states A1 through A6 are 8.31, 8.39, 8.38, 8.50, 8.13, and 8.20, respectively.

\paragraph{Magnesium.}
Magnesium is a vital element involved in various physiological processes and an essential cofactor in more than 300 enzymes, with a normal range of 1.5 to 2.5 mEq/L in healthy adults. Hypomagnesemia (serum Mg levels $<$ 1.5 mEq/L) can occur in critically ill patients, including sepsis patients, and is associated with prolonged ICU stay, increased need for mechanical ventilation, and increased mortality. In our study, we find that, in the average case, serum magnesium levels in all sepsis states are within the normal range. The average values of serum magnesium levels for states A1 through A6 are 2.06, 2.07, 2.02, 2.10, 2.04, and 2.29, respectively. We note that recent studies have shown that the administration of magnesium sulfate increases lactate clearance in critically ill patients with severe sepsis\cite{noormandi2020effect}, improve cerebral perfusion in patients with sepsis-associated encephalopathy (SAE)\cite{shaban2019effect}, and may be used to open up small vessels, to reduce organ failure for patients with severe sepsis and septic shock\cite{pranskunas2011microcirculatory}. Future development of clinical trials may focus on comparing the effects of the administration of magnesium sulfate on the health outcomes for different sepsis states.

\paragraph{Chloride.}
Chloride is an essential anion of the extracellular fluid, representing two-thirds of all negative charges in plasma and accounting for nearly one-third of plasma tonicity. Normal Serum chloride concentrations range from 96 to 106. mEq/L. Abnormal chloride levels in the blood (hypo- and hyperchloremia) are observed in critically ill patients. However, evidence on the effects of hypo- and hyperchloremia on the clinical outcomes, such as length of stay and mortality rate, are sparse~\cite{filis2018hyperchloraemia}. A recent study shows that hyperchloremia is not significantly related to an increased mortality rate, and hypochloremia is associated with increased mortality in patients with severe sepsis or septic shock~\cite{oh2017increased}. Our results are consistent with these findings: we observe a higher percentage of hypochloremia in MODS groups, which are the groups associated with a higher mortality rate. We also observe that hyperchloremia does not directly correlate with MODS groups. The percentage of \textcolor{black}{cases} with hypochloremia for states A1 through A6 are 7\%, 7\%, 5\%, 15\%, 12\%, and 10\%, respectively, and the percentage of \textcolor{black}{cases} hyperchloremia for states A1 through A6 are 39\%, 27\%, 36\%, 30\%, 35\%, and 35\%, respectively.

\iffalse 
Future development of clinical trials may focus on comparing the effects of chloride levels on clinical outcomes in different sepsis states.
\fi

\paragraph{Sodium.} The normal sodium level in the blood is 135 to 145 mEq/L. Hypernatremia (serum sodium concentration $>$ 145 mEq/L ) is an uncommon but important electrolyte abnormality in ICU patients. Hypernatremia also occurs in sepsis patients, but only a few studies\cite{de2019there,ni2016risk} have investigated the effect of serum sodium levels on the clinical outcomes in sepsis patients. Studies have shown that patients admitted with hypernatremia are significantly more likely to have sepsis \cite{de2019there} and that hypernatremia is strongly associated with worse outcomes in sepsis \cite{ni2016risk}. However, we notice that sepsis patients with hypernatremia only constitutes 7.3\% in the cohort and that although we find that state A4 (the state with highest mortality rate) constitutes the highest portion of patients with hypernatremia,  state A5 (the state with second-highest mortality rate) has a lower portion of patients with hypernatremia than the cohort. 

The percentage of \textcolor{black}{cases} with hypernatremia for states A1 through A6 are 7.3\%, 8.1\%, 4.1\%, 10.8\%, 4.5\%, and 8.8\%, respectively.

\paragraph{Potassium.}
Potassium is one of the electrolytes mostly present in intracellular fluid. The normal value of serum potassium is 3.5 to 5.0 mEq/L. Potassium homeostasis is important for negative resting membrane potential, neuromuscular, and cardiac excitability. Abnormal potassium has adverse effects on the heart: both hypo and hyperkalemia cause cardiac arrhythmia. Hypokalaemia also causes muscle paralysis, including respiratory muscles and GIT. Potassium abnormality can occur in critically ill patients in ICU due to organ derangement and some medications, and is associated with an increased complication rate and mortality risk \iffalse \cite{gennari1998hypokalemia} \fi. Our study found that the average potassium levels do not vary across sepsis states and are within the normal range in our cohort. The average potassium concentration for states A1 through A6 are 4.08, 4.22, 4.14, 4.31, 4.06, and 4.34, respectively.

\subsection*{SOFA and SIRS scores.}\label{sec:sofa_n_sirs}

\paragraph{Systemic Inflammatory Response Syndrome (SIRS).}

\textit{Sepsis} was first defined as a  systemic inflammatory response syndrome (SIRS) \cite{bone1992definitions}. SIRS is the clinical presentation of the host response to inflammation. It manifests in four symptoms, temperature $\geq$ 38 degree Celsius or $\leq$ 36 degree Celsius, respiratory rate $\geq$ 20 breaths/minute or  PaCO\textsubscript{2} $<$ 32 mm of Hg, heart rate $>$ 90 beats/minute, white blood count $>$12000/mm3 or $<$ 4000/mm\textsuperscript{3} or bands $>$ 10\%. Two or more symptoms of SIRS indicate a SIRS positive case. However, it was argued that SIRS is not an adequate measure, since a sepsis-related symptom may be observed without infection. Due to its non-specific issue, the diagnostic metric of sepsis was first replaced by sepsis-2 \cite{levy20032001}, and finally changed to Sequential Organ Failure Assessment Score (SOFA).

\paragraph{Sequential Organ Failure Assessment\cite{singer2016third} (SOFA) Score.}\label{sec:SOFA}

SOFA measures the functionality of six organ systems -- respiratory, coagulation, cardiovascular, neurological, liver, and renal, shown in \textcolor{black}{Table} \ref{tab:SOFA}, each of which is measured by PaO\textsubscript{2}/FiO\textsubscript{2} ratio, platelet count, mean arterial pressure (MAP), Glasgow coma score (GCS), bilirubin, and creatinine or urine output, respectively. Each system is assigned a score from 0 to 4. The worst condition represents the highest score. The range of the SOFA score is 0-24. It has been shown that the SOFA score is a good predictor of mortality in intensive care units~\cite{jentzer2018predictive,jones2009sequential}. 
%As the SOFA score increase, the risk of mortality also increases. 

\iffalse

\subsubsection{Glasgow Coma Scale (GCS).}
GCS is the most commonly used measure for bedside assessment of brain injury. It is also used for assessing the consciousness level of critically ill patients, including sepsis patients. 

%By using GCS, usually we can see the consciousness level of the patients which helps to guess the improvement or deterioration of patient’s condition. 

\subsubsection{Mechanical Ventilation.\cite{hickey2020mechanical,pham2017mechanical}}

\paragraph{Indications of mechanical ventilation.}
\begin{itemize}
    \item Airway protection in a patient who is obtunded, or has a dynamic airway, \textit{e.g.}, from trauma or oropharyngeal infection.
    \item Hypercapnic respiratory failure due to a decrease in minute ventilation.
    \item Hypoxemic respiratory failure due to a failure of oxygenation.
    \item Cardiovascular distress, whereby mechanical ventilation can offload the energy requirements of breathing.
    \item Expectant course, \textit{e.g.}, anticipated patient decline or impending transfer.
\end{itemize}

\fi

\subsection*{Performance analysis for the feature selection methods.}\label{sec:feature_sel}

We compare the top 15 features selected by (i)$Q_j(P_k)$ method measuring the ratio of inter-cluster inertia at the i-th feature $B_i$ to the total inertia at the i-th feature $T_i$, (ii) $Q'_j(P_k)$ method measuring the ratio of the between-cluster inertia at i-th feature $B_i$ to the total between-cluster inertia $\sum^p_{i=1}B_i$, (iii) and the feature that has a lower probability of overlapping between clusters.

%
%
%We compare the top 15 features selected by: (i) $Q_j(P_k)$ method, which calculates the percentage of the between-cluster at the i-th feature inertia is of the total inertia at the i-th feature, (ii) $Q'_j(P_k)$ method which calculates the percentage of the between-cluster inertia at i-th feature is of the total between-cluster inertia, and (iii) the variation test which selects the feature that has a lower probability of overlapping between clusters 
%
The selected features are shown in \textcolor{black}{Table} \ref{tab:features}. We observe that the first five features selected by  $Q_j(P_k)$ and $Q'_j(P_k)$ are identical -- SGOT, SGPT, PaO2/FiO2, PaO2, and Platelet Count, in the same order, except for PaO2 and Platelets. For the rest of the features selected by $Q_j(P_k)$ and $Q'_j(P_k)$, PT, WBC Count, Arterial Lactate, Age, and HR were selected by both $Q_j(P_k)$ and $Q'_j(P_k)$. Among  the 10 features mentioned above, 8 are selected by the variation test as well. We present a Venn diagram for the inclusion-exclusion comparison between $Q_j(P_k)$, $Q'_j(P_k)$, and the variation test in Fig  \ref{fig:ven_diag}. We observe that the three methods are largely consistent in selecting representative features from the clusters.

\subsection*{Consistency analysis for archetypal analysis.}\label{sec:consistency}

\textcolor{black}{We regarded each time point as a snapshot of patient status and used archetypal analysis to find distinct states, and higher-order transition probabilities across states identified through archetypal analysis. The validity of transitions primarily relies on the stability or consistency of finding sepsis states from archetypal analysis. If the identification of sepsis state for each time point is significantly different for different runs, the validity of sepsis state identification and that of the constructed temporal models suffers. To test the stability of archetypal analysis, we used archetypal analysis on the sepsis dataset 20 times to compare the consistency of results between runs. Archetypal analysis is an unsupervised technique. Thus, we chose Normalized Mutual Information (NMI) \cite{estevez2009normalized} and Adjusted Rand Index (ARI) \cite{milligan1986study} as the consistency metric and measured how well the identified state for each time point has a one-to-one mapping relationship across runs. The average value of NMI and ARI are 0.9961 and 0.9732 with standard deviations 0.0011 and 0.0052, respectively. }

\clearpage

\begin{figure*}[h]
   \centering
   
        \begin{subfigure}[h]{0.5\textwidth}
        \caption{ }\label{fig:elbow}
        \includegraphics[width=8.5cm]{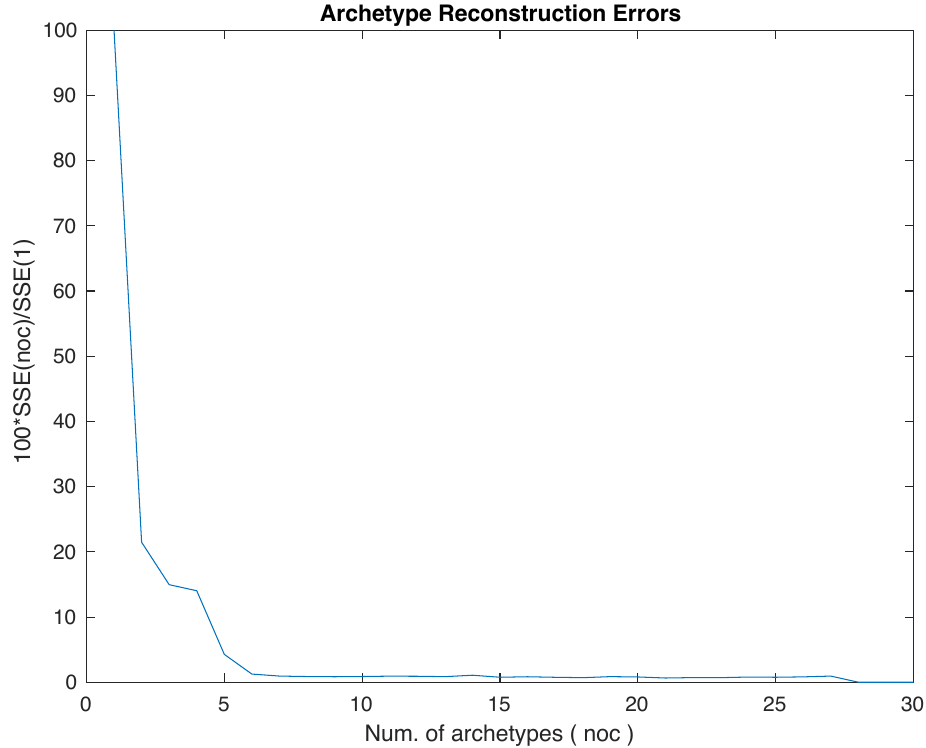}
        \end{subfigure}~
        \begin{subfigure}[h]{0.5\textwidth}
        \caption{ }
        \includegraphics[width=8.5cm]{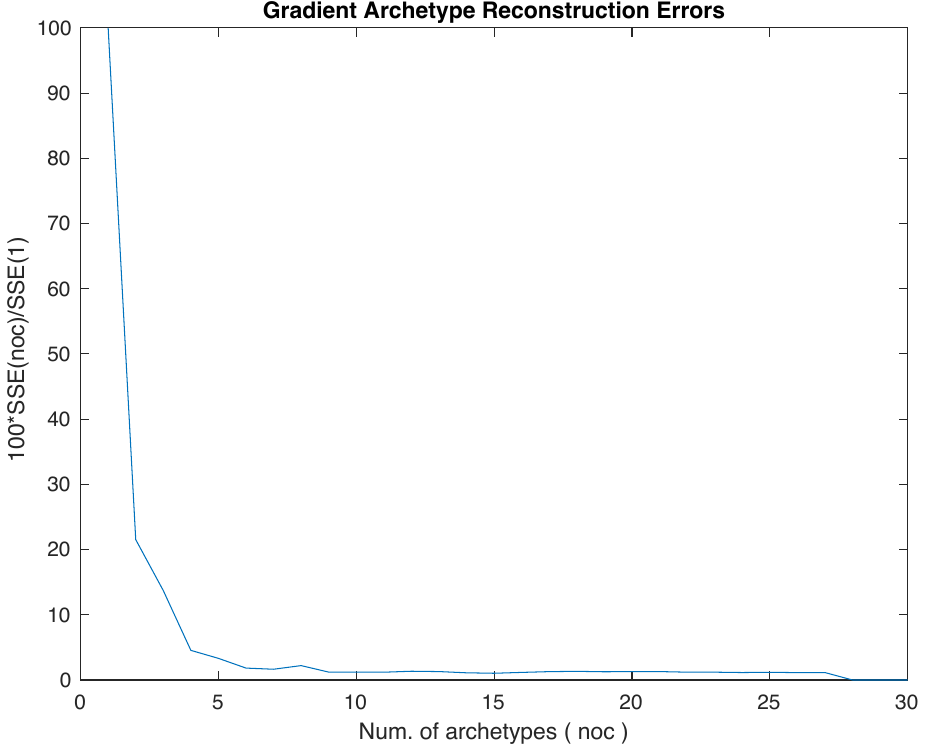}
        \end{subfigure}
    
        \caption{  {\textbf{(a)} \textcolor{black}{Elbow method for finding optimal number of archetypes.} \textbf{(b)} \textcolor{black}{Elbow method for finding optimal number of gradient archetypes.}}  }\label{fig:S1}
        
\end{figure*}

\begin{figure*}[h]
   \centering
   
        \begin{subfigure}[h]{0.5\textwidth}
        \caption{ }
        \includegraphics[width=8.5cm]{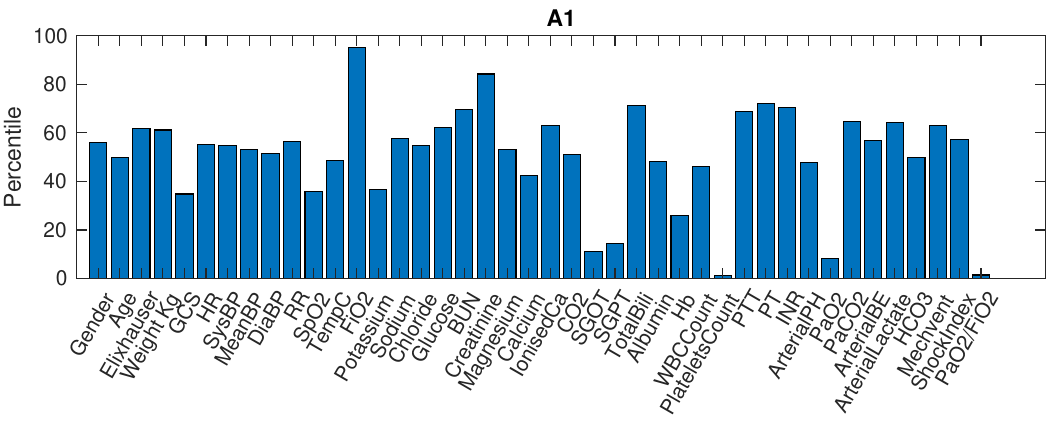}
        \end{subfigure}~
        \begin{subfigure}[h]{0.5\textwidth}
        \caption{}
        \includegraphics[width=8.5cm]{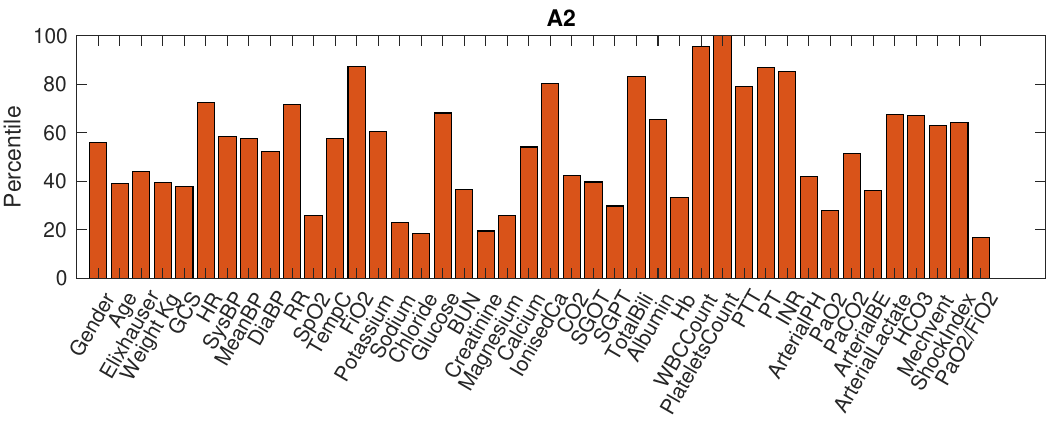}
        \end{subfigure}\\
        \begin{subfigure}[h]{0.5\textwidth}
        \centering
        \caption{}
        \includegraphics[width=8.5cm]{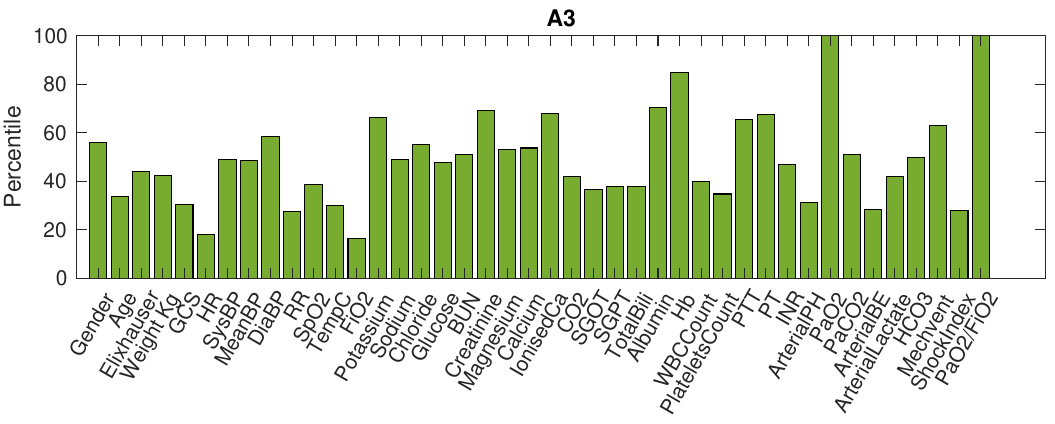}
        \end{subfigure}~
        \begin{subfigure}[h]{0.5\textwidth}
        \centering
        \caption{}
        \includegraphics[width=8.5cm]{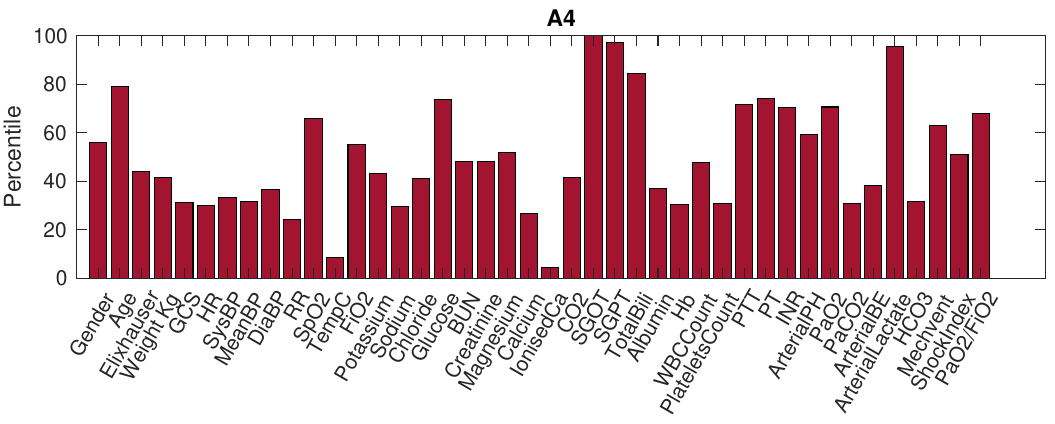}
        \end{subfigure}\\
        \begin{subfigure}[h]{0.5\textwidth}
        \centering
        \caption{}
        \includegraphics[width=8.5cm]{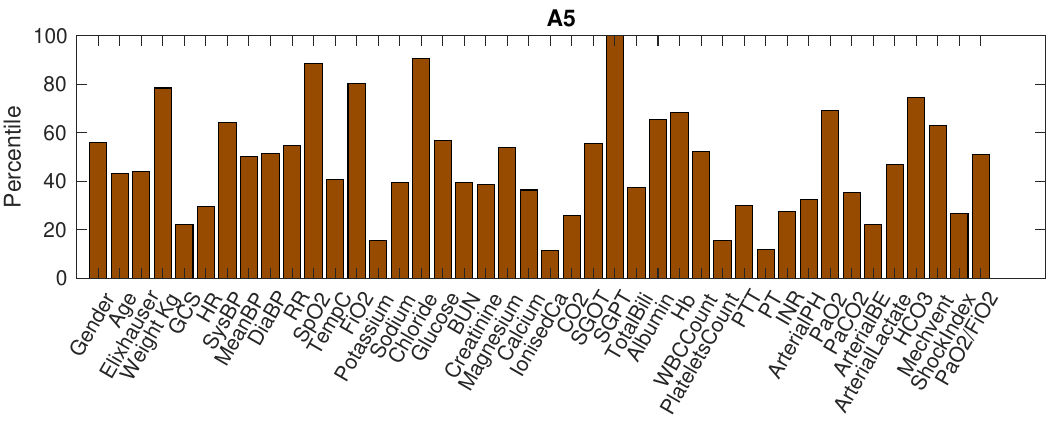}
        \end{subfigure}~
        \begin{subfigure}[h]{0.5\textwidth}
        \centering
        \caption{}
        \includegraphics[width=8.5cm]{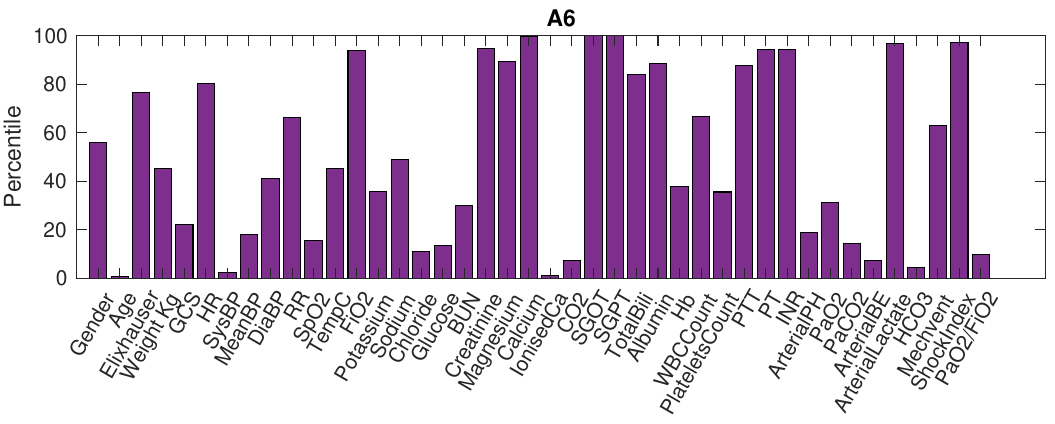}
        \end{subfigure}
    
        \caption{ Percentile value (y-axis) of each variable (x-axis) in an archetype, as compared to the overall cohort. }\label{fig:archetypes}
\end{figure*}

\begin{figure*}[h]
   \centering
   
        \begin{subfigure}[h]{1.0\textwidth}
        \caption{ }
        \centering
        \includegraphics[width=12cm]{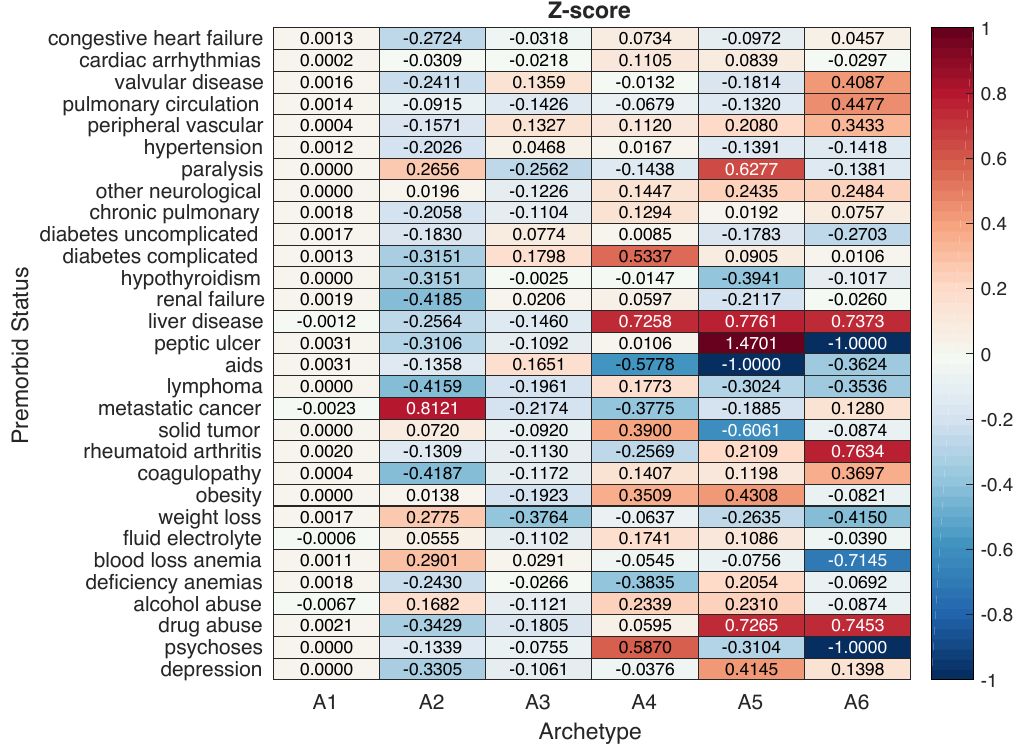}
        \end{subfigure}\\
         \begin{subfigure}[h]{1.0\textwidth}
         \centering
        \caption{ }
        \includegraphics[width=12cm]{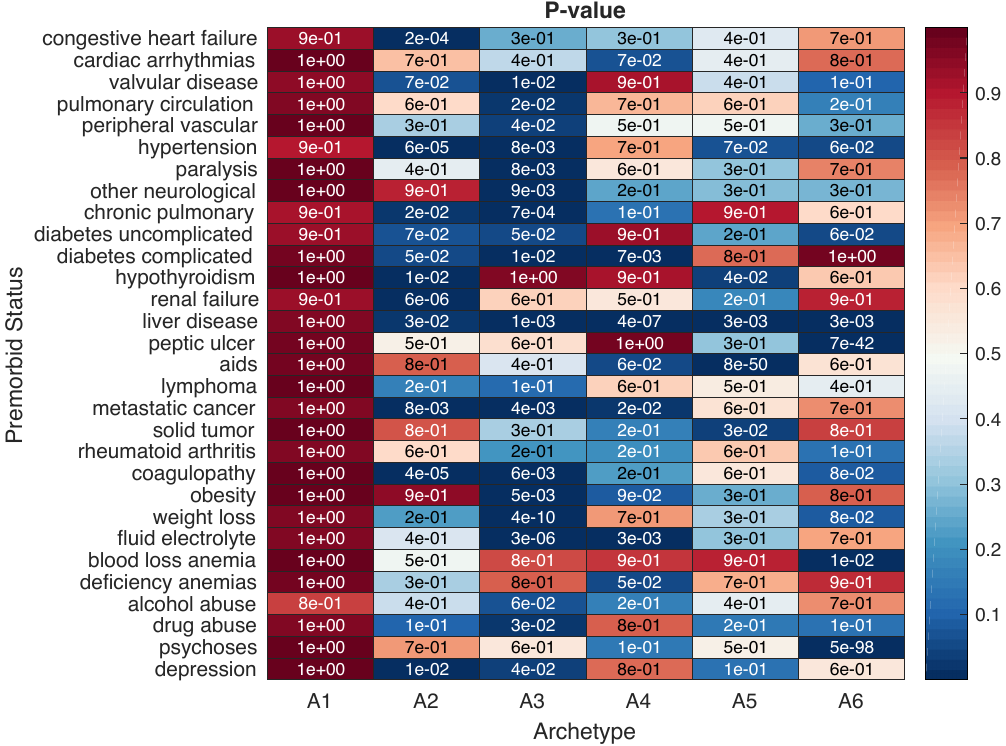}
        \end{subfigure}
        
        \caption{ \textcolor{black}{\textbf{(\textcolor{black}{A})} Z-score analysis of comorbidity profiles (row) of each sepsis type (column). Entries approaching red in intensity indicate that the comorbidity profiles are expressed in the corresponding sepsis states, and entries closer to blue indicate that the comorbidity profiles are suppressed in corresponding sepsis states. \textbf{(\textcolor{black}{B})} P-values for the pairwise two-sample t-test for the comorbidity profiles (row) of each sepsis type (column). Statistically Significant entries approach blue.} }\label{fig:etiological} 
\end{figure*}

  \begin{figure}[ht!]

   \centering
   
        \begin{subfigure}[h]{1.0\textwidth}
        \centering
        \caption{}\label{fig:2-order-Markov}
        \includegraphics[width=18.0cm]{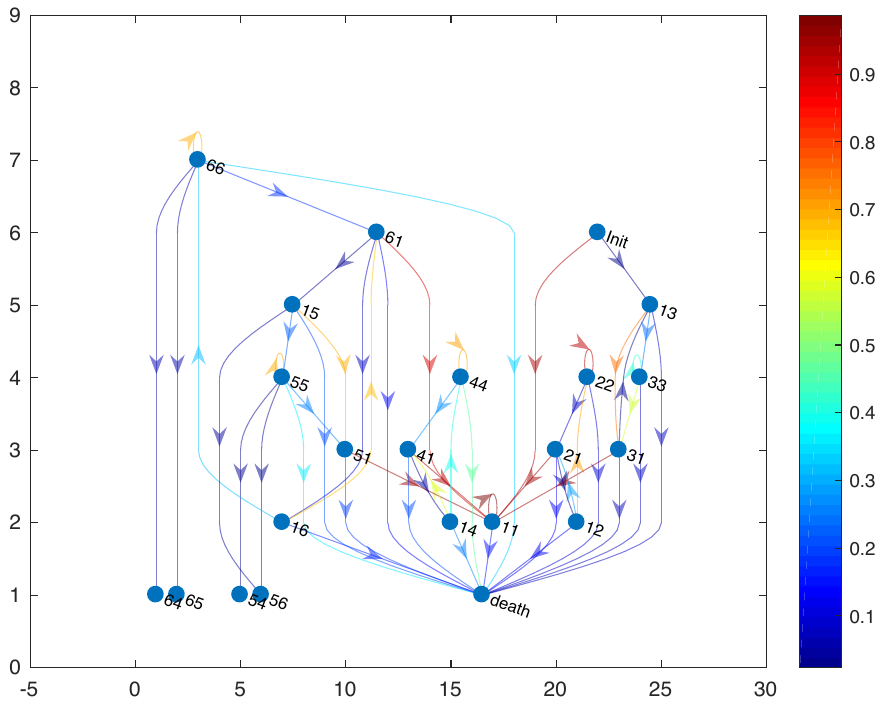}
        \end{subfigure}
    
    \caption{ {\textcolor{black}{\textbf{ (\textcolor{black}{A})} Second-order transition graph: Edges approaching red in color indicate higher transition probabilities, and edges approaching black indicate lower transition probabilities. }} }\label{fig:higher-order-markov}
    
\end{figure}

\begin{figure*}[h]
   \centering
   
         \begin{subfigure}[h]{1.0\textwidth}
        \caption{ }
        \centering
        \includegraphics[width=12cm]{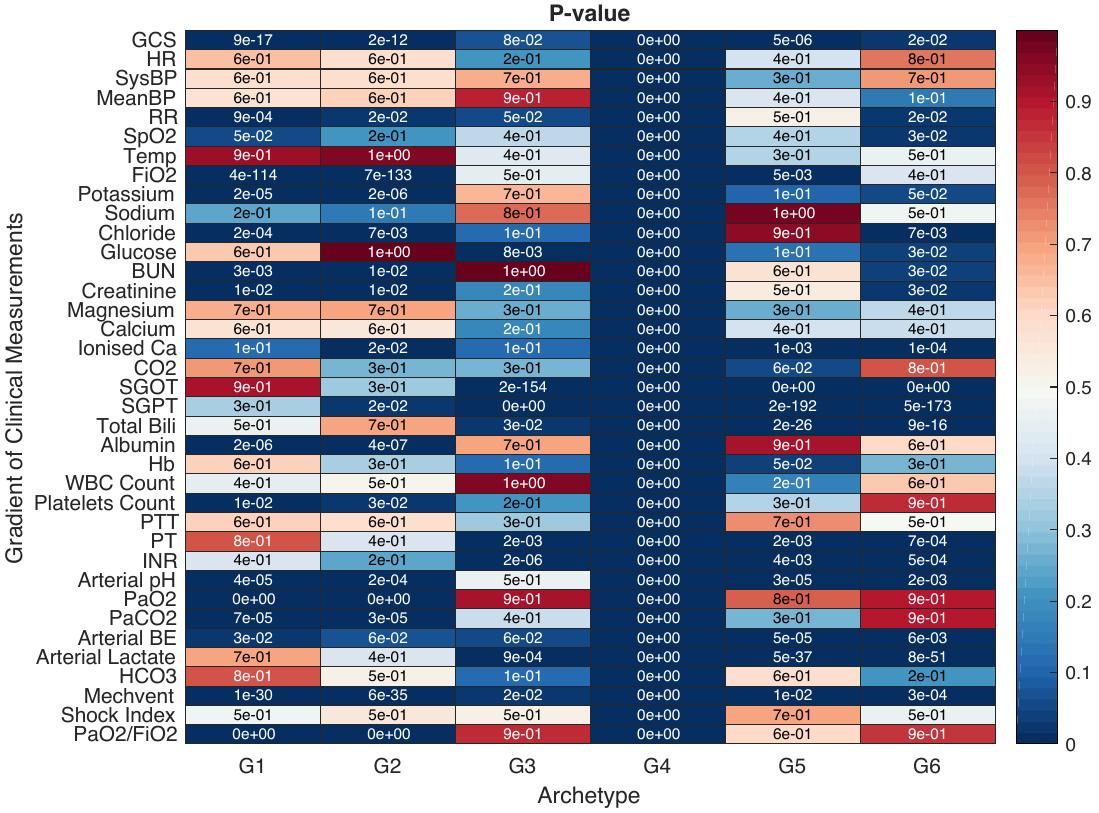}
        \end{subfigure}
    
        \caption{ \textcolor{black}{\textbf{(\textcolor{black}{A})} P-values for the pairwise two-sample t-test for the gradients of clinical measurements (row) of each gradient group (column). Statistically significant entries approach blue.} }\label{fig:pval-gradient} 
\end{figure*}

\begin{table}[h]
\caption{Sequential Organ Failure Assessment Score}\label{tab:SOFA}

\begin{tabular}{lllllll}
\hline
\textbf{SOFA}  & \textbf{Respiratory}                                                & \textbf{Cardiovascular}                                                                                                              & \textbf{Nervous}                                                          & \textbf{Hepatic}                                                    & \textbf{Renal}                                                                                & \textbf{Coagulation}                                                    \\ \hline
\textbf{score} & \textbf{\begin{tabular}[c]{@{}l@{}}PaO2/ FiO2\\ ratio\end{tabular}} & \textbf{\begin{tabular}[c]{@{}l@{}}Mean Arterial \\ Pressure/\\ vasopressors\end{tabular}}                                           & \textbf{\begin{tabular}[c]{@{}l@{}}Glasgow \\ coma \\ score\end{tabular}} & \textbf{\begin{tabular}[c]{@{}l@{}}Bilirubin,\\ mg/dl\end{tabular}} & \textbf{\begin{tabular}[c]{@{}l@{}}Creatinine, \\ mg/dl \\ (or urine \\ output)\end{tabular}} & \textbf{\begin{tabular}[c]{@{}l@{}}Platelets\\ ( $\times$ 10\textsuperscript{3}/mm\textsuperscript{3})\end{tabular}} \\ \hline
0              & $\geq$ 400                                                       & $\geq$ 70 mm/Hg                                                                                                                   & 15                                                                   &$<$ 1.2                                                            &$<$ 1.2                                                                                      & $\geq$ 150                                                           \\ \hline
1              & $<$ 400                                                       & $<$ 70 mm/Hg                                                                                                                   & 13 - 14                                                                   & 1.2 - 1.9                                                           & 1.2 - 1.9                                                                                     & $<$ 150                                                           \\ \hline
2              & $<$ 300                                                       & \begin{tabular}[c]{@{}l@{}}dopamine $\leq$ 5 or dobutamine\\ (any dose)\end{tabular}                                                      & 10 - 12                                                                   & 2 - 5.9                                                             & 2.0 - 3.4                                                                                     & $<$ 100                                                           \\ \hline
3              & $<$ 200                                                       & \begin{tabular}[c]{@{}l@{}}dopamine $>$ 5, epinephrine $\leq$ 0.1 \\ or norepinephrine $\leq$ 0.1\end{tabular}                        & 6 - 9                                                                     & 6 - 11.9                                                            & \begin{tabular}[c]{@{}l@{}}3.5 - 4.9 (or $<$\\ 500 ml/d)\end{tabular}                   & $<$ 50                                                            \\ \hline
4              & $<$ 100                                                       & \begin{tabular}[c]{@{}l@{}}dopamine $>$ 15, epinephrine $>$\\  0.1 or norepinephrine $>$ 0.1\end{tabular} & $<$ 6                                                               & $>$ 12                                                     & \begin{tabular}[c]{@{}l@{}}$>$ 5.0 (or $<$ 200\\ ml/d)\end{tabular}            & $<$ 20                                                            \\ \hline
\end{tabular}

\end{table}

\begin{table}[]
\caption{\textcolor{black}{Pairwise Hotelling t-square test across states.}} \label{fig:t2-test}
\scriptsize
\centering
\begin{tabular}{|c|c|c|c|c|c|c|c|}
\hline
\textbf{T2 value (p-value)} & \textbf{Overall}           & \textbf{A1}                & \textbf{A2}                & \textbf{A3}                & \textbf{A4}                & \textbf{A5}                & \textbf{A6}                \\ \hline
\textbf{Overall}            & N/A (N/A)                  & 2524.91 ( \textless 1e-6)  & 22307.48 ( \textless 1e-6) & 47071.87 ( \textless 1e-6) & 23924.82 ( \textless 1e-6) & 6806.29 ( \textless 1e-6)  & 20907.46 ( \textless 1e-6) \\ \hline
\textbf{A1}                 & 2524.91 ( \textless 1e-6)  & N/A (N/A)                  & 26138.32 ( \textless 1e-6) & 56868.21 ( \textless 1e-6) & 26715.05 (1e-6)            & 7706.59 ( \textless 1e-6)  & 25410.95 ( \textless 1e-6) \\ \hline
\textbf{A2}                 & 22307.48 ( \textless 1e-6) & 26138.32 ( \textless 1e-6) & N/A (N/A)                  & 48629.09 ( \textless 1e-6) & 38956.26 ( \textless 1e-6) & 13818.11 ( \textless 1e-6) & 32865.23 ( \textless 1e-6) \\ \hline
\textbf{A3}                 & 47071.87 ( \textless 1e-6) & 56868.21 ( \textless 1e-6) & 48629.09 ( \textless 1e-6) & N/A (N/A)                  & 35331.04 ( \textless 1e-6) & 10601.98 ( \textless 1e-6) & 29424.46 ( \textless 1e-6) \\ \hline
\textbf{A4}                 & 23924.82 ( \textless 1e-6) & 26715.05 ( \textless 1e-6) & 38956.26 ( \textless 1e-6) & 35331.04 ( \textless 1e-6) & N/A (N/A)                  & 2616.32 ( \textless 1e-6)  & 2258.40 ( \textless 1e-6)  \\ \hline
\textbf{A5}                 & 6806.29 ( \textless 1e-6)  & 7706.59 ( \textless 1e-6)  & 13818.11 ( \textless 1e-6) & 10601.98 ( \textless 1e-6) & 2616.32 ( \textless 1e-6)  & N/A (N/A)                  & 2378.98 ( \textless 1e-6)  \\ \hline
\textbf{A6}                 & 20907.46 ( \textless 1e-6) & 25410.95 ( \textless 1e-6) & 32865.23 ( \textless 1e-6) & 29424.46 ( \textless 1e-6) & 2258.40 ( \textless 1e-6)  & 2378.98 ( \textless 1e-6)  & N/A (N/A)                  \\ \hline
\end{tabular}
\end{table}

\begin{table}[]
\caption{ \textcolor{black}{Two-sample t-test for each variable for each sepsis state compared to overall populations.}} \label{fig:t-test}
\centering

\begin{tabular}{lllllll}
\hline
\textbf{p-value}           & \textbf{A1} & \textbf{A2} & \textbf{A3} & \textbf{A4} & \textbf{A5} & \textbf{A6} \\ \hline
\textbf{Demographic}         &             &             &             &             &             &             \\ \hline
Age                 & 0.95        & 5.36e-104   & 2.62e-001   & 5.64e-004   & 2.81e-037   & 5.78e-031   \\ \hline
Gender              & 0.95        & 1.04e-015   & 2.20e-007   & 4.16e-002   & 2.57e-001   & 3.74e-008   \\ \hline
\textbf{Vitals}     &             &             &             &             &             &             \\ \hline
HR                  & 0.41        & 5.96e-111   & 1.30e-019   & 3.98e-003   & 2.89e-004   & 2.09e-002   \\ \hline
SysBP               & 0.67        & 8.81e-004   & 5.0465e-003 & 3.89e-001   & 9.42e-001   & 7.53e-002   \\ \hline
MeanBP              & 0.51        & 7.53e-010   & 1.1781e-003 & 1.82e-002   & 3.56e-004   & 9.05e-001   \\ \hline
DiaBP               & 0.20        & 1.46e-016   & 1.8983e-009 & 2.18e-002   & 3.98e-004   & 8.64e-001   \\ \hline
Temp                & 0.98        & 4.97e-021   & 2.73e-007   & 6.84e-006   & 5.09e-002   & 6.38e-001   \\ \hline
RR                  & 0.77        & 4.29e-048   & 4.39e-020   & 4.79e-003   & 1.13e-002   & 1.24e-002   \\ \hline
\textbf{Lab Values} &             &             &             &             &             &             \\ \hline
GCS                 & 0.22        & 1.19e-006   & 1.64e-161   & 1.03e-068   & 8.08e-011   & 1.12e-005   \\ \hline
SpO2                & 0.96        & 2.99e-006   & 1.87e-005   & 2.06e-028   & 3.01e-005   & 1.32e-010   \\ \hline
FiO2                & 0.82        & 1.19e-003   & 0.00e+000   & 1.36e-024   & 1.61e-001   & 5.59e-001   \\ \hline
Sodium              & 0.41        & 7.64e-003   & 7.92e-014   & 1.01e-001   & 6.40e-001   & 6.30e-001   \\ \hline
Chloride            & 0.14        & 9.19e-020   & 4.4156e-003 & 4.85e-030   & 8.56e-004   & 5.38e-003   \\ \hline
Potassium           & 0.02        & 6.47e-031   & 5.1573e-015 & 1.54e-040   & 6.68e-001   & 2.01e-018   \\ \hline
Glucose             & 0.76        & 1.28e-006   & 8.69e-002   & 1.23e-013   & 9.82e-001   & 1.51e-011   \\ \hline
BUN                 & 0.31        & 9.83e-018   & 2.08e-014   & 6.06e-010   & 3.05e-002   & 9.28e-001   \\ \hline
Creatinine          & 0.99        & 2.49e-022   & 7.45e-001   & 2.03e-016   & 1.31e-004   & 6.55e-006   \\ \hline
Magnesium           & 0.93        & 1.13e-001   & 3.55e-013   & 8.52e-006   & 4.09e-001   & 1.03e-035   \\ \hline
Calcium             & 0.31        & 4.87e-006   & 3.7620e-009 & 9.46e-014   & 1.18e-004   & 8.77e-003   \\ \hline
Ionised Ca          & 0.63        & 9.93e-001   & 1.30e-005   & 6.17e-036   & 1.09e-015   & 5.05e-009   \\ \hline
CO2                 & 0.12        & 3.27e-006   & 1.32e-033   & 1.29e-020   & 1.15e-003   & 1.34e-017   \\ \hline
SGOT                & 6.19e-161   & 1.71e-003   & 1.12e-011   & 0.00e+000   & 0.00e+000   & 0.00e+000   \\ \hline
SGPT                & 1.05e-120   & 1.17e-002   & 4.6993e-012 & 0.00e+000   & 0.00e+000   & 0.00e+000   \\ \hline
Total Bilirubin     & 0.81        & 2.14e-005   & 1.0579e-009 & 5.48e-075   & 2.29e-003   & 5.82e-015   \\ \hline
Albumin             & 0.24        & 1.18e-017   & 2.11e-082   & 9.48e-002   & 6.38e-001   & 1.82e-001   \\ \hline
Hb                  & 0.79        & 2.77e-099   & 9.21e-014   & 2.90e-002   & 1.78e-008   & 1.42e-006   \\ \hline
WBC                 & 0.04        & 0.00e+000   & 7.8681e-027 & 1.05e-003   & 4.15e-002   & 9.49e-002   \\ \hline
Platelets           & 1.29e-042   & 0.00e+000   & 6.37e-001   & 1.68e-023   & 5.17e-006   & 4.41e-006   \\ \hline
aPTT                & 0.53        & 4.94e-001   & 9.25e-001   & 1.99e-025   & 8.19e-044   & 2.42e-007   \\ \hline
PT                  & 0.13        & 6.18e-002   & 1.5697e-003 & 8.92e-099   & 1.31e-003   & 1.52e-159   \\ \hline
INR                 & 0.06        & 9.81e-003   & 1.09e-002   & 4.02e-104   & 4.08e-082   & 2.18e-226   \\ \hline
Arterial PH         & 0.73        & 5.10e-007   & 2.51e-004   & 1.53e-057   & 2.62e-004   & 2.55e-016   \\ \hline
PaCO2               & 0.34        & 4.82e-001   & 2.10e-011   & 4.52e-002   & 3.32e-005   & 3.01e-004   \\ \hline
PaO2                & 1.22e-120   & 3.54e-002   & 0.00e+000   & 4.28e-001   & 3.46e-001   & 6.13e-002   \\ \hline
Arterial BE         & 0.42        & 5.76e-005   & 8.02e-001   & 8.40e-047   & 9.01e-014   & 9.25e-026   \\ \hline
Arterial lactate    & 6.81e-005   & 1.39e-006   & 9.14e-002   & 0.00e+000   & 8.51e-078   & 6.81e-178   \\ \hline
HCO3                & 0.36        & 5.98e-004   & 8.8338e-003 & 4.55e-032   & 3.27e-007   & 4.44e-024   \\ \hline
Shock Index         & 0.79        & 1.58e-039   & 4.17e-019   & 1.59e-037   & 1.16e-003   & 2.91e-002   \\ \hline
PaO2/FiO2           & 9.39e-244   & 9.19e-006   & 0.00e+000   & 3.04e-004   & 6.01e-001   & 4.42e-002   \\ \hline
\textbf{Others}     &             &             &             &             &             &             \\ \hline
Weight              & 0.33        & 1.90e-002   & 4.62e-027   & 1.14e-002   & 5.63e-005   & 4.86e-001   \\ \hline
Mechvent            & 1.79e-003   & 4.44e-004   & 0.00e+000   & 2.14e-004   & 4.84e-006   & 7.48e-001   \\ \hline
Comorbidity Count   & 0.09        & 6.92e-082   & 3.36e-027   & 1.76e-019   & 3.15e-002   & 1.62e-002   \\ \hline
\end{tabular}
\end{table}

\begin{table}[]

\caption{\textcolor{black}{Pairwise Two-sample t-test for each clinical variable between MODS group.}} \label{fig:t-test2}
\centering

\begin{tabular}{llll}

\hline
\textbf{p-value}     & \textbf{A4 - A5} & \textbf{A5 - A6} & \textbf{A4 - A6} \\ \hline
\textbf{Demographic} &                  &                  &                  \\ \hline
Age                  & 4.61e-20         & 1.28e-01         & 3.34e-14         \\ \hline
Gender               & 5.14e-02         & 5.83e-03         & 6.86e-09         \\ \hline
\textbf{Vitals}      &                  &                  &                  \\ \hline
HR                   & 9.43e-02         & 2.91e-01         & 6.42e-01         \\ \hline
SysBP                & 7.63e-01         & 8.84e-01         & 8.78e-01         \\ \hline
MeanBP               & 1.79e-04         & 8.77e-03         & 3.19e-01         \\ \hline
DiaBP                & 2.19e-02         & 3.36e-03         & 2.38e-01         \\ \hline
Temp                 & 5.54e-02         & 9.01e-02         & 1.61e-01         \\ \hline
RR                   & 3.99e-01         & 7.98e-01         & 5.18e-01         \\ \hline
\textbf{Lab Values}  &                  &                  &                  \\ \hline
GCS                  & 1.01e-01         & 1.00e-01         & 1.25e-04         \\ \hline
SpO2                 & 4.44e-01         & 5.86e-01         & 9.68e-01         \\ \hline
FiO2                 & 3.20e-03         & 5.69e-01         & 5.72e-05         \\ \hline
Sodium               & 2.82e-01         & 9.72e-01         & 2.67e-01         \\ \hline
Chloride             & 6.23e-02         & 5.55e-01         & 4.45e-03         \\ \hline
Potassium            & 1.73e-06         & 8.20e-07         & 5.64e-01         \\ \hline
Glucose              & 1.91e-02         & 1.06e-05         & 2.13e-12         \\ \hline
BUN                  & 9.38e-08         & 3.57e-02         & 5.97e-04         \\ \hline
Creatinine           & 6.47e-01         & 9.26e-01         & 7.23e-01         \\ \hline
Magnesium            & 1.43e-02         & 3.27e-04         & 3.42e-06         \\ \hline
Calcium              & 1.42e-06         & 2.77e-01         & 1.44e-05         \\ \hline
Ionised Ca           & 1.73e-01         & 2.47e-02         & 2.17e-01         \\ \hline
CO2                  & 2.54e-01         & 4.39e-03         & 2.48e-02         \\ \hline
SGOT                 & 9.94e-249        & 1.24e-192        & 1.11e-31         \\ \hline
SGPT                 & 2.41e-217        & 4.65e-03         & 2.52e-312        \\ \hline
Total Bilirubin      & 3.06e-08         & 3.59e-03         & 2.60e-02         \\ \hline
Albumin              & 2.07e-01         & 1.81e-01         & 7.37e-01         \\ \hline
Hb                   & 1.09e-04         & 3.38e-01         & 3.25e-03         \\ \hline
WBC                  & 2.25e-04         & 3.88e-04         & 8.11e-01         \\ \hline
Platelets            & 5.51e-01         & 6.24e-01         & 2.11e-01         \\ \hline
aPTT                 & 1.20e-01         & 4.00e-01         & 4.82e-01         \\ \hline
PT                   & 1.01e-01         & 2.60e-03         & 3.48e-10         \\ \hline
INR                  & 5.78e-05         & 2.51e-02         & 3.42e-14         \\ \hline
Arterial PH          & 2.58e-03         & 4.02e-02         & 4.33e-01         \\ \hline
PaCO2                & 9.46e-03         & 4.71e-01         & 5.33e-02         \\ \hline
PaO2                 & 2.29e-01         & 5.28e-01         & 3.15e-02         \\ \hline
Arterial BE          & 9.61e-01         & 3.08e-01         & 1.84e-01         \\ \hline
Arterial lactate     & 1.07e-05         & 7.32e-02         & 4.18e-03         \\ \hline
HCO3                 & 4.77e-01         & 1.28e-02         & 2.08e-02         \\ \hline
Shock Index          & 3.07e-01         & 3.88e-01         & 1.21e-02         \\ \hline
PaO2/FiO2            & 2.93e-02         & 3.56e-01         & 1.65e-01         \\ \hline
\textbf{Others}      &                  &                  &                  \\ \hline
Weight               & 4.84e-02         & 4.03e-03         & 9.29e-02         \\ \hline
Mechvent             & 8.41e-02         & 2.75e-02         & 1.52e-06         \\ \hline
Comorbidity Count    & 2.99e-02         & 9.66e-01         & 1.21e-02         \\ \hline
\end{tabular}

\end{table}

\begin{table*}[h]
\caption{List of the top 15 features selected by $Q_j(P_k)$ and $Q'_j(P_k)$ methods, and the features selected by variation test method. }\label{tab:features}
\centering
\begin{tabular}{llllll}
\hline
\textbf{Method \textbackslash Rank} & \textbf{1}      & \textbf{2}  & \textbf{3}  & \textbf{4}      & \textbf{5}  \\ \hline
\textbf{$Q_j(P_k)$}                       & SGOT            & SGPT        & PaO2/FiO2   & PaO2      & Platelets        \\ \hline
\textbf{$Q'_j(P_k)$}                       & SGOT            & SGPT        & PaO2/FiO2   &  Platelets            & PaO2    \\ \hline
\textbf{$Variation$ $Test$}              & SGOT            & SGPT        & PaO2/FiO2   & Platelets  & PaO2        \\ \hline
\textbf{Method \textbackslash Rank} & \textbf{6}      & \textbf{7}  & \textbf{8}  & \textbf{9}      & \textbf{10} \\ \hline
\textbf{$Q_j(P_k)$}                       & Arterial lactate         & FiO2         & WBC Count          & INR               & Mechvent         \\ \hline
\textbf{$Q'_j(P_k)$}                       & Glucose & Age       & PT    & HR            & PTT    \\ \hline
\textbf{$Variation$ $Test$}              & Arterial lactate & FiO2       & PT          & INR             & Mechvent    \\ \hline
\textbf{Method \textbackslash Rank} & \textbf{11}     & \textbf{12} & \textbf{13} & \textbf{14}     & \textbf{15} \\ \hline
\textbf{$Q_j(P_k)$}                       & PT        & GCS    & Age         & HR & Comorbidity  Count       \\ \hline
\textbf{$Q'_j(P_k)$}                       & WBC Count              & Weight         & BUN         & Arterial lactate              & DiaBP  \\ \hline
\textbf{$Variation$ $Test$}              & WBC Count        & GCS         & Creatinine  & N/A             & N/A         \\ \hline
\end{tabular}

\end{table*}

\begin{table}[]
\vspace{-1.0cm}
\caption{Definitions of Comorbidities.}\label{tab:Comorbidity}
\small

\begin{tabular}{@{}lll@{}}
\toprule
Comorbidity                                                                                                               & ICD-9-CM Codes                                                                                                                                                                 & \begin{tabular}[c]{@{}l@{}}DRG Screen: Case Does Not Have \\ the Following Disorders (DRG):\end{tabular} \\ \midrule
Congestive heart failure                                                                                                  & \begin{tabular}[c]{@{}l@{}}398.91, 402.11, 402.91, 404.11, 404.13, 404.91,\\ \quad404.93,428.0--428.9\end{tabular}                                                                   & Cardiac$^a$                                                                                                  \\
Cardiac arrhythmias                                                                                                       & \begin{tabular}[c]{@{}l@{}}426.10, 426.11, 426.13, 426.2--426.53,\\ \quad426.6--426.89, 427.0, 427.2, 427.31,\\ \quad427.60,427.9, 785.0, V45.0, V53.3\end{tabular}                        & Cardiac$^a$                                                                                                  \\
Valvular disease                                                                                                          & \begin{tabular}[c]{@{}l@{}}093.20--093.24, 394.0--397.1, 424.0--424.91,\\ \quad 746.3--746.6,V42.2,V43.3\end{tabular}                                                                    & Cardiac$^a$                                                                                                  \\
Pulmonary circulation disorders                                                                                           & 416.0--416.9, 417.9                                                                                                                                                             & Cardiac$^a$ or COPD                                                                                          \\
Peripheral vascular disorders                                                                                             & \begin{tabular}[c]{@{}l@{}}440.0--440.9, 441.2, 441.4, 441.7, 441.9,\\ \quad 443.1--443.9, 447.1,557.1,557.9, V43.4\end{tabular}                                                       & Peripheral vascular (130--131)                                                                           \\
\begin{tabular}[c]{@{}l@{}}Hypertension (combined)\\ Hypertension, uncomplicated\\ Hypertension, complicated\end{tabular} & \begin{tabular}[c]{@{}l@{}}\\401.1, 401.9\\ 402.10, 402.90, 404.10, 404.90, 405.11, 405.19,\\ \quad 405.91, 405.99\end{tabular}                                                        & \begin{tabular}[c]{@{}l@{}}Hypertension (134)\\ Hypertension (134) or cardiac$^a$ or renal$^a$ \end{tabular}      \\
Paralysis                                                                                                                 & 342.0--342.12, 342.9--344.9                                                                                                                                                      & Cerebrovascular (5, 14--17)                                                                              \\
Other neurological disorders                             
& \begin{tabular}[c]{@{}l@{}}331.9, 332.0, 333.4,333.5,334.0--335.9,340,\\ \quad 341.1--341.9,345.00--345.11, \\ \quad 345.40--345.51, 345.80-345.91, 348.1, \\ \quad 348.3, 780.3, 784.3\end{tabular} & Nervous system (1--35)                                                                                   \\
Chronic pulmonary disease                                                                                                 & \begin{tabular}[c]{@{}l@{}}490--492.8, 493.00--493.91, 494, 495.0--505,\\ \quad 506.4\end{tabular}                                                                                      & COPD (88) or asthma (96--88)                                                                             \\
Diabetes, uncomplicated$^b$                                                                                                   & 250.00--250.33                                                                                                                                                                  & Diabetes (294--295)                                                                                      \\
Diabetes, complicated$^b$                                                                                                     & 250.40--250.73, 250.90-250.93                                                                                                                                                   & Diabetes (294--295)                                                                                      \\
Hypothyroidism                                                                                                            & 243--244.2, 244.8, 244.9                                                                                                                                                        & Thyroin (290) or endocrine (300--301)                                                                    \\
Renal failure                                                                                                             & \begin{tabular}[c]{@{}l@{}}403.11, 403.91, 404.12, 404.92, 585, 586,\\ \quad V42.0,V45.1,V56.0,V56.8\end{tabular}                                                                    & Kidney transplant (302) or renal failure/dialysis (316--317)                                             \\
Liver disease                                                                                                             & \begin{tabular}[c]{@{}l@{}}070.32, 070.33, 070.54, 456.0, 456.1, 456.20,\\ \quad 456.21 571.0, 571.2, 571.3,\\ \quad 571.40--571.49, 571.5, 571.6, 571.8,\end{tabular}                      & Liver$^a$                                                                                                    \\
\begin{tabular}[c]{@{}l@{}}Peptic ulcer disease excluding\\ \quad bleeding\end{tabular}                                         & \begin{tabular}[c]{@{}l@{}}531.70, 531.90, 532.70, 532.90, 533.70,\\ \quad 533.90,534.70,534.90, V12.71\end{tabular}                                                                 & GI hemorrhage or ulcer (174--178)                                                                        \\
AIDS$^b$                                                                                                                      & 042--044.9                                                                                                                                                                      & HIV (488--490)                                                                                           \\
Lymphoma                                                                                                                  & \begin{tabular}[c]{@{}l@{}}200.00--202.38, 202.50--203.01,203.8--203.81,\\ \quad 238.6, 273.3,V10.71,V10.72,V10.79\end{tabular}                                                         & Leukemia/lymphoma$^a$                                                                                        \\
Metastatic cancer$^b$                                                                                                         & 196.0--199.1                                                                                                                                                                    & Cancer$^a$                                                                                                   \\
\begin{tabular}[c]{@{}l@{}}Solid tumor without\\ \quad metastasis$^b$\end{tabular}                                                  & \begin{tabular}[c]{@{}l@{}}140.0--172.9,174.0--175.9,179--195.8,\\ \quad V10.00--V10.9\end{tabular}                                                                                      & Cancer$^a$                                                                                                   \\
\begin{tabular}[c]{@{}l@{}}Rheumatoid arthritis/collagen\\ vascular diseases\end{tabular}                                 & \begin{tabular}[c]{@{}l@{}}701.0, 710.0--710.9, 714.0--714.9,\\ \quad 720.0--720.9, 725\end{tabular}                                                                                    & Connective tissue (240--241)                                                                             \\
Coagulopathy                                                                                                              & 2860--2869, 287.1, 287.3--287.5                                                                                                                                                  & Coagulation (397)                                                                                        \\
Obesity                                                                                                                   & 278.0                                                                                                                                                                          & Obesity procedure (288) or nutrition/metabolic (296--298)                                                \\
Weight loss                                                                                                               & 260--263.9                                                                                                                                                                      & Nutrition/metabolic (296--298)                                                                           \\
Fluid and electrolyte disorders                                                                                           & 276.0--276.9                                                                                                                                                                    & Nutrition/metabolic (296--298)                                                                           \\
Blood loss anemia                                                                                                         & 2800                                                                                                                                                                           & Anemia (395--396)                                                                                        \\
Deficiency anemias                                                                                                        & 280.1--281.9, 285.9                                                                                                                                                             & Anemia (395--396)                                                                                        \\
Alcohol abuse                                                                                                             & \begin{tabular}[c]{@{}l@{}}291.1, 291.2, 291.5, 291.8, 291.9,\\ \quad 303.90-303.93,305.00--305.03, V113\end{tabular}                                                                 & Alcohol or drug (433--437)                                                                               \\
Drug abuse                                                                                                                & \begin{tabular}[c]{@{}l@{}}292.0, 292.82--292.89,292.9,304.00--304.93,\\ \quad 305.20-305.93\end{tabular}                                                                              & Alcohol or drug (433--437)                                                                               \\
Psychoses                                                                                                                 & 295.00--298.9, 299.10--299.11                                                                                                                                                    & Psychose (430)                                                                                           \\
Depression                                                                                                                & 300.4, 301.12, 309.0, 309.1, 311                                                                                                                                               & Depression (426)                                                                                         \\ \bottomrule
\end{tabular}

\footnotesize{\quad ICD-9-CM, International Classification of Diseases, 9th Revision, Clinical Modification; DRG, diagnosis-related
group; COPD, chronic obstructive pulmonary disease; GI, gastrointestinal; AIDS, acquired immune deficiency
syndrome; HIV, human immunodeficiency virus.\\
\quad $^a$ Definitions of DRG groups: Cardiac: DRGs 103--108, 110--112, 115--118, 120-127, 129, 132--133, 135-143; Renal:
DRGs 302-305, 315--333; Liver: DRGs 199-202, 205-208; Leukemia/lymphoma: DRGs 400-414, 473, 492; Cancer:
DRGs 10, 11, 64, 82, 172, 173, 199, 203, 239, 257--260, 274, 275, 303, 318, 319, 338, 344, 346, 347, 354, 355, 357, 363,
366, 367, 406--414. \\
$^b$ A hierarchy was established between the following pairs of comorbidities: If both uncomplicated diabetes and
complicated diabetes are present, count only complicated diabetes. If both solid tumor without metastasis and
metastatic cancer are present, count only metastatic cancer.\\
$^*$This table is adopted from A., Steiner et al. \cite{elixhauser1998comorbidity}.}\\
\end{table}

\end{document}